\documentclass[a4paper,11pt]{article}
\usepackage{bm}
\usepackage[normalem]{ulem}
\usepackage[left=2.5cm,right=2.5cm,top=2cm,bottom=2.5cm]{geometry}
\usepackage{epsfig,amsfonts,amssymb}
\usepackage{xcolor}
\usepackage{amsmath}
\usepackage{framed}
\usepackage{cite}
\usepackage{varioref}
\colorlet{shadecolor}{lightgray}
\usepackage{hyperref}
\usepackage{mathrsfs}
\usepackage{mathtools}
\usepackage{parskip}
\usepackage{subcaption}

\def \be {\begin{equation}}
\def \ee {\end{equation}}
\def \bea {\begin{eqnarray}}
\def \eea {\end{eqnarray}}
\def \nn {\nonumber}

\labelformat{equation}{Eq.~(#1)} 

\allowdisplaybreaks

\begin{document}

\baselineskip 24pt

\begin{center}

{\Large \bf Double soft graviton factors from the gravitational Wilson line
%Double soft graviton factor from the generalized Wilson line
}

\end{center}

\vskip .6cm
\medskip

\vspace*{4.0ex}

\baselineskip=18pt

\centerline{\large \rm Karan Fernandes$^{a,b}$, Feng-Li Lin$^{a,b}$ and Chris D. White$^{c}$}

\vspace*{4.0ex}
\centerline{\large \it ~$^a$Department of Physics, National Taiwan Normal University, Taipei, 11677, Taiwan}

\centerline{\large \it ~$^b$Center of Astronomy and Gravitation, National Taiwan Normal University, Taipei 11677, Taiwan}

\centerline{\large \it ~$^c$Centre for Theoretical Physics, School of Physical and Chemical Sciences,}
\centerline{\large \it Queen Mary University of London, Mile End Road, London E1 4NS, UK}

\vspace*{1.0ex}
\centerline{\small E-mail: karanfernandes86@gmail.com, fengli.lin@gmail.com, christopher.white@qmul.ac.uk}

\vspace*{5.0ex}

\centerline{\bf Abstract} \bigskip

The description of low-energy (``soft") gravitons using universal theorems continues to attract attention. In this paper, we consider the emission of two soft gravitons, using a previously developed formalism that describes (next-to) soft graviton emission in terms of generalised Wilson lines (GWLs). Based on Schwinger's proper time methods, the GWL allows for a systematic accounting of graviton emission from external hard particles in the amplitude, as well as from three-graviton vertices located off the individual worldlines. By combining these effects, previously derived results for the leading double soft graviton theorem are recovered. Still, the formalism allows us to go further in deriving new universal double soft graviton terms at subleading order in the momentum expansion. We further demonstrate how gauge invariance can be utilized to account for double soft graviton emissions within the non-radiative amplitude, including the effects of non-zero initial positions of the hard particles. Our results can be packaged into an exponential dressing operator, and we comment on possible applications to the effective field theory for binary scattering processes.

\vfill \eject

\baselineskip 18pt

\tableofcontents

\section{Introduction}

The soft factorization of scattering amplitudes has provided rich insights into the infrared structure of gauge and gravitational theories. This is realized through soft theorems for a particular limit of scattering amplitudes of which one or more particles have asymptotically vanishing energy and momentum. In this limit, the scattering amplitudes factorize into the product of a soft factor and the remaining amplitudes involving only the hard processes~\cite{Weinberg:1965nx,Low:1954kd,Low:1958sn,Burnett:1967km,Cachazo:2014fwa,DiVecchia:2015oba,DiVecchia:2015jaq,DiVecchia:2016amo,Chakrabarti:2017ltl, Marotta:2020oob}. Soft theorems provide the universal soft-factorization of amplitudes; together with the Kinoshita–Lee–Nauenberg theorem \cite{Kinoshita:1962ur, Lee:1964is} and the full IR-factorization theorems (soft and collinear), they explain why IR-divergent parts factorize and cancel in inclusive high-energy observables, to all orders in perturbation theory  \cite{Bloch:1937pw, Collins:1989gx, Becher:2014oda}. Over the past decade, soft theorems have been shown to be equivalent to Ward identities of asymptotic symmetries and late-time radiative observables in perturbative gauge and gravitational theories \cite{Strominger:2017zoo, McLoughlin:2022ljp}. In gravitational theories, the single soft theorems up to leading and subleading order arise as Ward identities of BMS symmetries~\cite{He:2014laa, Kapec:2014opa, Campiglia:2014yka, Hamada:2018vrw}, while the consecutive double soft graviton theorem follows from Ward identities associated with generalized BMS symmetries \cite{Distler:2018rwu}. Likewise,  the Fourier transform of the leading soft graviton factor recovers the linear memory effect \cite{Strominger:2014pwa, DiVecchia:2022nna}. In contrast, the subleading soft factors recover the spin memory effect and tail contributions \cite{Pasterski:2015tva, Laddha:2018rle, Laddha:2018myi, Saha:2019tub, Fernandes:2020tsq} and the non-linear memory associated with collinear double soft gravitons \cite{Fernandes:2024xqr} in the late time gravitational waves from scattering processes. We also note that an analytic expression of the one-loop corrected post-Minkowski waveform from binary black hole encounters can be determined from its soft expansion, and agrees with the soft graviton factor expansion up to sub-sub-leading order~\cite{Elkhidir:2023dco,Brandhuber:2023hhy,Herderschee:2023fxh,Georgoudis:2023lgf,Bini:2023fiz,Georgoudis:2023eke,Georgoudis:2023ozp,Bini:2024rsy}.

One can also understand the important role of factorization in establishing the infrared finiteness of physical observables in gauge and gravitational theories through resummation techniques~\cite{Bloch:1937pw,Yennie:1961ad,Sterman:1986aj,Catani:1989ne,Korchemsky:1993uz,Korchemsky:1992xv,Bauer:2001yt,Bauer:2000yr,Bauer:2002nz,Becher:2006nr,Dixon:2008gr,Laenen:2008gt,Becher:2009qa,Naculich:2011ry,White:2011yy,Melville:2013qca,Luna:2016idw,Becher:2014oda,Bonocore:2015esa,Rothstein:2016bsq}, which provide the following structure.
A given amplitude $\mathcal{A}^{(n)}$ with $n$ hard external particles can be decomposed as a the product of an infrared divergent soft function $\mathcal{S}_n$, and an infrared finite hard process $\mathcal{H}_n$ \footnote{In general, the factorized form also involves an additional jet function which describe collinear divergences of the theory. However, as we are interested in gravitational theories in which (hard) collinear singularities vanish~\cite{Akhoury:2011kq}, we need not consider this complication.}
\be\label{An_0}
\mathcal{A}^{(n)} = \mathcal{S}_n \times \mathcal{H}_n. 
\ee
This simple factorised form holds in the so-called eikonal approximation, in which all emitted radiation has strictly zero 4-momentum. The soft function can then be given an operator interpretation in terms of Wilson lines describing the emission of soft gauge bosons and their exchanges between the external particles of the amplitude. For instance, in gravitationally interacting theories, with coupling $\kappa^2 = 8 \pi G_N$ in terms of the Newton constant $G_N$, we have
\begin{align}
\mathcal{S}_n &= \left\langle 0 \left| \prod_{i=1}^n \Phi_i\left(0\,, \infty\right) \right| 0 \right\rangle\;, \label{sf.lead}
\end{align}
where
\begin{equation}
\Phi_i\left(a\,, b\right)  = \mathcal{P} \exp \left( i \kappa \int_a^b ds \; p_i^{\mu} p_i^{\nu} h_{\mu \nu}(p_i s) \right)\; \label{wl.lead}
\end{equation}
is the appropriate Wilson line operator describing soft graviton emission~\cite{Naculich:2011ry,White:2011yy}, along a straight-line classical trajectory $x_i^\mu=p_i^\mu s$. Both factorisation and Wilson lines have straightforward physical interpretations. The former arises from the fact that soft radiation has an infinite Compton wavelength, and thus cannot resolve the details of the underlying hard interaction. Wilson lines then arise, given that hard particles emitting soft radiation cannot recoil, and thus can only change by a phase. A Wilson line is then the only gauge covariant possibility. Another way to understand the above results is that the momentum space description of the Wilson line in \ref{wl.lead} recovers the exponential of the leading Weinberg soft graviton factor, and consequently, \ref{sf.lead} has the form of a soft coherent dressing of the hard amplitude. This soft function was argued to provide infrared divergences to all loop orders in~\cite{Naculich:2011ry,White:2011yy}. Hence, exponentiated soft factors also arise from the infrared factorization properties of scattering amplitudes, ensuring the finiteness of physical observables to all perturbative orders in gauge and gravitational theories. 

\begin{figure}
\begin{center}
    \scalebox{0.8}{\includegraphics{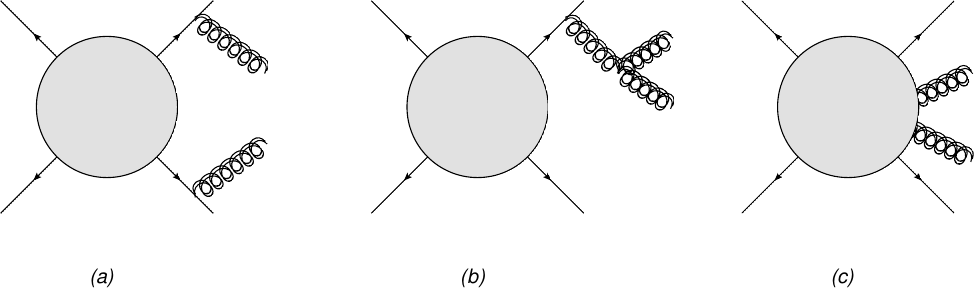}}
    \caption{(a) Example contribution to the (generalised) soft function, in which gravitons are emitted from hard (Wilson) lines only; (b) example contribution involving a multigraviton vertex in the bulk; (c) contribution involving emission of soft gravitons from within the hard interaction.}
    \label{fig:external}
\end{center}
\end{figure}

Motivated by the link between infrared singularities and resummation, \cite{Laenen:2008gt,White:2011yy} developed a formalism for describing radiation beyond the strict soft approximation (see \cite{Akhoury:2011kq,Akhoury:2013yua,White:2011yy,Melville:2013qca,Luna:2016idw,Beneke:2021ilf,Beneke:2021umj,Beneke:2021aip,Beneke:2022pue,Beneke:2022ehj,Bonocore:2020xuj,Bonocore:2021qxh,Grunberg:2009yi,Soar:2009yh, Moch:2009hr,Moch:2009mu,deFlorian:2014vta,Presti:2014lqa,vanBeekveld:2023gio,Agarwal:2023fdk,Buonocore:2023mne,Bonocore:2015esa,Bonocore:2016awd,Gervais:2017yxv,Gervais:2017zky,Gervais:2017zdb,Laenen:2020nrt,DelDuca:2017twk,vanBeekveld:2019prq,Bonocore:2014wua,Bahjat-Abbas:2018hpv,Bahjat-Abbas:2019fqa,Engel:2021ccn,Bonocore:2021cbv,Engel:2023ifn,Kolodrubetz:2016uim,Moult:2016fqy,Feige:2017zci,Beneke:2017ztn,Beneke:2018rbh,Beneke:2019kgv,Bodwin:2021epw,Moult:2019mog,Beneke:2019oqx,Boughezal:2016zws,Moult:2017rpl,Chang:2017atu,Moult:2018jjd,Beneke:2018gvs,Ebert:2018gsn,Beneke:2019mua,Moult:2019uhz,Beneke:2020ibj} for alternative approaches and related works in both gauge theory and gravity, and \cite{White:2022wbr} for a pedagogical review). They used Schwinger proper time methods to express the propagators for hard particles in a background gauge field as (first-quantised) path integrals over their trajectories. Each path integral can be systematically carried out about the classical trajectory, such that the leading contributions correspond to the eikonal approximation described above. The first-order corrections then correspond to the emission of {\it next-to-soft} radiation, which is described by a generalized Wilson line (GWL). Consequently, \ref{An_0} receives corrections at subleading order in the momentum expansion, where the resulting formula has the schematic form
\begin{align}
\mathcal{A}^{(n)} &= \tilde{S}_n \times \mathcal{H}_n \times(1 + R_n) \label{amp.sl}.
\end{align}
Here, the soft function of \ref{sf.lead} has been replaced by a generalised soft function, which can be formally defined as a vacuum expectation value (VEV) of generalised Wilson lines, analogous to the leading soft case. It is well-known that VEVs of Wilson lines exponentiate, and the same can be proven for generalised Wilson lines~\cite{Laenen:2008gt,White:2011yy}. Thus, the generalised soft function assumes the exponential form
\begin{align}
\tilde{S}_n &= \exp\left[\sum_{G^{\text{E}}} G^{\text{E}} + \sum_{G^{\text{NE}}} G^{\text{NE}} +\cdots \right],
\label{gensoft}
\end{align}
where the first term in the exponent corresponds to the eikonal (leading soft) approximation, and the second to the next-to-eikonal (next-to-soft) behaviour. Broadly speaking, there are two types of contribution to the (next-to) soft function. Firstly, there are contributions involving only emissions of gauge bosons from the Wilson lines themselves, as exemplified by fig~\ref{fig:external}(a). At the next-to-soft level and beyond, these contributions also involve multiple gauge bosons emitted from the same point along the Wilson line. Secondly, there are contributions involving multiple boson vertices located off the Wilson lines, such as that shown in fig~\ref{fig:external}(b). The latter terms arise naturally upon calculating VEVs such as that of \ref{sf.lead}, given that there is an implicit factor of $e^{iS}$, where $S$ is the bulk action for the theory of interest. 

As well as a generalisation of the soft function, \ref{amp.sl} also includes a remainder term $R_n$, that appears at subleading order in the soft expansion. As discussed in \cite{Laenen:2008gt,White:2011yy}, the remainder term arises from soft graviton emissions from inside the hard interaction, with an example shown in fig~\ref{fig:external}(c). 
This thus formally breaks the factorization of the total amplitude into hard and soft contributions, which is expected at subleading order in the soft expansion: gauge bosons acquire a finite Compton wavelength, allowing them to resolve the underlying hard process. However, such contributions can be fixed on general terms, given that they are linked by gauge invariance with external emission effects. Historically, this result became known as the Low-Burnett-Kroll (LBK) theorem \cite{Low:1954kd,Low:1958sn,Burnett:1967km} (see \cite{DelDuca:1990gz} for an extension to massless particles, and \cite{Gross:1968in} for the first generalisation to gravity), and the final result for internal emission contributions assumes the form 
of angular momentum operators acting on the hard amplitude. 

Although phrased in terms of (generalised) Wilson lines, \ref{amp.sl} indeed reproduces known (next-to) soft theorems from the more formal literature (see e.g. \cite{White:2014qia} for a detailed discussion of single soft theorems). To begin with, we can consider \ref{amp.sl} without the remainder, and on expanding the soft function, the eikonal contribution in the exponential $G^{\text{E}}$ recovers the leading soft theorem for scattering amplitudes. Keeping also next-to-eikonal terms, one should recover the known simultaneous double soft factor for scattering amplitudes~\cite{Chakrabarti:2017ltl,Distler:2018rwu}. Whilst partial progress has been made in previous work~\cite{White:2011yy,Bonocore:2021qxh}, which we review in Appendix \ref{app1}, a full derivation of this result requires carefully keeping track of all contributions, including the three-graviton vertex contributions exemplified by fig.~\ref{fig:external}(b). We will provide such a derivation in this paper, showing full equivalence between the results obtained using the generalised Wilson line approach and those in \cite{Chakrabarti:2017ltl,Distler:2018rwu}.

Having reproduced previous results, the GWL approach will allow us to go further. That is, we may also include up to two soft emissions from inside the hard interaction (fig.~\ref{fig:external}(c)), via a suitably generalised form of the LBK approach. The result will be a universal expression for the subleading consecutive double soft factor (in the soft expansion), and the technical details of our calculation will be presented in Appendix \ref{app2}. To our knowledge, this result has not been previously derived, and it is interesting to examine the structure of our results. One might assume that all terms derived using the LBK theorem should be contained in the remainder function. However, a key feature of the double soft factor is that it involves products of single soft factors, which would agree with the expansion of exponentiated subleading single soft factors. Thus, in Section 4, we use the form of the gauge invariant double soft graviton factor, up to subleading order, to motivate a new soft function for the amplitude, which is a double soft graviton dressing involving exponentiated subleading soft factors. 

Our paper thus identifies new universal terms in the exponent of the (next-to) soft function of factorized amplitudes in gravitational theories, up to double soft graviton emissions. Furthermore, we expect the exponentiation of all types of contribution in fig.~\ref{fig:external} to carry over to the entire multiple soft graviton expansion in the GWL formalism. We will discuss possible applications of our result to radiative observables in effective field theories for black hole scattering.  

The structure of our paper is as follows. In section~\ref{sec2} we introduce the technicalities of the GWL approach in more detail, and derive explicit results for the external emission contributions involving two (next-to) soft gravitons, as exemplified by fig.~\ref{fig:external}(a) and (b). In section~\ref{sec3}, we consider internal emission contributions involving up to two gravitons, and fix their form by imposing gauge invariance. In section~\ref{sec4}, we combine all contributions and motivate the existence of a general exponentiated operator that acts on the hard amplitude to generate all internal and external emission contributions. We discuss the implications of our results and conclude in section~\ref{sec5}. Appendix \ref{app1} reviews the derivation of GWL up to two soft emissions exemplified in fig~\ref{fig:external}(a). Appendix~\ref{app2} contains a detailed derivation of the subleading double soft factor.

\section{Double soft graviton factor from GWL expectation values} \label{sec2}

Following the scheme outlined above, our aim in this section is to collect all external emission contributions up to two gravitons, including all effects up to next-to-soft level in the momentum expansion of the emitted radiation. We begin by reviewing the necessary details of the generalised Wilson line (GWL) approach developed and explored in \cite{Laenen:2008gt,White:2011yy, Bonocore:2021qxh}. To obtain an explicit form for this, one may perform a conventional weak-field expansion of the metric according to
\begin{equation}
    g_{\mu \nu} = \eta_{\mu \nu} + 2 \kappa h_{\mu \nu},\quad
    \kappa^2 = 8 \pi G_N,
    \label{graviton}
\end{equation}
where $G_N$ is Newton's constant, and we will write the right-hand side schematically as $\eta+h$ in what follows. In general, $h_{\mu\nu}$ should represent both hard and soft gravitons. For our purpose, we will assume that $h_{\mu\nu}$ represents only the soft graviton from now on. 

\begin{figure}
\begin{center}
\scalebox{0.8}{\includegraphics{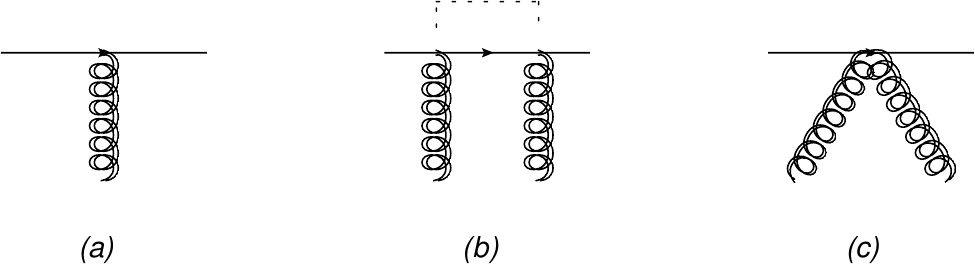}}
\caption{Diagrammatic interpretation of the generalised Wilson exponent of \ref{esf}--\ref{2sf3.int1}: (a) emission of a single graviton; (b) correlated emission of two gravitons from different locations; (c) emission of two gravitons from the same point.}
\label{fig:vertices}
\end{center}
\end{figure}

The starting point is to write the amplitude for $n$ hard particles with momenta $\{p_n\}$, each of which can emit further (next-to) soft radiation, in the form
\begin{equation}
{\cal A}^{(n)}(\{x_i, p_i\})=\int {\cal D} h_{\mu\nu}\, H(\{x_i\}; h)\,
e^{iS_{\rm EH}[\eta + h]}\prod_{j=1}^n
\langle p_j|(\hat{S}-i\epsilon)^{-1}|x_j\rangle\;.
\label{ampfac}
\end{equation}  
where the Einstein-Hilbert action
\begin{equation}
S_{EH}[g] = \frac{1}{2 \kappa^2}\int d^4 x \sqrt{-g} R\;,  
\end{equation}
and $H(\{x_i\}; h)$ is a {\it hard function} that produces particles at definite positions $\{x_i\}$, and whose precise definition may be found in \cite{Laenen:2008gt,White:2011yy}. Associated with each external leg $j$ is the propagator for a hard particle from an initial state with position $x_j$ to a state of definite final momentum $p_j$, where $\hat{S}$ is the operator representing interactions with a background soft graviton field. The $i\epsilon$ term implements the usual Feynman boundary conditions. Finally, there is a path integral over the soft graviton field $h_{\mu\nu}^{(s)}$, weighted appropriately by its Einstein-Hilbert action. As explained in \cite{Laenen:2008gt,White:2011yy}, one may use Schwinger proper time methods to express each propagator as a path integral over the trajectories of each hard particle. Each such path integral may then be carried out perturbatively about the classical straight-line trajectory
\begin{equation}
y^\mu_i(t)=p^\mu_i t+x_i^\mu,
\label{traj}
\end{equation}
which corresponds to the strict eikonal approximation in which all emitted radiation is soft. Subleading corrections to the trajectory amount, in momentum space, to a systematic expansion in the momentum of emitted radiation. Thus, keeping only the first set of subleading corrections is equivalent to the next-to-soft approximation. Carrying out this analysis, the upshot is that \ref{ampfac} assumes the form
\begin{equation}
{\cal A}^{(n)}(\{x_i, p_i\})=\int {\cal D} h_{\mu\nu}\, H(\{x_i\}; h)\,
e^{iS_{\rm EH}[\eta + h]} \prod_{j=1}^n
e^{-ix_j\cdot p_j} f(x_j,p_j; h),
\label{ampfac2}
\end{equation} 
where $f(x_j,p_j; h)$ is a suitable generalisation of the Wilson line operator appearing in \ref{wl.lead}. Results for the generalised Wilson line can then be obtained order-by-order in the graviton $h_{\mu\nu}$, where it turns out to be only necessary to expand to a finite order in $\kappa$ at a given order in the soft expansion, as reviewed in appendix \ref{app1}. One then has
\begin{align}
 f(x_i,p_i; h) 
 = \exp\big[\delta_1(x_i,p_i; h) + \delta_2^{(1)}(x_i,p_i; h)+ \delta_2^{(2)}(x_i,p_i; h)+ \delta_2^{(3)}(x_i,p_i; h)\big] \;,
\label{esf}
\end{align}
where the various contributions to the exponent are given by 
\begin{align}
&\delta_1(x_i,p_i; h) =   i \kappa  \int_0^{\infty} dt\, \left( h_{\mu \nu}(y_i(t)) p_i^{\mu}  p_i^{\nu}  + i \frac{t}{2} \eta_{\alpha \beta}\partial_{i}^{\alpha} \partial_{i}^{\beta}h_{\mu \nu}(y_i(t))p_i^{\mu}  p_i^{\nu} \right. \notag\\
&\left. \qquad \qquad \qquad \qquad \qquad \qquad +i  \left(\partial_i^{\mu}h_{\mu \nu}(y_i(t)) p_i^{\nu} - \frac{1}{2}  \eta_{\mu \nu}\partial_i^{\mu}h(y_i(t)) p_i^{\nu} \right) \right)  \;, \label{1sf.int} \\
&\delta_2^{(1)}(x_i,p_i; h) =  -{i \kappa^2 \over 2}  \int_0^{\infty} dt \int_0^{\infty} dt'\, \eta_{\alpha \beta}\partial_i^{\alpha}h_{\mu \nu}(y_i(t)) \partial_i^{\beta} h_{\rho \sigma}(y_i(t')) p_i^{\sigma} p_i^{\rho} p_i^{\mu} p_i^{\nu} \text{min}(t,t')\;, 
\label{2sf1.int} \\
&\delta_2^{(2)}(x_i,p_i; h) =   - 2 i \kappa^2    \int_0^{\infty} dt \int_0^{\infty} dt'\, h_{\mu \nu}(y_i(t)) h_{\rho \sigma}(y_i(t')) p_i^{\mu} p_i^{\rho} \eta^{\nu \sigma} \delta(t - t')\;,
\label{2sf2.int}\\
&\delta_2^{(3)}(x_i,p_i; h) =  -i \kappa^2   \int_0^{\infty} dt \int_0^{\infty} dt'  \, \left(\partial_i^{\sigma} h_{\mu \nu}(y_i(t)) h_{\rho \sigma}(y_i(t')) p_i^{\mu} p_i^{\nu} p_i^{\rho} \Theta(t - t') \right. \notag\\
&\left. \qquad\qquad\qquad \qquad\qquad \qquad\qquad  +  h_{\mu \nu}(y_i(t)) \partial_i^{\nu} h_{\rho \sigma}(y_i(t')) p_i^{\mu} p_i^{\rho} p_i^{\sigma} \Theta(t' - t) \right)\;.
\label{2sf3.int1}
\end{align}
Here $\partial_i^{\mu} = \frac{\partial}{\partial y_i^{\mu}}$, with $y_i^\mu$ defined according to \ref{traj}. We also remind the reader that the parameters $t$ and $t'$ are distance parameters along each generalised Wilson line. Different terms can then be furnished with a diagrammatic interpretation. \ref{1sf.int}, for example, corresponds to the emission of a single soft graviton from the worldline of the hard particle, as depicted in fig~\ref{fig:vertices}(a). We also note that while the first term in \ref{1sf.int} is the leading single soft graviton contribution, all other terms are subleading in the soft expansion and can contribute through off-shell gravitons when considering three-graviton vertices located off the Wilson lines. By contrast, \ref{2sf1.int} and \ref{2sf3.int1} represent emission of two gravitons from different locations, but where the emissions are somehow correlated, as in fig~\ref{fig:vertices}(b). Finally \ref{2sf2.int}, which contains a delta function setting $t=t'$, corresponds to the emissions of two gravitons from a single point on the worldline, as shown in fig.~\ref{fig:vertices}(c).

To recover and generalise known soft theorems, we must evaluate amplitudes in momentum rather than position space. To this end, 
we may express the gravitons appearing in \ref{1sf.int} - \ref{2sf3.int1} as 
\begin{align}
h_{\mu \nu}(y) &= \int \frac{d^4 k}{(2 \pi)^4} h_{\mu \nu}(k) e^{i k \cdot y}
\label{h.mom}
\end{align}
where, in a slight abuse of notation, we label the momentum modes of the graviton field by $h_{\mu\nu}(k)$, such that the argument of the function denotes whether we are in position or momentum space. In what follows, we will adopt the short-hand notation
\begin{equation}
\int_k\equiv \int \frac{d^4 k}{(2\pi)^4}.
\label{intkdef}
\end{equation}
As the graviton $h_{\mu\nu}$ is soft in our consideration of nearly eikonal scattering, this $ k$-integration is over a soft region with an IR cutoff corresponding to the maximal resolution of the detector.

Then, upon substituting \ref{h.mom} in \ref{esf}, we can evaluate the integrals over time by using an appropriate $i \epsilon$ prescription to ensure well-posed asymptotic solutions. For instance, for the first term in \ref{1sf.int} we have
\begin{align}
i \kappa \lim_{\epsilon \to 0} \int_0^{\infty} dt p_i^{\mu}  p_i^{\nu} \left[ \int_k  h_{\mu \nu}(k)  e^{i k (p_i t + x_i) - \epsilon t} \right]  =  - \kappa \int_{k} \frac{h_{\mu \nu}(k) p_i^{\mu}  p_i^{\nu}}{p_i \cdot k} e^{ik\cdot x_i}.
\label{delta1res}
\end{align}

If we are only interested in external emission contributions, we may take the initial positions of each hard particle $x_i^\mu\rightarrow 0$, and we can then recognise the second line as the known eikonal Feynman rule for soft graviton emission, which thus correctly arises as the lowest-order contribution in the GWL framework.  One may carry out the Fourier transforms of the other contributions in a similar manner. The individual generalised Wilson line factors can then be combined into a single exponential factor, such that \ref{ampfac2} becomes (again taking $x_i^\mu\rightarrow 0$)
\begin{equation}
{\cal A}^{(n)}(\{x_i=0, p_i\})=\int {\cal D} h_{\mu\nu}\, H(\{x_i=0\}; h)\,
e^{iS_{\rm EH}(h_{\mu\nu})} F(\{p_i\}; h),
\label{ampfac3}
\end{equation}
where the overall dressing factor on the right-hand side is given by
\begin{align}
F(\{p_i\};h) &= \exp \left[ \phi_1(\{p_i \};h) + \phi_2(\{p_i \};h) \right]\;, 
\label{sf.res}
\end{align}
with one- and two-graviton contributions
\begin{align}
\phi_1(\{p_i \};h) & =  - \kappa \sum_{i=1}^n \int_{k} h_{\mu \nu}(k)\left(\frac{p_i^{\mu}  p_i^{\nu}}{p_i \cdot k} + \frac{k^2 p_i^{\mu}  p_i^{\nu}}{2 (p_i \cdot k)^2} - \frac{p^{(\mu} k^{\nu)}}{p_i \cdot k} + \frac{\eta^{\mu \nu}}{2} \right)   \;, \label{1s.sf}\\
\phi_2\{p_i \};h) &=  \kappa^2 \sum_{i=1}^n\int_k\int_l \left[\frac{1}{2} \frac{h_{\mu \nu}(k) h_{\rho \sigma}(l) }{p_i\cdot (k+l)} \frac{k\cdot l \, p_i^{\mu} p_i^{\nu} p_i^{\rho} p_i^{\sigma}}{(p_i\cdot k) (p_i\cdot l)} +  2 \frac{h_{\mu \nu}(k) h_{\rho \sigma}(l)}{p_i\cdot (k+l)} p_i^{\mu} p_i^{\rho} \eta^{\nu \sigma} \right. \notag\\
&\left. \qquad\qquad\qquad  -   \frac{h_{\mu \nu}(k) h_{\rho \sigma}(l)}{p_i \cdot (k+l)} \left( \frac{p_i^{\mu} p_i^{\nu} p_i^{\rho} k^{\sigma}}{p_i\cdot k}+  \frac{p_i^{\mu} l^{\nu} p_i^{\rho} p_i^{\sigma}}{p_i\cdot l} \right) \right] \;. \label{2s.sf}
\end{align}

Our results match those of ref.~\cite{Bonocore:2021qxh}. For external emission contributions, one may ignore the dependence of the hard function $H(\{x_i=0\};h)$ on the graviton in \ref{ampfac3}.
Comparing the result with \ref{amp.sl} and recalling that external emission contributions correspond to the generalised soft function, we find
\begin{equation}
\tilde{S}_n=\int {\cal D} h_{\mu\nu}
e^{iS_{\rm EH}[\eta + h]} F(\{p_i\}; h).
\label{Stildedef}
\end{equation}

Carrying out the path integral over the soft graviton $h_{\mu\nu}$ will generate all possible Feynman diagrams in which soft gravitons are either connected directly to the Wilson lines themselves (via the dressing factor $F(\{p_i\};h)$), or to multigraviton vertices in the bulk (via the Einstein-Hilbert action). In order to compute the latter contributions explicitly, we must therefore expand the Einstein-Hilbert action to a sufficient perturbative order in the graviton. In the $(-,+,+,+)$ convention for the metric, we have the Ricci tensor and Ricci scalar given by
\begin{align}
R_{\mu \nu} = \Gamma^{\alpha}_{\mu \nu, \alpha} - \Gamma^{\alpha}_{\mu \alpha, \nu} + \Gamma^{\beta}_{\mu \nu}\Gamma^{\alpha}_{\beta  \alpha} - \Gamma^{\beta}_{\mu \alpha}\Gamma^{\alpha}_{\beta \nu} \,, \qquad R = g^{\mu \nu} R_{\mu \nu}\,,
\end{align}
with the Christoffel symbol
$$\Gamma_{\mu \nu}^{\alpha} = \frac{1}{2}g^{\alpha \beta} \left(g_{\mu \beta, \nu} + g_{\nu \beta, \mu} - g_{\mu \nu , \beta}\right)\,.$$
We also assume the de Donder gauge 
\begin{equation}
 G_{\mu} = \partial^{\alpha} h_{\mu \alpha} - \frac{1}{2} \partial_{\mu} h = 0.
\end{equation}
This amounts to adding the Lagrangian density $\eta^{\mu \nu} G_{\mu} G_{\nu}$ to the Einstein-Hilbert action. By expanding this gauge-fixed action $S^{\rm dD}_{EH}$, we find the following result up to three-graviton terms:
\begin{align}
S^{\rm dD}_{EH}[\eta + h] =  S^{(2)}_{EH}[\eta + h]+  S^{(3)}_{EH}[\eta + h] + \mathcal{O}(h^4)\;, \notag
\end{align}
with
\begin{align}
 S^{(2)}_{EH}[\eta + h] &=  \int d^4 x \frac{\eta^{\mu \nu}}{4} \left(\partial_{\mu} h(x) \partial_{\nu} h(x) - 2 \partial_{\mu} h_{\alpha \beta}(x) \partial_{\nu} h^{\alpha \beta}(x)\right)\;, \label{eh.quad}\\
 S^{(3)}_{EH}[\eta + h] &= 2  \kappa \int d^4 x \left[h^{\mu \nu}(x) \left( \frac{1}{2} \partial_{\mu} h_{\alpha \beta}(x) \partial_{\nu} h^{\alpha \beta}(x) + \partial^{\alpha} h_{\mu \beta}(x) \partial_{\alpha} h_{\nu \gamma}(x)\eta^{\beta \gamma} - \frac{1}{2} \partial^{\alpha} h_{\mu \nu}(x) \partial_{\alpha} h(x) \right. \right.  \notag\\
&\left. \left.  \;   -  \partial_{\alpha} h_{\beta \mu}(x) \partial_{\nu} h^{\alpha \beta}(x) \right) + \frac{1}{8} \eta^{\mu \nu}  h(x)\left(\partial_{\mu} h(x) \partial_{\nu} h(x) - 2 \partial_{\mu} h_{\alpha \beta}(x) \partial_{\nu} h^{\alpha \beta}(x)\right) \right] \,, \label{eh.trip}
\end{align}
where all derivatives are taken with respect to $x$. From the quadratic graviton term in the Einstein-Hilbert action \ref{eh.quad}, we have the following equation for the propagator
\begin{equation}
- i P^{\mu \nu \alpha \beta} \frac{\partial^2}{\partial x^2} \Delta_{\alpha \beta \rho \sigma} (x,y) = I^{\mu \nu}_{\rho \sigma} \delta^{4}(x - y)\;,
\end{equation}
where
\begin{align}
P^{\mu \nu \alpha \beta} &= \frac{1}{2}\left( \eta^{\mu \alpha} \eta^{\nu \beta} + \eta^{\mu \beta} \eta^{\nu \alpha} - \eta^{\mu \nu} \eta^{\alpha \beta} \right)\;, \notag\\
I^{\mu \nu}_{\rho \sigma} &= \frac{1}{2} \left(\delta^{\mu}_{\rho} \delta^{\nu}_{\sigma} + \delta^{\mu}_{\sigma} \delta^{\nu}_{\rho}\right)\;.
\end{align}
This leads to the well-known momentum-space expression 
\begin{equation}
\Delta_{\mu \nu \alpha \beta} (x,y) = - i \int_{\bar{k}} \frac{P_{\mu \nu \alpha \beta}}{\bar{k}^2} e^{i \bar{k} \cdot (x - y)}\;.
\label{grav.prop}
\end{equation}

On similarly Fourier transforming the three-graviton contribution in \ref{eh.trip}, we find
\begin{align}
& S^{(3)}_{EH}[\eta + h] =  - 2  \kappa \int_{K_1} \int_{K_2} \int_{K_3} (2\pi)^4 \delta^4(K_1 + K_2 + K_3) h_{\mu \nu}(K_1) h_{\alpha \beta}(K_2) h_{\gamma \delta}(K_3)\notag\\
&{\rm \bf Sym} \left\{ \frac{P}{3} \left( \frac{1}{2} K_{2}^{ \mu} K_{3}^{\nu} \eta^{\alpha \gamma}  \eta^{\beta \delta} \right) + \frac{P}{3} \left( K_2 \cdot K_3 \eta^{\mu \alpha} \eta^{\nu \gamma}  \eta^{\beta \delta} \right) - \frac{P}{6} \left( \frac{1}{2} K_2 \cdot K_3 \eta^{\mu \alpha} \eta^{\nu \beta}  \eta^{\gamma \delta} \right) \right. \notag\\
& \left. - \frac{P}{6} \left( K_{2}^{ \delta} K_{3}^{\nu} \eta^{\alpha \gamma}  \eta^{\mu \beta} \right)  + \frac{P}{3}\left(\frac{1}{8}K_2 \cdot K_3 \eta^{\mu \nu}  \eta^{\alpha \beta}  \eta^{\gamma \delta} \right) -  \frac{P}{3} \left( \frac{1}{4} K_2 \cdot K_3  \eta^{\mu \nu}  \eta^{\alpha \gamma}  \eta^{\beta \delta} \right) \right\}\,,
\label{3g.m}
\end{align} 
where $P$ denotes the number of independent permutations of $\{\mu\,,\nu\,, K_1\}\,,\{\alpha\,,\beta\,, K_2\}$ and $\{\gamma\,,\delta\,, K_3\}$, and ${\rm \bf Sym}\Big\{ \cdots \Big\}$ the symmetrization over the indices $\{\mu \nu\}\,,\{\alpha\,,\beta\}$ and $\{\gamma\,,\delta\}$. Hence, the last two lines of  \ref{3g.m} are completely symmetrized in the indices and under the exchange of gravitons. 

We now have all the ingredients needed to calculate the generalised soft function of \ref{Stildedef}. To do this, we simply have to generate all possible Feynman diagrams and combine the above vertices and propagators as needed. We note further that, as has already been proven in refs.~\cite{Laenen:2008gt,White:2011yy}, the generalised soft function has an exponential form, where the exponent features only connected diagrams. There are thus three sources of contribution, including up to two graviton emissions:
\begin{enumerate}
    \item[(i)] Emissions of single gravitons from all external lines (by fig.~\ref{fig:vertices}(a));
    \item[(ii)] Emissions of correlated pairs of gravitons from all external lines (by fig.~\ref{fig:vertices}(b) and (c));
    \item[(iii)] Emissions of single gravitons from any external line, which then split into a pair of gravitons via the three-graviton vertex (fig.~\ref{fig:external}(b)).
\end{enumerate}

Contributions of types (i) and (ii) are simply given by the dressing factor of \ref{sf.res}. By a direct calculation, we find that the three-graviton vertex graph (type (iii)) gives the contribution $\bar{\Delta}_2(\{p_i \};h) 
+ \bar{\Upsilon}_{2}(\{p_i \};h)$ to the soft function, with %takes the form
\begin{align}
&\bar{\Delta}_2(\{p_i \};h) 
= - \frac{\kappa^2}{2} \sum_{i=1}^n \int_{kl}  \frac{1}{k\cdot l\, p_i\cdot(k+l)} h_{\mu \nu}(k) h_{\rho \sigma} (l) \left[ - \eta^{\mu \rho} \eta^{\nu \sigma} (p_i\cdot k)(p_i\cdot l) + p_i^{\mu} p_i^{\nu} k^{\rho} k^{\sigma} +  l^{\mu} l^{\nu} p_i^{\rho} p_i^{\sigma} \right. \notag\\
& \left. \qquad \qquad \qquad \qquad  - 2 p_i^{\mu} p_i^{\rho} l^{\nu} k^{\sigma} + \eta^{\nu \sigma} \left( 2 (k \cdot l) p_i^{\mu} p_i^{\rho} + p_i \cdot(k- l) l^{\mu} p_i^{\rho} + p_i \cdot(l - k) p_i^{\mu} k^{\rho}\right) \right]\;,
\label{2s.gp}\\
&\bar{\Upsilon}_{2}(\{p_i \};h) = - \frac{\kappa^2}{2} \sum_{i=1}^n \int_{kl}  \frac{1}{(p_i\cdot(k+l))^2} h_{\mu \nu}(k) h_{\rho \sigma} (l)  \left[ \eta^{\mu \rho} \eta^{\nu \sigma} \left((p_i\cdot k)^2 + (p_i\cdot l)^2+ (p_i\cdot k)(p_i\cdot l) \right) \right. \notag\\
& \left. \qquad\qquad\quad\qquad  + p_i^{\mu} p_i^{\nu} k^{\rho} k^{\sigma} +  l^{\mu} l^{\nu} p_i^{\rho} p_i^{\sigma} - 2 p_i^{\mu} p_i^{\rho} l^{\nu} k^{\sigma} + 2 \eta^{\nu \sigma} \left( (k \cdot l) p_i^{\mu} p_i^{\rho} - (p_i \cdot l) l^{\mu} p_i^{\rho} - (p_i \cdot k) p_i^{\mu} k^{\rho}\right) \right]\,. \label{2s.gpsl}
\end{align}
We note that the three-graviton vertex contribution in \ref{2s.gp} by itself can be compared against the result in \cite{Chakrabarti:2017ltl}, and we find exact agreement. Additionally, we also include another three-graviton vertex contribution in \ref{2s.gpsl}, which is subleading relative to \ref{2s.gp} in the soft expansion, and will be relevant later in deriving the double soft graviton factors.

In summary, the generalized soft function up to the two-graviton level is then given by
\begin{equation}
\tilde{S}_n(\{p_i \};h) = \exp\Big[\phi_1(\{p_i \};h) + \phi_2(\{p_i \};h) + \bar{\Delta}_2(\{p_i \};h) + \bar{\Upsilon}_{2}(\{p_i \};h)\Big],
\label{sf.1}
\end{equation}
with ingredients as given in \ref{1s.sf}, \ref{2s.sf}, \ref{2s.gp} and \ref{2s.gpsl}, and the subscripts denoting the number of gravitons in the final state. 
This completes our derivation of external emission contributions to the soft graviton amplitude.

In the following, we will find it convenient to define
\begin{align}
\bar{\Delta}_2  (\{p_i \}; h) = \Delta_2 (\{p_i \}; h) + \psi(\{p_i\};h) 
\label{2sgp.d}
\end{align}
with
\begin{align}
&\Delta_2 (\{p_i \}; h) = - {\kappa^2 \over 2} \sum_{i=1}^n \int_{kl}  \frac{1}{k\cdot l\, p_i\cdot(k+l)} h_{\mu \nu}(k) h_{\rho \sigma} (l) \left[ \eta^{\mu \rho} \eta^{\nu \sigma} \left((p_i\cdot k)^2 + (p_i\cdot l)^2+ (p_i\cdot k)(p_i\cdot l) \right) \right. \notag\\
& \left. \quad  + p_i^{\mu} p_i^{\nu} k^{\rho} k^{\sigma} +  l^{\mu} l^{\nu} p_i^{\rho} p_i^{\sigma} - 2 p_i^{\mu} p_i^{\rho} l^{\nu} k^{\sigma} + 2 \eta^{\nu \sigma} \left( (k \cdot l) p_i^{\mu} p_i^{\rho} - (p_i \cdot l) l^{\mu} p_i^{\rho} - (p_i \cdot k) p_i^{\mu} k^{\rho}\right) \right]
\label{2s4.mom} 
\end{align}
and 
\begin{align} 
\psi(\{p_i \}; h)  = {\kappa^2 \over 2} \sum_{i=1}^n \int_{kl}  \frac{1}{k\cdot l} h_{\mu \nu}(k) h_{\rho \sigma} (l) \left[ \eta^{\mu \rho} \eta^{\nu \sigma} p_i \cdot (k+l) -   \eta^{\nu \sigma} \left(  l^{\mu} p_i^{\rho} + p_i^{\mu} k^{\rho}\right)\right]\;.  \qquad
\label{diff}
\end{align}
The three-graviton vertex contribution in \ref{2s4.mom} has the form given in \cite{Distler:2018rwu}, which differs from \ref{2s.gp} and the corresponding expression in \cite{Chakrabarti:2017ltl} by the term in \ref{diff}. However, as will be shown in \ref{ss2}, the two expressions in \ref{2s4.mom} and \ref{2s.gp} are equivalent when acting on the hard amplitude $T(\{p_n\})$ by invoking momentum and angular momentum conservation. In this way, \ref{2s.gp} and  \ref{2s4.mom} provide equivalent contributions to the double soft graviton factor, and certain properties of \ref{2s4.mom} will be useful in deriving subleading double soft factor contributions. The physical interpretation of \ref{diff} is that it results upon expanding the off-shell graviton denominator in the three-graviton vertex contribution. Indeed, one may verify that \ref{diff} results from \ref{2s4.mom} upon multiplying the latter by  
\begin{displaymath}
1 + \frac{1}{2}\frac{(k+l))^2}{p \cdot(k+l)}.
\end{displaymath}

For the above analysis, we were entitled to ignore potentially non-zero displacements $x_i^\mu$ of each hard line, given that we were only considering external emission contributions, which by definition are represented by the generalized soft function. If non-zero displacements ($x^\mu_i \ne 0$) are included, then one may verify that eq.~\ref{sf.1} ceases to be gauge invariant\footnote{The gauge invariance of the double soft graviton factor of \ref{2s4.mom} for $x^\mu_i=0$ has been checked in \cite{Distler:2018rwu}. Unsurprisingly, the three-graviton vertex contribution \ref{2s.gp} is essential to maintain the gauge invariance of the double soft graviton factor with $x_i=0$. This contribution is missing in \cite{Fernandes:2024xqr}, so that the gauge invariance for $x_i=0$ can be maintained only in the collinear limit of the two emitted soft gravitons.}. To recover the gauge invariant double soft factor, as noted in~ \cite{Chakrabarti:2017ltl}, we will need to account for the internal emission contributions exemplified by fig.~\ref{fig:external}(c). In the following section, we will derive these using an appropriate generalisation of Low's theorem~\cite{Low:1954kd, Low:1958sn}, with $x_i \neq 0$. In the process, we will also find a new subleading double soft graviton factor.

\section{Subleading double soft graviton factor from Low's theorem} \label{sec3}

In the previous section, we have considered external emission contributions to the (next-to) soft graviton amplitude, starting from the factorisation formula of \ref{ampfac3}. In this section, we proceed to calculate the remaining contributions, which have various sources, as discussed in the previous works of \cite{Laenen:2008gt,White:2011yy}.
\begin{itemize}
\item Each hard particle could have an arbitrary initial position $x_i^\mu$, which must then be integrated over. 
\item The hard function itself has a dependence on the graviton $h_{\mu\nu}$. Expanding out this dependence leads to a diagrammatic interpretation in terms of graviton emissions from inside the hard interaction (fig.~\ref{fig:external}(c)).
\end{itemize}
Reinstating the initial positions and summing over them, \ref{ampfac3} is then revised to 
\bea
\mathcal{A}^{(n)}(\{p_i\}) &=& \int \prod_{i = 1}^n dx_i \; e^{-i \sum_{i= 1}^n p_i \cdot x_i} \mathcal{A}^{(n)}(\{x_i, p_i\})\;, \nn
\\ 
&=& \int \prod_{i = 1}^n dx_i \; e^{-i \sum_{i= 1}^n p_i \cdot x_i} \int \mathcal{D}h_{\mu \nu}\, H(\{x_i\};h) \; e^{i S_{EH}[\eta +h]} 
\prod_{j=1}^n f(x_j,p_j;h), 
\label{amp}
\eea

There are two effects of having non-zero initial positions in \ref{amp}. The first -- and most visible -- is the overall factor of $e^{-i \sum_{i= 1}^n p_i \cdot x_i} $ in the integral. The second is that the generalised Wilson lines $f(x_j,p_j;h)$ will now include factors such as
\begin{equation}
    e^{i k\cdot x_j},\quad e^{i  (k+l) \cdot x_j}
    \label{expfacs}
\end{equation}
where $k$ and $l$ are emitted graviton momenta. Such factors arise upon transforming the generalised Wilson line exponent to momentum space, as has already been seen in \ref{delta1res}. Upon Taylor expanding such contributions in $k$ or $l$, they will indeed contribute subleading terms in the soft expansion, which will be important to keep track of in what follows. 

Let us now turn to the dependence of the hard interaction on the soft graviton field. As pioneered in the early works of~\cite{Low:1958sn, Burnett:1967km} and generalised in the more recent works of \cite{DelDuca:1990gz,Laenen:2008gt, White:2011yy, Bern:2014vva}, one can isolate all internal emission contributions arising from the hard interaction by Taylor expanding in the graviton field:
\bea
H(\{x_i\}; h ) &=&  T(\{x_i \} ) + \int d^4x \; N^{\mu\nu}(\{x_i \}; x) h_{\mu\nu}(x) \nn
\\
 && + \int d^4x \int d^4y \; N^{\mu\nu\rho\sigma}(\{x_i \}; x, y) h_{\mu\nu}(x) h_{\rho\sigma}(y) \nn
\\
&& + \int d^4x \; L^{\mu\nu\rho\sigma}(\{x_i \}; x) h_{\mu\nu}(x) h_{\rho\sigma}(x)  +{\cal O}(h^3)\;, \label{hard_w_soft}
\eea
Here $T(\{x_i \} )$ is the hard amplitude with no gravitons, $N^{\mu\nu}(\{x_i \}; x)$ is the correction to the hard amplitude due to one emission, while the corrections for two gravitons come from $N^{\mu\nu\rho\sigma}(\{x_i \}; x, y)$ (two distinct emissions) and $L^{\mu\nu\rho\sigma}(\{x_i \}; x)$ (gravitons emitted from the same spacetime point). In the following, we will need the coefficients appearing on the right-hand side of \ref{hard_w_soft} in momentum space. For simplicity, we will then denote their Fourier transforms by the same symbol, but replace the coordinate arguments with the momentum ones, e.g., $T(\{x_i\}) \rightarrow T(\{p_i\})$, $N^{\mu\nu}(\{x_i\}; x) \rightarrow N^{\mu\nu}(\{p_i\}; k)$, etc. We will also use the property that the on-shell soft gravitons satisfy the de Donder gauge conditions
\be
k^{\mu}h_{\mu \nu}(k) = 0 \;; \quad \eta^{\mu \nu} h_{\mu \nu}(k) = 0\,.
\label{gc}
\ee

Using the momentum space representation for gravitons, the single graviton coefficient in \ref{hard_w_soft} can be written as
\begin{align}
\int d^4x \; N^{\mu\nu}(\{x_i \}; x) h_{\mu\nu}(x) &= \int d^4x \int_{k k'}\; N^{\mu\nu}(\{x_i \}; k') h_{\mu\nu}(k) e^{i (k + k')\cdot x}\notag\\
&= \int_k \; N^{\mu\nu}(\{x_i \}; -k) h_{\mu\nu}(k).
\label{n1.int}
\end{align}
Similarly, one finds 
\begin{align}
 \int d^4x \int d^4y \; N^{\mu\nu\rho\sigma}(\{x_i \}; x, y) h_{\mu\nu}(x) h_{\rho\sigma}(y) 
 &= \int_{k}\int_l \; N^{\mu\nu\rho\sigma}(\{x_i \}; -k, -l) h_{\mu\nu}(k) h_{\rho\sigma}(l) \,, \label{n.int}\\
\int d^4x \; L^{\mu\nu\rho\sigma}(\{x_i \}; x) h_{\mu\nu}(x) h_{\rho\sigma}(x) 
&= \int_{k}\int_l \; L^{\mu\nu\rho\sigma}(\{x_i \}; -k-l) h_{\mu\nu}(k) h_{\rho\sigma}(l) \label{l.int} \,,
\end{align}
such that substituting the resulting form for the hard interaction back into \ref{amp} yields
\begin{align}
\mathcal{A}^{(n)}(\{p_i\}) &= \int \mathcal{D}h_{\mu \nu}
\int \prod_{i = 1}^n dx_i \; e^{-i \sum_{i= 1}^n p_i \cdot x_i}  \, \left[
T(\{x_i \} ) + \int_k \; N^{\mu\nu}(\{x_i \}; -k) h_{\mu\nu}(k)\right.
\notag\\
&\left.+ \int_{k}\int_l \; N^{\mu\nu\rho\sigma}(\{x_i \}; -k, -l) h_{\mu\nu}(k) h_{\rho\sigma}(l)
 + \int_{k}\int_l \; L^{\mu\nu\rho\sigma}(\{x_i \}; -k-l) h_{\mu\nu}(k) h_{\rho\sigma}(l)
\right]\notag\\
&\times e^{i S_{EH}[\eta +h]} 
\prod_{j=1}^n f(x_j,p_j;h)+\ldots, 
\label{fullamp}
\end{align}
where the ellipsis denotes terms involving more than two graviton emissions. The interpretation of this formula is as follows. Carrying out the path integral over the graviton field $h_{\mu\nu}$ generates all possible Feynman diagrams involving external soft emission contributions (from the generalised Wilson lines and bulk vertices in the Einstein-Hilbert action), and internal soft emission contributions. The latter are represented by the factor in the square brackets, whose various terms can be interpreted as additional vertices located ``within" the hard interaction. Finally, the integral over positions $x^\mu_i$ performs a Fourier transform to momentum space, where it is important to keep track of the additional position dependence of the generalised Wilson lines, as noted above in \ref{expfacs}. At this point, the coefficients appearing in the square brackets in \ref{fullamp} are unknown. However, the key point of \cite{Low:1958sn, Burnett:1967km,DelDuca:1990gz,Laenen:2008gt, White:2011yy, Bern:2014vva} is that they can be fixed by evaluating the amplitude and imposing gauge invariance. To this end, let us write \ref{fullamp} as
\begin{align}
&\mathcal{A}^{(n)}(\{p_i\}) = \int \mathcal{D}h_{\mu \nu} \;e^{i S_{EH}[\eta +h]}\left( \mathcal{A}_0(\{p_i\}) + \mathcal{A}_1(\{p_i\}\,,h) + \mathcal{A}_2(\{p_i\}\,,h) + \mathcal{O}(h^3)\right)\,,
\label{amp.se}
\end{align}
where ${\cal A}_r$ collects contributions involving $r$ (subleading) soft graviton emissions. We can then analyse each of these amplitudes in turn. As a precursor, the no-emission term is simply given by
\be
\mathcal{A}_0(\{p_i\}) = \int \prod_{i = 1}^n dx_i \; e^{-i \sum_{i= 1}^n p_i \cdot x_i} \; T(\{x_i \} ) = T (\{p_i \} )\;
\label{a0.fin}
\ee
which, as expected, is simply the momentum-space form of the non-radiative amplitude. Explicit calculation of the one- and two-graviton emission contributions will reproduce known results for single and double soft theorems, but will also allow us to extend the double soft results to subleading order in the soft expansion, as we will show in the following.

\subsection{Single emission factor}

Upon expanding the generalised Wilson line factors in \ref{fullamp}, the total amplitude term with one graviton is 
\begin{align}
\mathcal{A}_1(\{p_i\}\,,h) = \int \prod_{i = 1}^n dx_i \; e^{-i \sum_{i= 1}^n p_i \cdot x_i} \int_{k}h_{\mu \nu}(k)  \left(  - \kappa  \sum_{i= 1}^n  \frac{p_i^{\mu}  p_i^{\nu}}{p_i \cdot k} T(\{x_i \} ) e^{ i k \cdot x_i} + N^{\mu\nu}(\{x_i \}; -k) \right)\;.
\label{a1.int}
\end{align}
Replacing $k \to -k$ and performing the Fourier transform on all the hard positions, we then arrive at the result
\begin{align}
\mathcal{A}_1(\{p_i\},h) &= \int_k h_{\mu \nu}(-k) \mathcal{M}^{\mu \nu}(\{p_i\}\,, k)\,, \label{a1.fin0}
\end{align}
where we have defined the amplitude stripped of its external graviton factor:
\begin{align}
\mathcal{M}^{\mu \nu}(\{p_i\}\,, k) &= \kappa  \sum_{i= 1}^n \frac{p_i^{\mu}  p_i^{\nu}}{p_i \cdot k} T(\{p_i +  k \} )  +  N^{\mu\nu}(\{p_i \}; k) \,.
\label{m1.fin}
\end{align}
Here and in what follows, we have defined the notation
\begin{equation}
\{p_i+k\}\equiv\{p_1,\cdots,p_i+k,\cdots,p_n\},
\end{equation}
where the shift of a single hard momentum arises from the above-noted position dependence of each generalised Wilson line operator. 
\ref{m1.fin} is the same relation as has been previously considered in e.g. \cite{White:2011yy,Bern:2014vva}, and can be used to recover the subleading single soft factor contributions involving angular momentum terms. To this end, one first imposes gauge invariance of the amplitude via the gravitational Ward identity~\cite{Brout:1966oea}
\begin{equation}
k_{\mu}\mathcal{M}^{\mu \nu}(\{p_i\}\,, k) = 0,
\label{Ward}
\end{equation}
which in turn implies
\be\label{soft_rel_1}
\kappa \sum_{i=1}^n p_i^{\nu} T(\{p_i+k\}) + k_{\mu} N^{\mu\nu}(\{p_i\}; k) =0\;.
\ee
Next, one can Taylor expand \ref{soft_rel_1} in the emitted graviton momentum, $k$. To leading order, this implies the momentum conservation identity
\be
\sum_{j=1}^n p_j^{\nu} \; T(\{p_i\}) = 0 \,. \label{mom.con}
\ee
This is also the gauge invariance of the amplitude, up to single graviton emissions, in the absence of any initial positions for the external hard particles. The next two orders of the expansion imply 
\be\label{gc_1_1}
 N^{\mu\nu}_n =- \kappa \sum_{i=1}^n p_i^{\nu} \partial_i^{\mu} T_n
\ee
and 
\be\label{gc_1_2}
\partial_k^{(\rho}N_n^{\mu)\nu} = - {\kappa \over 2}\sum_{i=1}^n  p_i^{\nu} \partial_i^{\rho}\partial^{\mu}_i T_n\;, 
\ee
where here and from now on, we will adopt the short-hand notations
\be
T_n\equiv T(\{p_i\})\;, \quad N^{\mu\nu}_n\equiv N^{\mu\nu}(\{p_i\};0)\;, \quad \partial_i^{\mu}\equiv {\partial \over \partial p_{i\mu}}\;, \quad \partial^{\mu}_k \equiv {\partial \over \partial k_{\mu}}\;.
\ee
We will also define (anti-)symmetrised quantities via
\begin{align}
A^{(\mu}B^{\nu)} ={1\over 2} (A^{\mu} B^{\nu}+A^{\nu}B^{\mu})\;,\qquad 
A^{[\mu}B^{\nu]} ={1\over 2} (A^{\mu} B^{\nu}-A^{\nu}B^{\mu}).
\label{symdef}
\end{align}
In \ref{gc_1_1}, we may decompose $N_n^{\mu \nu}$ into its symmetric and antisymmetric parts, where the latter may be recognised as containing the orbital angular momentum operator
\be \label{am.def}
J_i^{\mu\nu} =  i \big( p_i^{\mu} \partial_i^{\nu} - p_i^{\nu} \partial_i^{\mu}\big)\;
\ee
associated with each external line. Given that we are working with scalar hard particles, this is also the total angular momentum, and thus the antisymmetric part of $N_n^{\mu \nu}$ vanishes as a consequence of angular momentum conservation: 
\be
N^{[\mu\nu]}_n = \frac{i\kappa}{2} \sum_{i=1}^n J_i^{\mu \nu}\; T_n = 0 \,.
\label{am.con}
\ee
Hence, $N_n^{\mu \nu}$ is symmetric in its indices. Furthermore, by antisymmetrizing the indices $\rho$ and $\nu$ in \ref{gc_1_2} and using the symmetry of $N^{\mu \nu}$, one obtains 
\be\label{gc_1_no}
\partial_k^{[\rho}N^{\nu]\mu}_n = - {i \kappa \over 2} \sum_{i=1}^n J_i^{\rho\nu} \partial_i^{\mu} T_n\;.
\ee
Hence, by performing a soft series expansion of \ref{m1.fin}, and on using \ref{gc_1_1}, \ref{gc_1_2} and  \ref{gc_1_no}, one finds \cite{Bern:2014vva}
\be
\mathcal{M}^{\mu \nu}(\{p_i\}\,, k)=  \sum_{i= 1}^n \left[\left(\mathcal{M}_{\text{LO}}\right)_{i;k}^{\mu \nu} + \left(\mathcal{M}_{\text{NLO}}\right)_{i;k}^{\mu \nu} + \left(\mathcal{M}_{\text{NNLO}}\right)_{i;k}^{\mu \nu}\right] T_n  + \mathcal{E}^{\mu \nu}_{\text{NLO};k} + \mathcal{E}^{\mu \nu}_{\text{NNLO};k} + \mathcal{O}(k^2)\;, \label{m1.res}
\ee
where we have defined the result in terms of its contributions up to next-to-next-to-leading order (NNLO) in the soft expansion in $k$:
\begin{align}
\left(\mathcal{M}_{\text{LO}}\right)_{i;k}^{\mu \nu} &= \kappa \frac{p_i^{\mu} p_i^{\nu}}{p_i \cdot k}\,,  \label{m1.lo}\\
\left(\mathcal{M}_{\text{NLO}}\right)_{i;k}^{\mu \nu} &= -i \kappa \frac{k_{\alpha}}{ p_i \cdot k}  p_i^{(\nu} J_i^{\mu) \alpha}\,,  \label{m1.nlo}\\
\left(\mathcal{M}_{\text{NNLO}}\right)_{i;k}^{\mu \nu} &=  \kappa \frac{k_{\alpha} k_{\beta}}{2 p_i \cdot k} J_{i}^{\alpha (\mu} J_{i}^{\nu) \beta}\,. \label{m1.nnlo}
\end{align}
The remaining terms $\mathcal{E}^{\mu \nu}_{\text{NLO};k}$ and $\mathcal{E}^{\mu \nu}_{\text{NNLO};k}$ in \ref{m1.res} are additional pieces that show up at subleading and the sub-subleading order computation, and are given by 
\begin{align}
\mathcal{E}^{\mu \nu}_{\text{NLO};k} &= -i \kappa \sum_{i= 1}^n \frac{k_{\alpha}}{ p_i \cdot k}  p_i^{[\nu} J_i^{\mu] \alpha}  T_n\,,
\label{m1.rem1} \\
\mathcal{E}^{\mu \nu}_{\text{NNLO};k} &= \frac{\kappa}{2}  \sum_{i= 1}^n \left[\eta^{\mu \nu} k_{\alpha} - k^{\mu} \delta^{\nu}_{\alpha} - \frac{p_i^{\mu} k^{\nu}}{p_i \cdot k} k_{\alpha}\right] \partial_i^{\alpha} T_n\,.
\label{m1.rem2}
\end{align}
These vanish upon contracting with the external graviton field:
\begin{equation}
\mathcal{E}^{\mu \nu}_{\text{NLO},k} h_{\mu \nu}(-k) =\mathcal{E}^{\mu \nu}_{\text{NNLO},k} h_{\mu \nu}(-k) = 0, 
\label{epsvanish}
\end{equation}
by virtue of the on-shell condition for gravitons of \ref{gc} as well as $h_{\mu\nu}=h_{\nu\mu}$, and thus will not contribute to the final result for the single-emission amplitude. Finally,  the $\mathcal{O}(k^2)$ terms in \ref{m1.res} are kinematically subleading, and thus need not be considered in the following.
On substituting \ref{m1.res} in \ref{a1.fin0} with the above considerations, we arrive at the result
\begin{align}
\mathcal{A}_1(\{p_i\}\,,h) &= \int_k\, h_{\mu \nu}(-k) \sum_{i= 1}^n \left[ \left(\mathcal{M}_{\text{LO}}\right)_{i;k}^{\mu \nu} + \left(\mathcal{M}_{\text{NLO}}\right)_{i;k}^{\mu \nu} + \left(\mathcal{M}_{\text{NNLO}}\right)_{i;k}^{\mu \nu}  + \mathcal{O}(k^2) \right] T_n\,, \notag\\
& =\int_k \kappa  \sum_{i= 1}^n h_{\mu \nu}(-k) \left[ \frac{p_i^{\mu} p_i^{\nu}}{p_i \cdot k}   - \frac{i k_{\alpha}}{p_i \cdot k}  p_i^{(\mu} J_i^{\nu) \alpha}  + \frac{k_{\alpha} k_{\beta}}{2 p_i \cdot k} J_{i}^{\alpha (\mu } J_{i}^{\nu) \beta} + \mathcal{O}(k^2) \right] T_n\,.
\label{a1.fin}
\end{align}
This is indeed the known single soft graviton factorization of the amplitude to sub-subleading order~\cite{Cachazo:2014fwa}.

\subsection{Double emission factor up to next-to-soft level} \label{ss2}

Having verified that the single soft emission factor is correctly reproduced by the GWL formalism even up to NNLO in the momentum expansion, let us now turn to the case of two graviton emissions. We will reproduce the known result for the leading double soft graviton theorem \cite{Chakrabarti:2017ltl,Distler:2018rwu} before extending our analysis to derive a new result at the next-to-soft level. We begin by considering $\mathcal{A}_2(\{p_i\}\,,h)$, which upon expanding all relevant contributions in \ref{fullamp} has the expression
\begin{align}
&\mathcal{A}_2(\{p_i\}\,,h) = \int \prod_{i = 1}^n dx_i \; e^{-i \sum_{i= 1}^n p_i \cdot x_i} \int_{k l} h_{\mu \nu}(k) h_{\rho \sigma}(l) \left( \frac{\kappa^2}{2}  \sum_{i= 1}^n \sum_{j= 1}^n  e^{ i k \cdot x_i} e^{ i l \cdot x_j} \frac{p_i^{\mu}  p_i^{\nu}}{p_i \cdot k} \frac{p_j^{\rho}  p_j^{\sigma}}{p_j \cdot l} T(\{x_i \} )  \right. \notag\\
&\left. \qquad - \frac{\kappa}{2} \sum_{i= 1}^n \frac{p_i^{\mu}  p_i^{\nu}}{p_i \cdot k} N^{\rho \sigma}(\{x_i \}; -l) e^{ i k \cdot x_i} - \frac{\kappa}{2} \sum_{j= 1}^n \frac{p_j^{\rho}  p_j^{\sigma}}{p_j \cdot l} N^{\mu\nu}(\{x_i \}; -k) e^{ i l \cdot x_j}  \right. \notag\\
& \left. \qquad \qquad  + N^{\mu\nu\rho\sigma}(\{x_i \}; -k, -l) + L^{\mu\nu\rho\sigma}(\{x_i \}; -k-l)   \right. \notag\\
& \left. \qquad \qquad  - \frac{\kappa^2}{2} \sum_{i= 1}^n e^{i (k+l) \cdot  x_i} \left(    \frac{\Upsilon_i^{\mu \nu \rho \sigma}(p_i, k, l)}{(p_i\cdot (k+l))^2}  + \frac{\alpha_i^{\mu \nu \rho \sigma}(p_i, k, l)}{p_i\cdot (k+l)} + \psi_i^{\mu \nu \rho \sigma}(p_i, k, l)\right)  T(\{x_i \} )   \right)\,,
\label{a2.int}
\end{align}
where we have defined

\begin{align}
&\Upsilon_i^{\mu \nu \rho \sigma}(p_i, k, l) =   \eta^{\rho (\mu} \eta^{\nu) \sigma} \left((p_i\cdot k)^2 + (p_i\cdot l)^2+ (p_i\cdot k)(p_i\cdot l) \right) + p_i^{\mu} p_i^{\nu} k^{\rho} k^{\sigma} +  l^{\mu} l^{\nu} p_i^{\rho} p_i^{\sigma}     \phantom{ \frac{p_i^{\mu} }{p_i \cdot k}} \notag\\
& \phantom{ \frac{p_i^{\mu} }{p_i \cdot k}} + 2 (k \cdot l) p_i^{(\mu} \eta^{\nu)(\rho)} p_i^{\sigma)} - 2 p_i^{(\mu}l^{\nu)} p_i^{(\rho}  k^{\sigma)} - 2 (p_i \cdot l) l^{(\mu} \eta^{\nu) (\rho}  p_i^{\sigma)} - 2 (p_i \cdot k) p_i^{(\mu} \eta^{\nu) (\rho}  k^{\sigma)} \;,
\label{up.def}
\end{align}
\begin{align}
&\alpha_i^{\mu \nu \rho \sigma}(p_i, k, l) =  - \frac{k\cdot l}{(p_i\cdot k) (p_i\cdot l)} p_i^{\mu} p_i^{\nu} p_i^{\rho} p_i^{\sigma} - 2 p_i^{(\mu}\eta^{\nu) (\rho} p_i^{\sigma)}   + 2 \frac{p_i^{\mu} p_i^{\nu}}{p_i\cdot k} p_i^{(\rho} k^{\sigma)} + 2 \frac{p_i^{\rho} p_i^{\sigma}}{p_i\cdot l} p_i^{(\mu} l^{\nu)}   \notag\\
& \quad  + \frac{1}{k\cdot l}  \left[  \eta^{\rho (\mu} \eta^{\nu) \sigma} \left((p_i\cdot k)^2 + (p_i\cdot l)^2+ (p_i\cdot k)(p_i\cdot l) \right) + p_i^{\mu} p_i^{\nu} k^{\rho} k^{\sigma} +  l^{\mu} l^{\nu} p_i^{\rho} p_i^{\sigma}     \phantom{ \frac{p_i^{\mu} }{p_i \cdot k}} \right.\notag\\
& \left. \phantom{ \frac{p_i^{\mu} }{p_i \cdot k}}  - 2 p_i^{(\mu}l^{\nu)} p_i^{(\rho}  k^{\sigma)} - 2 (p_i \cdot l) l^{(\mu} \eta^{\nu) (\rho}  p_i^{\sigma)} - 2 (p_i \cdot k) p_i^{(\mu} \eta^{\nu) (\rho}  k^{\sigma)} \right]\;,
\label{alph.def}
\end{align}
and 
\begin{align}
\psi_i^{\mu \nu \rho \sigma} (p_i, k, l) = \frac{1}{k \cdot l} \left[-\eta^{\rho (\mu} \eta^{\nu) \sigma} p_i \cdot (k+l) +  l^{(\mu} \eta^{\nu) (\sigma}  p_i^{\rho)} + p_i^{(\mu} \eta^{\nu) (\sigma}  k^{\rho)}\right]\;. \label{psi.def}
\end{align}
These three functions are associated with $\bar{\Upsilon}_{2}$ of \ref{2s.gpsl}, $\Delta_2$ of \ref{2s4.mom}, and $\psi$ of \ref{diff}, respectively. 

We note that under $k \to -k$ and $l \to -l$, we have the symmetry properties 
\begin{align}
\Upsilon_i^{\mu \nu \rho \sigma}(p_i, -k, -l) & = \Upsilon_i^{\mu \nu \rho \sigma}(p_i, k, l)   \label{upsym} \\
\alpha_i^{\mu \nu \rho \sigma}(p_i, -k, -l) = \alpha_i^{\mu \nu \rho \sigma}(p_i, k, l), &\quad  \psi_i^{\mu \nu \rho \sigma}(p_i, -k, -l) = - \psi_i^{\mu \nu \rho \sigma}(p_i, k, l).
\label{alphasym}
\end{align}
Additionally, $\Upsilon_i^{\mu \nu \rho \sigma}(p_i, k, l)$ and $\alpha_i^{\mu \nu \rho \sigma}(p_i, k, l)$ have the following useful relations when contracted with the soft momenta $k$ and $l$:
\begin{align}
k_{\mu}\Upsilon_i^{\mu \nu \rho \sigma}(p_i, k, l) &= p_i \cdot (k+l) \left((p_i\cdot l) k^{(\rho}\eta^{\sigma) \nu} - l^{\nu} k^{(\rho} p_i^{\sigma)}\right) + (k \cdot l) \left(p_i\cdot(k-l)\eta^{\nu (\rho} p_i^{\sigma)} + p_i^{\rho}p_i^{\sigma} l^{\nu}\right)\;, \label{k.ups}\\
l_{\rho}\Upsilon_i^{\mu \nu \rho \sigma}(p_i, k, l) &= p_i \cdot (k+l) \left((p_i\cdot k) l^{(\mu}\eta^{\nu) \sigma} - k^{\sigma} l^{(\mu} p_i^{\nu)}\right) + (k \cdot l) \left(p_i\cdot(l-k)\eta^{\sigma (\mu} p_i^{\nu)} + p_i^{\mu}p_i^{\nu} k^{\sigma}\right)\;, \label{l.ups}\\
k_{\mu} l_{\rho} \Upsilon_i^{\mu \nu \rho \sigma}(p_i, k, l) & = (p_i\cdot k)(p_i\cdot l) (k \cdot l) \eta^{\sigma \nu}\;, \label{kl.ups}
\end{align}
and
\begin{align}
\frac{k_{\mu}\alpha_i^{\mu \nu \rho \sigma}(p_i, k, l)}{p_i \cdot(k+l)} &= \frac{p_i^{\rho}p_i^{\sigma}}{p_i \cdot l} l^{\nu} - p_{i}^{(\rho} \eta^{\sigma) \nu} -\frac{1}{k \cdot l} \left( l^{\nu} p_{i}^{(\rho} k^{\sigma)} - (p_i\cdot l) \eta^{\nu (\rho} k^{\sigma)} \right)  \,, \label{k.alph} \\
 \frac{l_{\rho}\alpha_i^{\mu \nu \rho \sigma}(p_i, k, l)}{p_i \cdot(k+l)} &= \frac{p_i^{\mu}p_i^{\nu}}{p_i \cdot k} k^{\sigma} - p_{i}^{(\mu} \eta^{\nu) \sigma} -\frac{1}{k \cdot l} \left( k^{\sigma} p_{i}^{(\mu} l^{\nu)} - (p_i\cdot k) \eta^{\sigma (\mu} l^{\nu)} \right) \,, \label{l.alph}\\
  k_{\mu} l_{\rho} \alpha_i^{\mu \nu \rho \sigma}(p_i, k, l)  & = 0\,. \label{kl.alph}
\end{align}

We can now consider $\mathcal{A}_2(\{p_i\}\,,h)$ in \ref{a2.int} after transforming the integrated momenta via $k \to -k$ and $l \to -l$, along with the Fourier transform on the hard particle positions, finding
\begin{align}
\mathcal{A}_2(\{p_i\}\,,h) &= \int_{k l} h_{\mu \nu}(-k) h_{\rho \sigma}(-l) \mathcal{M}^{\mu \nu \rho \sigma}(\{p_i\}\,, k\,,l),\label{a2.fin}
\end{align}
where the relevant amplitude stripped of external graviton fields is
\begin{align}
& \mathcal{M}^{\mu \nu \rho \sigma} (\{p_i\}\,, k\,,l) = \frac{\kappa^2}{2}  \sum_{i= 1}^n \left( \sum_{j= 1}^n  \frac{p_i^{\mu}  p_i^{\nu}}{p_i \cdot k} \frac{p_j^{\rho}  p_j^{\sigma}}{p_j \cdot l} T(\{p_i + k\,, p_j + l \} )  + \left(\frac{\alpha_i^{\mu \nu \rho \sigma}(p_i, k, l)}{p_i\cdot (k+l)} + \psi_i^{\mu \nu \rho \sigma}(p_i, k, l) \right. \right. \notag\\
& \left. \left. \phantom{\frac{\alpha_i}{p_i}}  -\frac{\Upsilon_i^{\mu \nu \rho \sigma}(p_i, k, l)}{(p_i\cdot (k+l))^2} \right) T(\{p_i + k + l \} ) \right)  +  \frac{\kappa}{2} \left(\sum_{i= 1}^n \frac{p_i^{\mu}  p_i^{\nu}}{p_i \cdot k} N^{\rho \sigma}(\{p_i + k \}; l)  + \sum_{j= 1}^n \frac{p_j^{\rho}  p_j^{\sigma}}{p_j \cdot l} N^{\mu\nu}(\{p_j + l \}; k)\right)  \notag\\
&  \qquad \qquad \qquad \qquad     + N^{\mu\nu\rho\sigma}(\{p_i \}; k, l)  + L^{\mu\nu\rho\sigma}(\{p_i \}; k+l)  \;.
\label{m2.ft}
\end{align}

We can view this result as the double graviton generalization of \ref{m1.fin}. Similar to the previous case, we can further expand with respect to the soft momenta, yielding
\begin{align}
& \mathcal{M}^{\mu \nu \rho \sigma} (\{p_i\}\,, k\,,l) = \frac{\kappa^2}{2}  \sum_{i= 1}^n \sum_{j= 1}^n  \frac{p_i^{\mu}  p_i^{\nu}}{p_i \cdot k} \frac{p_j^{\rho}  p_j^{\sigma}}{p_j \cdot l} \left[1 + k_{\alpha} \partial_i^{\alpha} +  l_{\beta} \partial_j^{\beta}  + \frac{k_{\alpha} k_{\beta}}{2} \partial_i^{\alpha} \partial_i^{\beta} + \frac{l_{\alpha} l_{\beta}}{2} \partial_j^{\alpha} \partial_j^{\beta} + k_{\alpha} l_{\beta} \partial_i^{\alpha} \partial_j^{\beta} \right]T_n \notag\\
& \; +  \frac{\kappa}{2} \sum_{j= 1}^n \frac{p_j^{\rho}  p_j^{\sigma}}{p_j \cdot l} \left(N_n^{\mu \nu} + k_{\alpha} \partial_k^{\alpha} N_n^{\mu \nu}+ l_{\beta}\partial_j^{\beta}N_{n}^{\mu \nu}\right) + \frac{\kappa}{2} \sum_{i= 1}^n \frac{p_i^{\mu}  p_i^{\nu}}{p_i \cdot k} \left( N_n^{\rho \sigma} + l_{\beta} \partial_l^{\beta} N_n^{\rho \sigma} + k_{\alpha}\partial_i^{\alpha}N_{n}^{\rho \sigma} \right) \notag\\
& \qquad + \frac{\kappa^2}{2} \sum_{i= 1}^n \frac{\alpha_i^{\mu \nu \rho \sigma}(p_i, k, l)}{p_i(k+l)}\left(T_n + (k+l)_{\alpha} \partial_i^{\alpha}T_n\right)      - \frac{\kappa^2}{2} \sum_{i= 1}^n \frac{\Upsilon_i^{\mu \nu \rho \sigma}(p_i, k, l)}{(p_i\cdot (k+l))^2} T_n  
\notag\\
& \qquad \qquad +  \frac{\kappa^2}{2} \sum_{i= 1}^n \psi_i^{\mu \nu \rho \sigma}(p_i, k, l)\left(T_n + (k+l)_{\alpha} \partial_i^{\alpha}T_n\right)  + L^{\mu\nu\rho\sigma}_n + N^{\mu\nu\rho\sigma}_n + \mathcal{O}(k\,,l)\;,
\label{m2.ex0}
\end{align}
where we introduced the shorthand notation
$$ N^{\mu\nu\tau\sigma}_n = N^{\mu\nu\tau\sigma}(\{p_i \}; 0,0) \quad  {\rm and} \quad  L^{\mu\nu\tau\sigma}_n = L^{\mu\nu\tau\sigma}(\{p_i \}; 0)\,, $$
and $\mathcal{O}(k\,,l)$ denotes subleading terms in the soft expansion that do not need to be considered further. 

We can now demonstrate that the two terms involving $\psi_i^{\mu \nu \rho \sigma}(p_i, k, l)$ in \ref{m2.ex0} in fact vanish. For the term without any derivatives of the hard amplitude, we have
\begin{align}
 \frac{\kappa^2}{2} \sum_{i= 1}^n \psi_i^{\mu \nu \rho \sigma}(p_i, k, l) T_n & =  \frac{\kappa^2}{2 k \cdot l} \sum_{i= 1}^n \left[-\eta^{\rho (\mu} \eta^{\nu) \sigma} p_i \cdot (k+l) +  l^{(\mu} \eta^{\nu) (\sigma}  p_i^{\rho)} + p_i^{(\mu} \eta^{\nu) (\sigma}  k^{\rho)}\right] T_n  = 0 
 \label{mc}
 \end{align}
 upon using the momentum conservation of the hard amplitude. For the term with a derivative acting on the hard amplitude, we first shift functions in \ref{m2.ex0}, being entitled to do so, given that these are as yet undetermined:
\begin{equation}
L_n^{\mu \nu \rho \sigma} +  N_n^{\mu \nu \rho \sigma} \to L_n^{\mu \nu \rho \sigma} +  N_n^{\mu \nu \rho \sigma} - J_n^{\mu \nu \rho \sigma}\;.
\end{equation}
We choose the additional coefficients on the right-hand side to be given by
\begin{align}
J_n^{\mu \nu \rho \sigma}  = \frac{\kappa^2}{2 k \cdot l} \sum_{i= 1}^n p_i \cdot (k+l) \left[-\eta^{\rho (\mu} \eta^{\nu) \sigma} (k+l)_{\beta} \partial_i^{\beta} +  l^{(\mu} \eta^{\nu) (\sigma}  \partial_i^{\rho)} + k^{(\rho} \eta^{\sigma) (\mu}  \partial_i^{\nu)} \right] T_n\;,
\label{j.def}
\end{align}
which then amounts to a redefinition of the other coefficients.
\ref{m2.ex0} then gives the following combination:
\begin{align}
& \frac{\kappa^2}{2} \sum_{i= 1}^n \psi_i^{\mu \nu \rho \sigma}(p_i, k, l) (k+l)_{\alpha} \partial_i^{\alpha} T_n - J_n^{\mu \nu \rho \sigma} \notag\\
& =  - i \frac{\kappa^2}{2 k \cdot l} (k+l)_{\alpha} \sum_{i= 1}^n \left[-\eta^{\rho (\mu} \eta^{\nu) \sigma} J_i^{\alpha \beta} (k+l)_{\beta} +  l^{(\mu} \eta^{\nu) (\sigma}  J_i^{\rho) \alpha} + k^{(\rho} \eta^{\sigma) (\mu}  J_i^{\nu) \alpha} \right] T_n = 0\;,
\label{amc}
 \end{align}
on using the conservation of angular momentum of the hard amplitude. Thus, the contributions from $\psi_i^{\mu \nu \rho \sigma}(p_i, k, l)$ in \ref{m2.ex0}, up to first derivatives of the hard amplitude, will vanish. Note that \ref{mc} realizes the point noted below \ref{diff}, namely that the two leading double soft factors in \cite{Chakrabarti:2017ltl} and \cite{Distler:2018rwu}, while differing in their explicit form, are equivalent to one another due to momentum conservation of the hard amplitude. The implication of \ref{amc} is that this difference will also not affect the subleading double soft factor, which will be derived in this subsection.

Let us now proceed to further analyse \ref{m2.ex0}, without the unnecessary contribution from $\psi_i^{\mu \nu \rho \sigma}(p_i, k, l)$. On contracting with $k_{\mu}$, $l_{\rho}$ and $k_{\mu} l_{\rho}$, we arrive at the gauge conditions
\be
k_{\mu}  \mathcal{M}^{\mu \nu \rho \sigma} (\{p_i\}\,, k\,,l) = 0 \;,\qquad  l_{\rho} \mathcal{M}^{\mu \nu \rho \sigma} (\{p_i\}\,, k\,,l) = 0 \;, \qquad  k_{\mu} l_{\rho} \mathcal{M}^{\mu \nu \rho \sigma} (\{p_i\}\,, k\,,l) = 0
\label{m2gc}
\ee
with 
\begin{align}
&k_{\mu}  \mathcal{M}^{\mu \nu \rho \sigma}(\{p_i\}\,, k\,,l) = \frac{\kappa^2}{2} \sum_{i= 1}^n p_i^{\nu} \sum_{j= 1}^n   \frac{p_j^{\rho}  p_j^{\sigma}}{p_j \cdot l}\left[1 + k_{\alpha} \partial_i^{\alpha} +  l_{\beta} \partial_j^{\beta}  + \frac{k_{\alpha} k_{\beta}}{2} \partial_i^{\alpha} \partial_i^{\beta} + \frac{l_{\alpha} l_{\beta}}{2} \partial_j^{\alpha} \partial_j^{\beta} + k_{\alpha} l_{\beta} \partial_i^{\alpha} \partial_j^{\beta} \right]T_n\notag\\
& + \frac{\kappa}{2} \sum_{j= 1}^n \frac{p_j^{\rho}  p_j^{\sigma}}{p_j \cdot l} \left( k_{\mu} N_n^{\mu \nu} + k_{\mu} k_{\alpha} \partial_k^{\alpha} N_n^{\mu \nu} + l_{\beta}\partial_j^{\beta} k_{\mu} N_{n}^{\mu \nu} \right) + \frac{\kappa}{2} \sum_{i= 1}^n p_i^{\nu} \left( N_n^{\rho \sigma} + l_{\beta} \partial_l^{\beta} N_n^{\rho \sigma} + k_{\alpha}\partial_i^{\alpha}N_{n}^{\rho \sigma} \right) \notag\\
& + \frac{\kappa^2}{2} \sum_{i= 1}^n \frac{k_{\mu}\alpha_i^{\mu \nu \rho \sigma}(p_i, k, l)}{p_i \cdot(k+l)} \left[ T_n + (k_{\alpha} + l_{\alpha}) \partial_i^{\alpha}T_n\right] + k_{\mu} \left(N^{\mu\nu\rho\sigma}_n  + L^{\mu\nu\rho\sigma}_n    - \frac{\kappa^2}{2} \sum_{i= 1}^n \frac{\Upsilon_i^{\mu \nu \rho \sigma}(p_i, k, l)}{(p_i\cdot (k+l))^2} T_n   \right) \notag\\
& \qquad + \mathcal{O}(k^2,l^2) \,,
\label{kM.exp}
\end{align}
\begin{align}
&l_{\rho}  \mathcal{M}^{\mu \nu \rho \sigma}(\{p_i\}\,, k\,,l) = \frac{\kappa^2}{2} \sum_{j= 1}^n p_j^{\sigma} \sum_{i= 1}^n \frac{p_i^{\mu}  p_i^{\nu}}{p_i \cdot k} \left[1 + k_{\alpha} \partial_i^{\alpha} +  l_{\beta} \partial_j^{\beta}  + \frac{k_{\alpha} k_{\beta}}{2} \partial_i^{\alpha} \partial_i^{\beta} + \frac{l_{\alpha} l_{\beta}}{2} \partial_j^{\alpha} \partial_j^{\beta} + k_{\alpha} l_{\beta} \partial_i^{\alpha} \partial_j^{\beta} \right]T_n \notag\\
& \; + \frac{\kappa}{2} \sum_{i= 1}^n \frac{p_i^{\mu}  p_i^{\nu}}{p_i \cdot k} \left( l_{\rho} N_n^{\rho \nu} + l_{\rho} l_{\beta} \partial_l^{\beta} N_n^{\rho \sigma} + k_{\alpha}\partial_i^{\alpha} l_{\rho}  N_{n}^{\rho \sigma} \right) + \frac{\kappa}{2} \sum_{j= 1}^n p_j^{\sigma} \left(N_n^{\mu \nu} + k_{\alpha} \partial_k^{\alpha} N_n^{\mu \nu}  + l_{\beta}\partial_j^{\beta}N_{n}^{\mu \nu}\right) \notag\\
&+ \frac{\kappa^2}{2} \sum_{i= 1}^n \frac{l_{\rho}\alpha_i^{\mu \nu \rho \sigma}(p_i, k, l)}{p_i \cdot(k+l)}  \left[ T_n + (k_{\alpha} + l_{\alpha}) \partial_i^{\alpha}T_n\right]  + l_{\rho} \left(N^{\mu\nu\rho\sigma}_n  + L^{\mu\nu\rho\sigma}_n    - \frac{\kappa^2}{2} \sum_{i= 1}^n \frac{\Upsilon_i^{\mu \nu \rho \sigma}(p_i, k, l)}{(p_i\cdot (k+l))^2} T_n  \right)  \notag\\
& \qquad + \mathcal{O}(k^2,l^2) \,,
\label{lM.exp}
\end{align}
and
\begin{align}
&k_{\mu} l_{\rho} \mathcal{M}^{\mu \nu \rho \sigma} (\{p_i\}\,, k\,,l) = \frac{\kappa^2}{2}  \sum_{i= 1}^n \sum_{j= 1}^n   p_i^{\nu} p_j^{\sigma}
\left[1 + k_{\alpha} \partial_i^{\alpha} +  l_{\beta} \partial_j^{\beta}  + \frac{k_{\alpha} k_{\beta}}{2} \partial_i^{\alpha} \partial_i^{\beta} + \frac{l_{\alpha} l_{\beta}}{2} \partial_j^{\alpha} \partial_j^{\beta} + k_{\alpha} l_{\beta} \partial_i^{\alpha} \partial_j^{\beta} \right]T_n \notag\\
&\; + \frac{\kappa}{2} \sum_{i= 1}^n  p_i^{\nu} \left( l_{\rho} N_n^{\rho \nu} + l_{\rho} l_{\beta} \partial_l^{\beta} N_n^{\rho \sigma} + k_{\alpha}\partial_i^{\alpha} l_{\rho}  N_{n}^{\rho \sigma} \right)  + \frac{\kappa}{2}\sum_{j= 1}^n  p_j^{\sigma} \left( k_{\mu} N_n^{\mu \nu} + k_{\mu} k_{\alpha} \partial_k^{\alpha} N_n^{\mu \nu} + l_{\beta}\partial_j^{\beta} k_{\mu} N_{n}^{\mu \nu} \right)  \notag\\
& \qquad + k_{\mu} l_{\rho} \left(N^{\mu\nu\rho\sigma}_n + L^{\mu\nu\rho\sigma}_n  - \frac{\kappa^2}{2} \sum_{i= 1}^n \frac{\Upsilon_i^{\mu \nu \rho \sigma}(p_i, k, l)}{(p_i\cdot (k+l))^2} T_n \right) + \mathcal{O}(k^3,l^3) \,.
\label{kMl}
\end{align}
As with the single emission case, the above gauge conditions can be used to solve for the various coefficient functions describing the internal emission contributions. To this end, we start by noting that \ref{kM.exp}, for example, amounts to several conditions arising from linear independence of the different combinations of the soft momenta $l^\mu$ and $k^\mu$. Denoting these combinations schematically as follows, we find the conditions: 
\begin{align}
\frac{1}{l}& \;:\; \frac{\kappa^2}{2}\sum_{j= 1}^n   \frac{p_j^{\rho}  p_j^{\sigma}}{p_j \cdot l} \sum_{i= 1}^n p_i^{\nu} T_n = 0\;, \label{lm1.mk}\\
\frac{k}{l}& \;:\;  \frac{\kappa}{2} \sum_{j= 1}^n \frac{p_j^{\rho}  p_j^{\sigma}}{p_j \cdot l} k_{\alpha} \left( \kappa \sum_{i= 1}^n p_i^{\nu} \partial_i^{\alpha}T_n + N_n^{\alpha \nu}  \right) = 0\;, \label{klm1.mk}\\
\frac{k^2}{l}& \;:\;  \frac{\kappa}{2} \sum_{j= 1}^n \frac{p_j^{\rho}  p_j^{\sigma}}{p_j \cdot l} k_{\alpha} k_{\beta} \left( \frac{\kappa}{2} \sum_{i= 1}^n p_i^{\nu} \partial_i^{\alpha}\partial_i^{\beta}T_n  +  \partial_k^{\alpha} N_n^{\beta \nu}\right) = 0\;,  \label{k2lm1.mk}\\
1 & \;:\; \frac{\kappa}{2} \sum_{i= 1}^n p_i^{\nu} \left(\kappa \sum_{j= 1}^n   \frac{p_j^{\rho}  p_j^{\sigma}}{p_j \cdot l}  l_{\beta} \partial_j^{\beta}T_n  + N_n^{\rho \sigma}\right) + \frac{\kappa^2}{2} \sum_{i= 1}^n \frac{k_{\mu}\alpha_i^{\mu \nu \rho \sigma}(p_i, k, l)}{p_i \cdot(k+l)} T_n\;,  \label{c.mk}\\
k\,,l & \;:\; k_{\alpha} \left(\frac{\kappa^2}{2} \sum_{i= 1}^n p_i^{\nu} \sum_{j= 1}^n \frac{p_j^{\rho}  p_j^{\sigma}}{p_j \cdot l} l_{\beta}\partial_i^{\alpha} \partial_j^{\beta} T_n + \frac{\kappa}{2} \sum_{i= 1}^n p_i^{\nu} \partial_i^{\alpha}N_{n}^{\rho \sigma} +  \frac{\kappa}{2} \sum_{j= 1}^n \frac{p_j^{\rho}  p_j^{\sigma}}{p_j \cdot l} l_{\beta}\partial_j^{\beta} N_{n}^{\alpha \nu} \right. \notag\\
&\left.  \quad  \phantom{\frac{\kappa^2}{2} \sum_{i= 1}^n}+ N_n^{\alpha \nu \rho \sigma} + L_n^{\alpha \nu \rho \sigma}\right) + \frac{\kappa^2}{2} \sum_{i= 1}^n \frac{k_{\mu}\alpha_i^{\mu \nu \rho \sigma}(p_i, k, l)}{p_i \cdot(k+l)} k_{\alpha} \partial_i^{\alpha} T_n  - \frac{\kappa^2}{2} \sum_{i= 1}^n \frac{k_{\mu}\Upsilon_i^{\mu \nu \rho \sigma}(p_i, k, l)}{(p_i\cdot (k+l))^2} T_n   \notag\\
& + l_{\beta} \left(\frac{\kappa}{2} \sum_{i= 1}^n p_i^{\nu} \left( \frac{\kappa}{2}\sum_{j= 1}^n   \frac{p_j^{\rho}  p_j^{\sigma}}{p_j \cdot l} l_{\alpha} \partial_{j}^{\alpha} \partial_{j}^{\beta} T_n + \partial_l^{\beta} N_n^{\rho \sigma} \right) + \frac{\kappa^2}{2} \sum_{i= 1}^n \frac{k_{\mu}\alpha_i^{\mu \nu \rho \sigma}(p_i, k, l)}{p_i \cdot(k+l)}  \partial_i^{\beta} T_n\right) = 0\;.  \label{k.mk}
\end{align}
Likewise, \ref{lM.exp} provides the following constraints: 
\begin{align}
\frac{1}{k}& \;:\; \frac{\kappa^2}{2}\sum_{i= 1}^n   \frac{p_i^{\mu}  p_i^{\nu}}{p_i \cdot k} \sum_{i= j}^n p_j^{\sigma} T_n = 0\;, \label{km1.ml}\\
\frac{l}{k}& \;:\;  \frac{\kappa}{2} \sum_{i= 1}^n \frac{p_i^{\mu}  p_i^{\nu}}{p_i \cdot k} l_{\beta} \left( \kappa \sum_{j= 1}^n p_j^{\sigma} \partial_j^{\beta}T_n + N_n^{\beta \sigma}  \right) = 0 \;, \label{lkm1.ml}\\
\frac{l^2}{k}& \;:\;  \frac{\kappa}{2} \sum_{i= 1}^n \frac{p_i^{\mu}  p_i^{\nu}}{p_i \cdot k} l_{\alpha} l_{\beta} \left( \frac{\kappa}{2} \sum_{j= 1}^n p_j^{\sigma} \partial_i^{\alpha}\partial_i^{\beta}T_n  +  \partial_l^{\alpha} N_n^{\beta \sigma}\right) = 0 \;, \label{l2km1.ml}\\
1 & \;:\; \frac{\kappa}{2} \sum_{j= 1}^n p_j^{\sigma} \left(\kappa \sum_{i= 1}^n   \frac{p_i^{\mu}  p_i^{\nu}}{p_i \cdot k}  k_{\alpha} \partial_i^{\alpha}T_n  + N_n^{\mu \nu}\right)  + \frac{\kappa^2}{2}  \sum_{i= 1}^n\frac{l_{\rho}\alpha_i^{\mu \nu \rho \sigma}(p_i, k, l)}{p_i \cdot(k+l)} T_n = 0 \;, \label{c.ml}\\
l\,,k & \;:\; l_{\beta} \left(\frac{\kappa^2}{2} \sum_{j= 1}^n p_j^{\sigma} \sum_{i= 1}^n \frac{p_i^{\mu}p_i^{\nu}}{p_i \cdot k} k_{\alpha} \partial_i^{\alpha} \partial_j^{\beta}  T_n + \frac{\kappa}{2} \sum_{j= 1}^n p_j^{\sigma} \partial_j^{\beta}N_{n}^{\mu \nu} +  \frac{\kappa}{2} \sum_{i= 1}^n \frac{p_i^{\mu}  p_i^{\nu}}{p_i \cdot k} k_{\alpha}\partial_i^{\alpha} N_{n}^{\beta \sigma} \right. \notag\\
&\left.  \quad \phantom{\frac{\kappa^2}{2} \sum_{i= 1}^n} + N_n^{\mu \nu \beta \sigma} +  L_n^{\mu \nu \beta \sigma}\right) + \frac{\kappa^2}{2}  \sum_{i= 1}^n\frac{l_{\rho}\alpha_i^{\mu \nu \rho \sigma}(p_i, k, l)}{p_i \cdot(k+l)} l_{\alpha}\partial_i^{\alpha} T_n  - \frac{\kappa^2}{2} \sum_{i= 1}^n \frac{l_{\rho}\Upsilon_i^{\mu \nu \rho \sigma}(p_i, k, l)}{(p_i\cdot (k+l))^2} T_n \notag\\
& + k_{\alpha} \left(\frac{\kappa}{2} \sum_{j= 1}^n p_j^{\sigma} \left( \frac{\kappa}{2}\sum_{i= 1}^n   \frac{p_i^{\mu}  p_i^{\nu}}{p_i \cdot k} k_{\beta} \partial_{i}^{\alpha} \partial_{i}^{\beta} T_n +  \partial_k^{\alpha} N_n^{\mu \nu}\right)  + \frac{\kappa^2}{2}  \sum_{i= 1}^n\frac{l_{\rho}\alpha_i^{\mu \nu \rho \sigma}(p_i, k, l)}{p_i \cdot(k+l)}  \partial_i^{\alpha} T_n\right) = 0 \;. \label{l.ml}
\end{align}

While we may similarly collect all coefficients of \ref{kMl}, only the coefficient of $k_{\mu} l_{\rho}$ will be important in the following, i.e.,
\begin{align}
kl \;:\; k_{\mu} l_{\rho} &\left[\frac{\kappa^2}{2} \sum_{i,j= 1}^n p_i^{\nu} p_j^{\sigma}\partial_i^{\mu}\partial_j^{\rho} + \frac{\kappa}{2} \sum_{j= 1}^n p_j^{\sigma} \partial_j^{\rho}N_{n}^{\mu \nu}  + \frac{\kappa}{2} \sum_{i= 1}^n p_i^{\nu} \partial_i^{\mu}N_{n}^{\rho \sigma} \right. \notag\\
 &\left. \qquad \qquad + N_{n} ^{\mu \nu \rho \sigma} + L_{n} ^{\mu \nu \rho \sigma}  - \frac{\kappa^2}{2} \sum_{i= 1}^n \frac{\Upsilon_i^{\mu \nu \rho \sigma}(p_i, k, l)}{(p_i\cdot (k+l))^2} T_n     \right] = 0\;.
\label{nn.kl}
\end{align}

We can now solve all of the constraints for each order in the soft momentum expansion. The constraints \ref{lm1.mk} and \ref{km1.ml}, to lowest order, are manifestly satisfied by momentum conservation of the hard amplitude \ref{mom.con}. The constraints for $N^{\mu \nu}_n$ in \ref{klm1.mk} and \ref{lkm1.ml} are satisfied by the known linear order condition \ref{gc_1_1}, while \ref{k2lm1.mk} and \ref{l2km1.ml} are likewise satisfied due to \ref{gc_1_2}. 

It can be shown that the constraints \ref{c.mk} and \ref{c.ml} are satisfied as a consequence of momentum conservation. We will show this explicitly in the case of \ref{c.mk}, and identical steps can be used in the case of \ref{c.ml}. By using \ref{gc_1_1} and \ref{k.alph} in \ref{c.mk}, and using momentum conservation for the hard amplitude \ref{mom.con}, we arrive at
\be
\frac{\kappa^2}{2} \sum_{i= 1}^n p_i^{\nu} \sum_{j= 1}^n \left(\frac{p_j^{\rho}  p_j^{\sigma}}{p_j \cdot l}  l_{\beta} \partial_j^{\beta}T_n  - p_j^{\sigma} \partial_j^{\rho} T_n\right)  + \frac{\kappa^2}{2} \sum_{i= 1}^n  \frac{p_i^{\rho}p_i^{\sigma}}{p_i \cdot l} l^{\nu} T_n = 0\,,
\ee
which vanishes upon using the identity
\be
\sum_{i= 1}^n p_i^{\nu} \partial_j^{\beta} T_n = \partial_j^{\beta}\left(\sum_{i= 1}^n p_i^{\nu} T_n\right) - \partial_j^{\beta}\left(\sum_{i= 1}^n p_i^{\nu}\right)T_n = - \sum_{i= 1}^n \delta_{ij} \eta^{\beta \nu}T_n\;.
\label{bp}
\ee 
Hence, \ref{gc_1_1} can be used to show that \ref{c.mk} and \ref{c.ml} are manifestly satisfied. 

To summarize, we thus far have the following relations
\begin{align}
N_n^{\mu \nu} = - \kappa \sum_{i=1}^n p_i^{\nu}\partial_i^{\mu} T_n \,, \qquad & N_n^{\rho \sigma} = - \kappa \sum_{j=1}^n p_j^{\sigma}\partial_i^{\rho} T_n \,, \label{n.kl} \\
\partial_k^{(\alpha} N_n^{\mu) \nu} = -\frac{\kappa}{2} \sum_{i=1}^n p_i^{\nu} \partial_i^{\alpha} \partial_i^{\mu} T_n \,, \qquad & \partial_l^{(\beta} N_n^{\rho) \sigma} = -\frac{\kappa}{2} \sum_{j=1}^n p_j^{\sigma} \partial_j^{\beta} \partial_i^{\mu} T_n \,, \label{sn.kl} \\
\partial_k^{[\alpha} N_n^{\mu] \nu} = -\frac{\kappa}{2} \sum_{i=1}^n \left(p_i^{\mu} \partial_i^{\alpha} \partial_i^{\nu} - p_i^{\alpha} \partial_i^{\mu} \partial_i^{\nu}\right) T_n \,,\quad & \partial_l^{[\beta} N_n^{\rho] \sigma} = -\frac{\kappa}{2} \sum_{j=1}^n  \left(p_j^{\rho} \partial_i^{\beta} \partial_j^{\sigma} - p_j^{\beta} \partial_j^{\rho} \partial_j^{\sigma}\right) T_n \,. \label{an.kl}
\end{align}
These are simply the double-graviton versions of \ref{gc_1_1}, \ref{gc_1_2} and \ref{gc_1_no}. 

We will next derive $N_n^{\mu \nu \rho \sigma}$ and $L_n^{\mu \nu \rho \sigma}$ from \ref{k.mk} and \ref{l.ml}. We discuss the case of \ref{k.mk} in detail. We will simply note that the result for \ref{l.ml} follows from the interchange $\{k\,; \mu\,, \nu\} \leftrightarrow \{l\,; \rho\,,\sigma\}$, as a consequence of the symmetry under the interchange of the two gravitons and their indices.  The conditions \ref{k.mk} and \ref{l.ml} appear to have terms with both $k$ and $l$ coefficients,  potentially obstructing the manner in which we have solved for the constraints thus far as the coefficient of a common soft momentum up to a particular order. However, using \ref{n.kl} and \ref{k.alph}, one can explicitly simplify the terms appearing in the last lines of \ref{k.mk} to find 
\begin{align}
&l_{\beta} \left(\frac{\kappa}{2} \sum_{i= 1}^n p_i^{\nu} \left( \frac{\kappa}{2}\sum_{j= 1}^n   \frac{p_j^{\rho}  p_j^{\sigma}}{p_j \cdot l} l_{\alpha} \partial_{j}^{\alpha} \partial_{j}^{\beta} T_n + \partial_l^{\beta} N_n^{\rho \sigma} \right) + \frac{\kappa^2}{2} \sum_{i= 1}^n \frac{k_{\mu}\alpha_i^{\mu \nu \rho \sigma}(p_i, k, l)}{p_i \cdot(k+l)}  \partial_i^{\beta} T_n\right) = k_{\mu} \sum_{i=1}^n \gamma^{\mu \nu \rho \sigma}_{i,k} \label{l.mk}
\end{align}
with
\begin{align}
\gamma^{\mu \nu \rho \sigma}_{i,k} = \frac{\kappa^2}{2 k \cdot l} \left[ l^{\mu} l^{\nu} p_i^{(\rho} \partial_i^{\sigma)} + (p_i \cdot l) \eta^{\rho(\mu} \eta^{\nu) \sigma} l_{\beta}\partial_i^{\beta} -  2 l^{(\mu} \eta^{\nu) (\rho} p_i^{\sigma)} l_{\beta}\partial_i^{\beta} \right] T_n\;.
\label{k.rel}
\end{align}
In a similar way, using \ref{k.alph} it follows that we have
\begin{align}
\frac{\kappa^2}{2}  \sum_{i= 1}^n& \frac{k_{\mu} \alpha_i^{\mu \nu \rho \sigma}(p_i, k, l)}{p_i \cdot(k+l)} k_{\alpha} \partial_i^{\alpha} T_n = k_{\mu}  \sum_{i= 1}^n \beta_{i,k}^{\mu \nu \rho \sigma} \label{k.alph2}
\end{align}
with
\begin{align}
\beta_{i,k}^{\mu \nu \rho \sigma} &= \frac{\kappa^2}{4} \left[ \frac{p_i^{\rho}p_i^{\sigma}}{p_i \cdot l} l^{\nu} - p_{i}^{(\rho} \eta^{\sigma) \nu} -\frac{1}{k \cdot l} \left( l^{\nu} p_{i}^{(\rho} k^{\sigma)} - (p_i\cdot l) \eta^{\nu (\rho} k^{\sigma)} \right) \right] \partial_i^{\mu} T_n  \notag\\
& \qquad \quad  + \frac{\kappa^2}{4 k \cdot l} \left[ \frac{p_i^{\rho}p_i^{\sigma}}{p_i \cdot l} l^{\nu} l^{\mu} - p_{i}^{(\rho} \eta^{\sigma) \nu} l^{\mu} - l^{\nu} p_{i}^{(\rho} \eta^{\sigma) \mu} + (p_i\cdot l) \eta^{\nu (\rho} \eta^{\sigma) \mu} \right] k_{\alpha} \partial_i^{\alpha} T_n  \,. \label{k.rel2} 
\end{align}
We now approach \ref{k.mk} in the following way. All terms involving a double sum over particle labels ($i$ and $j$) are collected to define the $N_{n}^{\mu \nu \rho \sigma}$ constraint conditions, while all terms with a single sum over the external particles are collected to go with the undetermined function $L_{n}^{\mu \nu \rho \sigma}$. We will further split $L_{n}^{\mu \nu \rho \sigma}$ into two distinct contributions:
\begin{displaymath}
L_{n}^{\mu \nu \rho \sigma} = \bar{L}_{n}^{\mu \nu \rho \sigma} + \tilde{L}_{n}^{\mu \nu \rho \sigma},
\end{displaymath}
where $L_{n}^{\mu \nu \rho \sigma}$ will be related to the contributions from \ref{l.mk} and \ref{k.alph2} that involve derivatives on $T_n$, while $\tilde{L}_{n}^{\mu \nu \rho \sigma}$ will be related to the $\Upsilon_i^{\mu \nu \rho \sigma}$ term that only involves no derivative on $T_n$.
Thus on using \ref{l.mk} and \ref{k.alph2} in \ref{k.mk}, we can then identify the following three constraint relations
\begin{align}
& k_{\mu} N_{n,k}^{\mu \nu \rho \sigma} =  k_{\mu} \left(-\frac{\kappa^2}{2}\sum_{i,j= 1}^n p_i^{\nu} \frac{p_j^{\rho}  p_j^{\sigma}}{p_j \cdot l} l_{\beta}\partial_i^{\mu} \partial_j^{\beta} T_n - \frac{\kappa}{2}  \sum_{i= 1}^n p_i^{\nu} \partial_i^{\mu}N_{n}^{\rho \sigma} -  \frac{\kappa}{2} \sum_{j= 1}^n \frac{p_j^{\rho}  p_j^{\sigma}}{p_j \cdot l} l_{\beta}\partial_j^{\beta} N_{n}^{\mu \nu} \right) \;,  \label{nn.k}\\
& k_{\mu} \bar{L}^{\mu \nu \rho \sigma}_{n,k} = - \sum_{i=1}^n k_{\mu} \left(\gamma^{\mu \nu \rho \sigma}_{i,k} + \beta^{\mu \nu \rho \sigma}_{i,k}\right) \;, \label{ll.k}\\
&k_{\mu}\tilde{L}_{n,k}^{\mu\nu\rho\sigma} = \frac{\kappa^2}{2}  \sum_{i= 1}^n \frac{k_{\mu}\Upsilon_{i}^{\mu \nu \rho \sigma}(p_i, k, l)}{(p_i\cdot (k+l))^2} T_n \,. 
\label{llt.k}
\end{align}

In \ref{nn.k} - \ref{llt.k}, the additional suffix `$k$' in $N_{n,k}^{\mu \nu \rho \sigma}$, $\bar{L}_{n,k}^{\mu \nu \rho \sigma}$ and $\tilde{L}_{n,k}^{\mu \nu \rho \sigma}$ indicates that these constraints result from contracting with the soft momenta $k_{\mu}$. We note that the corresponding constraint relations for the soft momentum $l_{\rho}$ from \ref{l.ml}, those for $l_{\rho} N_{n,l}^{\mu \nu \rho \sigma}$, $l_{\rho} \bar{L}^{\mu \nu \rho \sigma}_{n,l}$ and $l_{\rho} \tilde{L}^{\mu \nu \rho \sigma}_{n,l}$, follow from replacing  $\{k\,; \mu\,, \nu\} \leftrightarrow \{l\,; \rho\,,\sigma\}$ in \ref{nn.k}- \ref{llt.k}.

We will first determine the general solutions for $\bar{L}_n^{\mu \nu \rho \sigma}$ and $N_n^{\mu \nu \rho \sigma}$ which satisfy \ref{nn.k}, \ref{ll.k}, as well as the constraints arising from contracting with $l_{\rho}$.  Given the properties
\be
k_{\mu}\gamma^{\mu \nu \rho \sigma}_{i,l} = 0 = l_{\rho}\gamma^{\mu \nu \rho \sigma}_{i,k} \,, \qquad k_{\mu}\beta^{\mu \nu \rho \sigma}_{i,l} = 0 = l_{\rho}\beta^{\mu \nu \rho \sigma}_{i,k} \,,
\label{der.con}
\ee
it follows that $k_{\mu} \bar{L}^{\mu \nu \rho \sigma}_{n,l} = 0$ and $l_{\rho} \bar{L}^{\mu \nu \rho \sigma}_{n,k} = 0$. As a consequence, the general solution for $\bar{L}_n^{\mu \nu \rho \sigma}$ satisfying \ref{ll.k}, and the constraint from contracting with $l_{\rho}$, is simply given by the sum of $\bar{L}^{\mu \nu \rho \sigma}_{n,k}$ and $\bar{L}^{\mu \nu \rho \sigma}_{n,l}$
\begin{align}
\bar{L}^{\mu \nu \rho \sigma}_n &= \bar{L}_{n,k}^{\mu \nu \rho \sigma} + \bar{L}_{n,l}^{\mu \nu \rho \sigma} \notag\\
 &= -\frac{\kappa^2}{4}\sum_{i=1}^n \left[ \frac{p_i^{\rho}p_i^{\sigma}}{p_i \cdot l} l^{\nu} - p_{i}^{(\rho} \eta^{\sigma) \nu} -\frac{1}{k \cdot l} \left( l^{\nu} p_{i}^{(\rho} k^{\sigma)} - (p_i\cdot l) \eta^{\nu (\rho} k^{\sigma)} \right) \right] \partial_i^{\mu} T_n  \notag\\
& - \frac{\kappa^2}{4 k \cdot l} \sum_{i=1}^n\left[ \frac{p_i^{\rho}p_i^{\sigma}}{p_i \cdot l} l^{\nu} l^{\mu} - 2 p_{i}^{(\rho} \eta^{\sigma) (\nu} l^{\mu)} + (p_i\cdot l) \eta^{\nu (\rho} \eta^{\sigma) \mu} \right] k_{\alpha} \partial_i^{\alpha} T_n \notag\\
& - \frac{\kappa^2}{2 k \cdot l} \sum_{i=1}^n\left[ l^{\mu} l^{\nu} p_i^{(\rho} \partial_i^{\sigma)} + (p_i \cdot l) \eta^{\rho(\mu} \eta^{\nu) \sigma} l_{\beta}\partial_i^{\beta} -  2 l^{(\mu} \eta^{\nu) (\rho} p_i^{\sigma)} l_{\beta}\partial_i^{\beta} \right] T_n \notag\\
& \quad  + \{k\,; \mu\,, \nu\} \leftrightarrow \{l\,; \rho\,,\sigma\}\;.
\label{l.sol}
\end{align}

We further note from \ref{l.sol} that $k_{\mu} l_{\rho} \bar{L}_n^{\mu \nu \rho \sigma} = 0$. This in turn implies that \ref{nn.kl} provides a constraint for $N_{n}^{\mu \nu \rho \sigma}$, whose solution we denote by $N_{n,kl} ^{\mu \nu \rho \sigma}$, and on using  \ref{n.kl} we find 
\begin{align}
 N_{n,kl}^{\mu \nu \rho \sigma} &= \frac{\kappa^2}{2} \sum_{i,j= 1}^n p_i^{\nu} p_j^{\sigma} \partial_i^{\mu} \partial_j^{\rho} T_n + \frac{\kappa^2}{2} \sum_{i= 1}^n \left(p_i^{\nu} \eta^{\mu \sigma} \partial_i^{\rho} + p_i^{\sigma} \eta^{\nu \rho} \partial_i^{\mu}\right) T_n\;.
\label{kl.mkl}
\end{align}

From \ref{nn.k} we have
\be
  N_{n,k}^{\mu \nu \rho \sigma} = \frac{\kappa^2}{2} \sum_{i,j= 1}^n p_i^{\nu} p_j^{\sigma} \partial_i^{\mu} \partial_j^{\rho} T_n + \frac{\kappa^2}{2} \sum_{i= 1}^n \left(\frac{p_i^{\rho}  p_i^{\sigma}}{p_i \cdot l} l^{\nu} \partial_i^{\mu} + p_i^{\nu} \eta^{\mu \sigma} \partial_i^{\rho}\right) T_n\,. \label{nn.k2}
\ee

The corresponding constraint solution from contracting $ N_{n}^{\mu \nu \rho \sigma}$ with $l_{\rho}$, i.e. $N_{n,l}^{\mu \nu \rho \sigma}$, would follow from applying $\{k\,; \mu\,, \nu\} \leftrightarrow \{l\,; \rho\,,\sigma\}$ on \ref{nn.k2} and has the result

\be
 N_{n,l}^{\mu \nu \rho \sigma} = \frac{\kappa^2}{2} \sum_{i,j= 1}^n p_i^{\nu} p_j^{\sigma} \partial_i^{\mu} \partial_j^{\rho} T_n + \frac{\kappa^2}{2} \sum_{i= 1}^n \left(\frac{p_i^{\mu}  p_i^{\nu}}{p_i \cdot k} k^{\sigma} \partial_i^{\rho} + p_i^{\sigma} \eta^{\nu \rho} \partial_i^{\mu}\right) T_n \,.
\label{nn.l2}
\ee

We can now note that we can obtain $ N_{n,kl}^{\mu \nu \rho \sigma}$ by either contracting \ref{nn.k2} with $l_{\rho}$, or \ref{nn.l2} with $k_{\mu}$: 
\be
l_{\rho}  N_{n,k}^{\mu \nu \rho \sigma} \equiv l_{\rho} \left(N_{n,kl} ^{\mu \nu \rho \sigma}\right)\;,\qquad  \qquad k_{\mu}  N_{n,l}^{\mu \nu \rho \sigma} \equiv k_{\mu} \left(N_{n,kl} ^{\mu \nu \rho \sigma}\right)  \,.
\ee

From this property, it follows that the general solution for $ N^{\mu\nu\rho\sigma}_n$ satisfying all constraints is given by
\begin{align}
 N^{\mu\nu\rho\sigma}_n &=  N_{n,k}^{\mu \nu \rho \sigma} +  N_{n,l}^{\mu \nu \rho \sigma} -  N_{n,kl}^{\mu \nu \rho \sigma}\;, \notag\\
&= \frac{\kappa^2}{2} \sum_{i,j= 1}^n p_i^{\nu} p_j^{\sigma} \partial_i^{\mu} \partial_j^{\rho} T_n + \frac{\kappa^2}{2} \sum_{i= 1}^n \left(\frac{p_i^{\rho}  p_i^{\sigma}}{p_i \cdot l} l^{\nu} \partial_i^{\mu} + \frac{p_i^{\mu}  p_i^{\nu}}{p_i \cdot k} k^{\sigma} \partial_i^{\rho}\right) T_n\;.
\label{n.sol}
\end{align}

Lastly, in the case of \ref{llt.k}, we can go further and demonstrate that 
\be
\tilde{L}_n^{\mu\nu\rho\sigma} \approx \frac{\kappa^2}{2}  \sum_{i= 1}^n \frac{\Upsilon_{i,k}^{\mu \nu \rho \sigma}(p_i, k, l)}{(p_i\cdot (k+l))^2} T_n \, ,
\label{lom2.up}
\ee
where the $\approx$ symbol indicates that they are equal up to terms that are pure gauge. This relation can be used to subtract $\tilde{L}_n^{\mu\nu\rho\sigma}$ from $L_n^{\mu\nu\rho\sigma}$ in \ref{m2.ex0} so that only $\bar{L}^{\mu \nu \rho \sigma}_n$ will appear in \ref{m2.ex0}.

To see \ref{lom2.up}, we proceed as before and derive the solutions for the constraints that result from contracting with the soft momenta. We use \ref{llt.k} to find the following constraints with respect to $k_{\mu}$, $l_{\rho}$ and $k_{\mu} l_{\rho}$
\begin{align}
\frac{\kappa^2}{2} \sum_{i= 1}^n \frac{k_{\mu}\Upsilon_i^{\mu \nu \rho \sigma}(p_i, k, l)}{(p_i \cdot(k+l))^2} T_n - k_{\mu}\tilde{L}_{n,k}^{\mu \nu \rho \sigma} &= 0\,, \label{tn.k}\\
\frac{\kappa^2}{2} \sum_{i= 1}^n \frac{l_{\rho}\Upsilon_i^{\mu \nu \rho \sigma}(p_i, k, l)}{(p_i \cdot(k+l))^2} T_n - l_{\rho}\tilde{L}_{n,l}^{\mu \nu \rho \sigma} &= 0 \,,\label{tn.l}\\
\frac{\kappa^2}{2} \sum_{i= 1}^n \frac{k_{\mu}l_{\rho}\Upsilon_i^{\mu \nu \rho \sigma}(p_i, k, l)}{(p_i \cdot(k+l))^2} T_n - k_{\mu}l_{\rho}\tilde{L}_{n,kl}^{\mu \nu \rho \sigma} &= 0 \,.\label{tn.kl}
\end{align}
Here the notations $\tilde{L}_{n,k}^{\mu \nu \rho \sigma}$, $\tilde{L}_{n,l}^{\mu \nu \rho \sigma}$ and $\tilde{L}_{n,kl}^{\mu \nu \rho \sigma}$ indicate the expressions after contracting with $k_{\mu}$, $l_{\rho}$ and $k_{\mu}l_{\rho}$ respectively. 
We can now proceed exactly as we did for the $\bar{L}^{\mu\nu\rho\sigma}_n$ solution. We first find the following constraint solutions for \ref{tn.k} - \ref{tn.kl}:
\begin{align}
\tilde{L}_{n,k}^{\mu \nu \rho \sigma} & = \frac{\kappa^2}{2} \sum_{i= 1}^n \frac{1}{(p_i \cdot(k+l))^2} \left[ \eta^{\mu (\rho}\eta^{\sigma) \nu} \left((p_i \cdot l)^2 + \frac{1}{2}(p_i \cdot l)(p_i \cdot k) \right) + l^{\mu} l^{\nu} p_i^{\rho} p_i^{\sigma} - 2 (p_i \cdot l) l^{(\mu} \eta^{\nu)(\rho}p_i^{\sigma)}  \right. \notag\\
&\left. + \frac{1}{2}\left( p_i^{\mu} \left( (p_i \cdot l) k^{(\rho} \eta^{\sigma) \nu} - l^{\nu} k^{(\rho}p_i^{\sigma)} \right) + (k \cdot l) p_i^{\mu} \eta^{\nu (\rho} p_i^{\sigma)} + 2 (p_i \cdot k) l^{[\mu} \eta^{\nu](\rho} p_i^{\sigma)} \right) \right], \label{ntk.sol}\\
\tilde{L}_{n,l}^{\mu \nu \rho \sigma} & = \frac{\kappa^2}{2} \sum_{i= 1}^n \frac{1}{(p_i \cdot(k+l))^2} \left[ \eta^{\mu (\rho}\eta^{\sigma) \nu} \left((p_i \cdot k)^2 + \frac{1}{2}(p_i \cdot l)(p_i \cdot k) \right) + k^{\rho} k^{\sigma} p_i^{\mu} p_i^{\nu} - 2 (p_i \cdot k) k^{(\rho} \eta^{\sigma)(\mu}p_i^{\nu)}  \right. \notag\\
&\left.  + \frac{1}{2}\left( p_i^{\rho} \left( (p_i \cdot k) l^{(\mu} \eta^{\nu) \sigma} - k^{\sigma} l^{(\mu}p_i^{\nu)} \right) + (k \cdot l) p_i^{\rho} \eta^{\sigma (\nu} p_i^{\mu)} + 2 (p_i \cdot l) k^{[\rho} \eta^{\sigma](\mu} p_i^{\nu)} \right) \right], \label{ntl.sol}\\
\tilde{L}_{n,kl}^{\mu \nu \rho \sigma} & = \frac{\kappa^2}{8} \sum_{i= 1}^n \frac{1}{(p_i \cdot(k+l))^2} \left[ \eta^{\mu \rho} (p_i \cdot l)(p_i \cdot k) + k^{\rho} p_i^{\mu} (p_i \cdot l) + l^{\mu} p_i^{\rho} (p_i \cdot k) + p_i^{\mu} p_i^{\rho} (k \cdot l)  \right] \eta^{\sigma \nu}\,, \label{ntkl.sol}
\end{align}
where, as with the solutions for $ \bar{N}^{\mu\nu\rho\sigma}_n$, we have the relations
\be
l_{\rho} \tilde{L}_{n,k}^{\mu \nu \rho \sigma} \equiv l_{\rho} \left(\tilde{L}_{n,kl} ^{\mu \nu \rho \sigma}\right)\;,\qquad  \qquad k_{\mu} \tilde{L}_{n,l}^{\mu \nu \rho \sigma} \equiv k_{\mu} \left(\tilde{L}_{n,kl} ^{\mu \nu \rho \sigma}\right)  \,.
\ee
On using \ref{ntk.sol} - \ref{ntkl.sol}, we may thus derive the following general solution which satisfies the constraints \ref{tn.k} - \ref{tn.kl}:
\begin{align}
\tilde{L}_n^{\mu \nu \rho \sigma} & = \tilde{L}_{n,k}^{\mu \nu \rho \sigma} + \tilde{L}_{n,l}^{\mu \nu \rho \sigma} - \tilde{L}_{n,kl}^{\mu \nu \rho \sigma} \notag\\
&= \frac{\kappa^2}{2}  \sum_{i= 1}^n \frac{\Upsilon_i^{\mu \nu \rho \sigma}(p_i, k, l)}{(p_i\cdot (k+l))^2}T_n + \mathcal{E}_{\Upsilon}^{\mu \nu \rho \sigma},
\label{ltf}
\end{align}
where
\begin{align}
\mathcal{E}_{\Upsilon}^{\mu \nu \rho \sigma} & = \frac{\kappa^2}{2} \sum_{i= 1}^n \frac{1}{(p_i \cdot(k+l))^2} \left[ 2 p_i^{(\mu}l^{\nu)} p_i^{(\rho}k^{\sigma)} - 2 (k \cdot l) p_i^{(\mu} \eta^{\nu) (\rho} p_i^{\sigma)} \right. \notag\\
&\left.  + \frac{1}{2}\left( p_i^{\mu} \left( (p_i \cdot l) k^{(\rho} \eta^{\sigma) \nu} - l^{\nu} k^{(\rho}p_i^{\sigma)} \right) + (k \cdot l) p_i^{\mu} \eta^{\nu (\rho} p_i^{\sigma)} + 2 (p_i \cdot k) l^{[\mu} \eta^{\nu](\rho} p_i^{\sigma)} \right) \right.\notag\\
&\left.  + \frac{1}{2}\left( p_i^{\rho} \left( (p_i \cdot k) l^{(\mu} \eta^{\nu) \sigma} - k^{\sigma} l^{(\mu}p_i^{\nu)} \right) + (k \cdot l) p_i^{\rho} \eta^{\sigma (\nu} p_i^{\mu)} + 2 (p_i \cdot l) k^{[\rho} \eta^{\sigma](\mu} p_i^{\nu)} \right) \right. \notag\\
& \left.  -\frac{1}{4}  \eta^{\sigma \nu} \left( \eta^{\mu \rho} (p_i \cdot l)(p_i \cdot k) + k^{\rho} p_i^{\mu} (p_i \cdot l) + l^{\mu} p_i^{\rho} (p_i \cdot k) + p_i^{\mu} p_i^{\rho} (k \cdot l)  \right)   \right] \label{eu.def}
\end{align}
has the property
$$ k_{\mu}\mathcal{E}_{\Upsilon}^{\mu \nu \rho \sigma} = 0 = l_{\rho}\mathcal{E}_{\Upsilon}^{\mu \nu \rho \sigma}$$
We note that we always have the freedom to add a `transverse term' to the amplitude, which vanishes on contracting with either the $k$ or $l$ graviton momenta. As a consequence, we can gauge away $\mathcal{E}_{\Upsilon}^{\mu \nu \rho \sigma}$, and thus \ref{ltf} can be written as \ref{lom2.up}.

We have now obtained explicit forms for all coefficients appearing in 
\ref{m2.ex0}, such that we may write explicit forms for the double soft-graviton amplitude, stripped of external graviton fields. Given the lengthy form of the result, it is convenient to report separate results at LO and NLO in the soft (momentum) expansion. That is, we can decompose the stripped amplitude as:
\begin{align}
\mathcal{M}^{\mu \nu \rho \sigma} (\{p_i\}\,, k\,,l) = \mathcal{M}_{\text{LO}}^{\mu \nu \rho \sigma} (\{p_i\}\,, k\,,l) + \mathcal{M}_{\text{NLO}}^{\mu \nu \rho \sigma} (\{p_i\}\,, k\,,l) + \mathcal{O}(k\,, l)\;,
\label{m2.fin}
\end{align}
with leading and subleading double soft factor contributions
\begin{align}
&\mathcal{M}_{\text{LO}}^{\mu \nu \rho \sigma}(\{p_i\}\,, k\,,l) = \frac{\kappa^2}{2}  \sum_{i= 1}^n \sum_{j= 1}^n  \frac{p_i^{\mu}  p_i^{\nu}}{p_i \cdot k} \frac{p_j^{\rho}  p_j^{\sigma}}{p_j \cdot l}T_n + \frac{\kappa^2}{2} \sum_{i= 1}^n \frac{\alpha_i^{\mu \nu \rho \sigma}(k\,,l)}{p_i(k+l)} T_n   \notag\\
&\quad + \frac{\kappa}{2} \sum_{j= 1}^n \frac{p_j^{\rho}  p_j^{\sigma}}{p_j \cdot l} \left(\sum_{i= 1}^n  \frac{p_i^{\mu}  p_i^{\nu}}{p_i \cdot k} k_{\alpha} \partial_i^{\alpha} T_n +   N_n^{\mu \nu}\right) + \frac{\kappa}{2} \sum_{i= 1}^n \frac{p_i^{\mu}  p_i^{\nu}}{p_i \cdot k}\left(\sum_{j= 1}^n \frac{p_j^{\rho}  p_j^{\sigma}}{p_j \cdot l} l_{\beta} \partial_j^{\beta}T_n + N_n^{\rho \sigma} \right)\;,
\label{m2.lox}
\end{align}
and 
\begin{align}
&\mathcal{M}_{\text{NLO};k,l}^{\mu \nu \rho \sigma} (\{p_i\}\,, k\,,l) = \frac{\kappa^2}{2}  \sum_{i= 1}^n \sum_{j= 1}^n  \frac{p_i^{\mu}  p_i^{\nu}}{p_i \cdot k} \frac{p_j^{\rho}  p_j^{\sigma}}{p_j \cdot l} k_{\alpha} l_{\beta} \partial_i^{\alpha} \partial_j^{\beta}T_n +  \frac{\kappa}{2} \sum_{j= 1}^n \frac{p_j^{\rho}  p_j^{\sigma}}{p_j \cdot l} l_{\beta}\partial_j^{\beta}N_{n}^{\mu \nu}  \notag\\
& + \frac{\kappa}{2} \sum_{i= 1}^n \frac{p_i^{\mu}  p_i^{\nu}}{p_i \cdot k} k_{\alpha}\partial_i^{\alpha}N_{n}^{\rho \sigma} 
+ \frac{\kappa}{2} \sum_{j= 1}^n \frac{p_j^{\rho}  p_j^{\sigma}}{p_j \cdot l} \left(\kappa \sum_{i= 1}^n  \frac{p_i^{\mu}  p_i^{\nu}}{p_i \cdot k} \frac{k_{\alpha} k_{\beta}}{2} \partial_i^{\alpha} \partial_i^{\beta} T_n +   k_{\alpha} \partial_k^{\alpha} N_n^{\mu \nu}\right) \notag\\
&+ \frac{\kappa}{2} \sum_{i= 1}^n \frac{p_i^{\mu}  p_i^{\nu}}{p_i \cdot k}\left(\kappa \sum_{j= 1}^n \frac{p_j^{\rho}  p_j^{\sigma}}{p_j \cdot l} \frac{l_{\alpha} l_{\beta}}{2} \partial_j^{\alpha} \partial_j^{\beta} T_n  + l_{\beta} \partial_l^{\beta} N_n^{\rho \sigma} \right) \notag\\
 & + \frac{\kappa^2}{2} \sum_{i= 1}^n \frac{\alpha_i^{\mu \nu \rho \sigma}(k\,,l)}{p_i(k+l)}(k+l)_{\alpha} \partial_i^{\alpha}T_n  + N^{\mu\nu\rho\sigma}_n + \bar{L}^{\mu\nu\rho\sigma}_n  \,,
 \label{m2.nlox}
\end{align}
respectively.
On using \ref{alph.def} and \ref{n.kl} in \ref{m2.lox}, we recover the leading order double soft graviton factorized amplitude, which we write as
\begin{equation}
\mathcal{M}_{\text{LO}}^{\mu \nu \rho \sigma}(\{p_i\}\,, k\,,l) = \left[\sum_{i,j= 1}^n \left(\mathcal{M}_{\text{LO}}\right)_{i,j;k,l}^{\mu \nu \rho \sigma}  + \sum_{i= 1}^n \left(\mathcal{M}_{\text{LO}}\right)_{i;k,l}^{\mu \nu \rho \sigma}\right] T_n,
\end{equation}
where we defined the double sum and single sum contributions
\begin{align}
&\left(\mathcal{M}_{\text{LO}}\right)_{i,j;k,l}^{\mu \nu \rho \sigma} = \frac{\kappa^2}{2} \left[\frac{p_i^{\mu}  p_i^{\nu}}{p_i \cdot k}\frac{p_j^{\rho}  p_j^{\sigma}}{p_j \cdot l}  - i\left(\frac{p_i^{\mu}  p_i^{\nu}}{p_i \cdot k} \frac{p_j^{(\sigma}J_j^{\rho) \alpha}l_{\alpha}}{p_j \cdot l} + \frac{p_j^{\rho}  p_j^{\sigma}}{p_j \cdot l}  \frac{p_i^{(\nu}J_i^{\mu) \alpha}k_{\alpha}}{p_i \cdot k}\right) \right] \label{mlo.1}
\end{align}
and
\begin{align}
&\left(\mathcal{M}_{\text{LO}}\right)_{i;k,l}^{\mu \nu \rho \sigma} = \frac{\kappa^2}{2}\frac{1}{p_i(k+l)}\left[ - \frac{k\cdot l}{(p_i\cdot k) (p_i\cdot l)} p_i^{\mu} p_i^{\nu} p_i^{\rho} p_i^{\sigma} - 2 p_i^{(\mu}\eta^{\nu) (\rho} p_i^{\sigma)}   + 2 \frac{p_i^{\mu} p_i^{\nu}}{p_i\cdot k} p_i^{(\rho} k^{\sigma)} + 2 \frac{p_i^{\rho} p_i^{\sigma}}{p_i\cdot l} p_i^{(\mu} l^{\nu)}  \right. \notag\\
& \left. \quad  + \frac{1}{k\cdot l}  \left(  \eta^{\rho (\mu} \eta^{\nu) \sigma} \left((p_i\cdot k)^2 + (p_i\cdot l)^2+ (p_i\cdot k)(p_i\cdot l) \right) + p_i^{\mu} p_i^{\nu} k^{\rho} k^{\sigma} +  l^{\mu} l^{\nu} p_i^{\rho} p_i^{\sigma} - 2 p_i^{(\mu}l^{\nu)} p_i^{(\rho}  k^{\sigma)}   \phantom{ \frac{p_i^{\mu} }{p_i \cdot k}} \right. \right.\notag\\
&\left. \left. \phantom{ \frac{p_i^{\mu} }{p_i \cdot k}}  - 2 (p_i \cdot l) l^{(\mu} \eta^{\nu) (\rho}  p_i^{\sigma)} - 2 (p_i \cdot k) p_i^{(\mu} \eta^{\nu) (\rho}  k^{\sigma)} \right) \right]\,,
\label{mlo.2}
\end{align} 
respectively. 

On the other hand, by using \ref{alph.def}, \ref{n.kl}-\ref{an.kl}, \ref{l.sol} and \ref{n.sol} in \ref{m2.nlox}, we find
\begin{align}
\mathcal{M}_{\text{NLO}}^{\mu \nu \rho \sigma}(\{p_i\}\,, k\,,l) &= \left[\sum_{i,j= 1}^n \left(\mathcal{M}_{\text{NLO}}\right)_{i,j;k,l}^{\mu \nu \rho \sigma} + \sum_{i,j= 1}^n \left(\mathcal{M}'_{\text{NLO}}\right)_{i,j;k,l}^{\mu \nu \rho \sigma}  + \sum_{i= 1}^n \left(\mathcal{M}_{\text{NLO}}\right)_{i;k,l}^{\mu \nu \rho \sigma} \right] T_n \;,
\label{m2.nlo}
\end{align}
where the double-sum contributions are 
\begin{align}
\left(\mathcal{M}_{\text{NLO}}\right)_{i,j;k,l}^{\mu \nu \rho \sigma}  &=  \frac{\kappa^2}{4} \left[ \left(\frac{p_i^{\mu}  p_i^{\nu}}{p_i \cdot k} \frac{l_{\alpha}l_{\beta}}{p_j \cdot l} J_j^{\alpha \rho}J_j^{\sigma \beta} + \frac{p_j^{\rho}  p_j^{\sigma}}{p_j \cdot l} \frac{k_{\alpha}k_{\beta}}{p_i \cdot k} J_i^{\alpha \mu}J_i^{\nu \beta} \right) \right. \notag\\
& \left. \qquad \quad  - \frac{k_{\alpha}}{p_i \cdot k} \frac{l_{\beta}}{p_j \cdot l} \left( p_i^{\nu}J_i^{\mu \alpha}  p_j^{\sigma}J_j^{\rho \beta} + p_j^{\sigma}J_j^{\rho \beta} p_i^{\nu}J_i^{\mu \alpha} \right)  \right]\;, \label{mnlo.1}\\
\left(\mathcal{M}'_{\text{NLO}}\right)_{i,j;k,l}^{\mu \nu \rho \sigma} &= -\frac{i \kappa^2}{4} \frac{1}{k \cdot l} \left[k_{\alpha} l_{\beta} \left(\frac{p_i^{\mu}  p_i^{\nu}}{p_i \cdot k} k^{(\rho}J_j^{\sigma) \beta} \partial_i^{\alpha} +  \frac{p_j^{\rho}  p_j^{\sigma}}{p_j \cdot l} l^{(\mu}J_i^{\nu) \alpha} \partial_j^{\beta}\right)\right.\notag\\
&\left.- \frac{k_{\alpha}}{2} \left(p_j^{\rho} l^{(\mu}J_i^{\nu) \alpha} \partial_j^{\sigma} +  p_j^{\sigma} l^{(\mu}J_i^{\nu) \alpha} \partial_j^{\rho} - l^{(\mu} p_i^{\nu)} J_i^{\rho \sigma} \partial_i^{\alpha} +p_i ^{\alpha} J_j^{\rho \sigma}  l^{(\mu} \partial_i^{\nu)} \right) \right. \notag\\
&\left. - \frac{l_{\beta}}{2} \left(p_i^{\mu} k^{(\rho}J_j^{\sigma) \beta} \partial_i^{\nu} +  p_i^{\nu} k^{(\rho}J_j^{\sigma) \beta} \partial_i^{\mu} -k^{(\rho} p_j^{\sigma)} J_i^{\mu \nu} \partial_j^{\beta} + p_j^{\beta}  J_i^{\mu \nu} k^{(\rho} \partial_j^{\sigma)} \right) \right] \,, \notag\\
& + \frac{\kappa^2}{4} \left[ \frac{p_j^{\rho}  p_j^{\sigma}}{p_j \cdot l} \left(\eta^{\mu \nu} k_{\alpha} - k^{\mu} \delta^{\nu}_{\alpha} - \frac{p_i^{\mu} k^{\nu}}{p_i \cdot k} k_{\alpha}\right) \partial_i^{\alpha} + \frac{p_i^{\mu}  p_i^{\nu}}{p_i \cdot k} \left(\eta^{\rho \sigma} l_{\alpha} - l^{\rho} \delta^{\sigma}_{\alpha} - \frac{p_j^{\rho} l^{\sigma}}{p_j \cdot l} l_{\alpha}\right) \partial_j^{\alpha} \right. \notag\\
& \left. \qquad  \qquad +\frac{i}{2} \left(p_j^{\sigma} J_i^{\mu \nu} \partial_j^{\rho} - \frac{p_j^{\rho} p_j^{\sigma}}{p_j \cdot l} J_i^{\mu \nu} l_{\beta}\partial_j^{\beta} + p_i^{\nu} J_j^{\rho \sigma} \partial_i^{\mu} - \frac{p_i^{\mu} p_i^{\nu}}{p_i \cdot k} J_j^{\rho \sigma} k_{\alpha}\partial_i^{\alpha}\right) \right]\label{mnlo.2}
\end{align}

and we have separated the double-sum terms into those that result from products of single soft factors \ref{mnlo.1} and those that do not in \ref{mnlo.2}. The latter contain terms having denominators with the mixed combination $(k\cdot l)^{-1}$. 
The single-sum terms in \ref{m2.nlo} are given by
\begin{align}
\left(\mathcal{M}_{\text{NLO}}\right)_{i;k,l}^{\mu \nu \rho \sigma} &= - i \kappa^2 \frac{1}{p_i \cdot (k+l)} \left\{ \frac{k_{\alpha} l_{\beta}}{k \cdot l} \left(\frac{1}{4} \eta^{\mu (\rho} \eta^{\sigma) \nu} p_i \cdot (k-l) +  l^{(\mu} \eta^{\nu)(\rho} p_i^{\sigma)} - k^{(\rho} \eta^{\sigma) (\mu} p_i^{\nu)} \right) J_i^{\alpha \beta} \right. \notag\\
&\left. \qquad \quad - \frac{(k+l)_{\alpha}}{2} \left[ p_i^{(\mu}\eta^{\nu) (\sigma} J_i^{\rho) \alpha} + p_i^{(\rho}\eta^{\sigma) (\nu} J_i^{\mu) \alpha}  + \frac{p_i^{\rho}  p_i^{\sigma}}{p_i \cdot l} \left(\frac{(k \cdot l)}{2 p_i\cdot k} p_i^{(\nu} J_i^{\mu) \alpha}  - 2 l^{(\nu} J_i^{\mu) \alpha}  \right)   \right. \right. \notag\\
 & \left. \left. \qquad  \qquad \quad   +  \frac{p_i^{\mu}  p_i^{\nu}}{p_i \cdot k} \left(\frac{(k \cdot l)}{2 p_i \cdot l}p_i^{(\sigma} J_i^{\rho) \alpha} - 2 k^{(\sigma} J_i^{\rho) \alpha} \right) - \frac{1}{k \cdot l} \left(l^{\mu}l^{\nu} p_i^{(\rho} J_i^{\sigma) \alpha}  +   k^{\rho}k^{\sigma} p_i^{(\mu} J_i^{\nu) \alpha}\right) \right.  \right. \notag\\
 &\left. \left. \qquad \qquad  \qquad \quad +\frac{1}{k \cdot l} \left( p_i^{(\mu}l^{\nu)} k^{(\sigma} J_i^{\rho) \alpha} + p_i^{(\rho}k^{\sigma)} l^{(\nu} J_i^{\mu) \alpha} \right) \right] \right\}
\notag\\
& - \frac{i \kappa^2}{2} \left[ \frac{1}{p_i \cdot l} \left(\frac{p_i^{\sigma} J_i^{\rho \nu} l^{\mu}}{2} + p_i^{\nu} \eta^{\mu (\rho} J_i^{\sigma)  \beta} l_{\beta} \right)  +  \frac{1}{p_i \cdot k} \left(\frac{ p_i^{\nu} J_i^{\mu \sigma} k^{\rho}}{2} +  p_i^{\sigma} \eta^{\rho (\mu} J_i^{\nu) \alpha} k_{\alpha} \right) \right. \notag\\
& \left.  \qquad  \qquad \quad  - \frac{p_i^{\mu}  p_i^{\nu}}{p_i \cdot k} \frac{1}{p_i \cdot l}k^{(\sigma} J_i^{\rho) \alpha} l_{\alpha} - \frac{p_i^{\rho}  p_i^{\sigma}}{p_i \cdot l}\frac{1}{p_i \cdot k} l^{(\nu} J_i^{\mu) \alpha} k_{\alpha} \right] \;.
\label{mnlo.3}
\end{align}
The detailed derivation of \ref{m2.nlo} is a bit involved and is given in Appendix \ref{app2}. Hence, on replacing \ref{m2.fin} in \ref{a2.fin}, we get
\begin{align}
&\mathcal{A}_2(\{p_i\}\,,h) = \int_{k l} h_{\mu \nu}(-k) h_{\rho \sigma}(-l) \left[\sum_{i,j= 1}^n \left(\mathcal{M}_{\text{LO}}\right)_{i,j;k,l}^{\mu \nu \rho \sigma} + \sum_{i,j= 1}^n \left(\mathcal{M}_{\text{NLO}}\right)_{i,j;k,l}^{\mu \nu \rho \sigma} \right. \notag\\
&\left. + \sum_{i= 1}^n \left(\mathcal{M}_{\text{LO}}\right)_{i;k,l}^{\mu \nu \rho \sigma} + \sum_{i= 1}^n \left(\mathcal{M}_{\text{NLO}}\right)_{i;k,l}^{\mu \nu \rho \sigma} + \sum_{i,j= 1}^n \left(\mathcal{M}'_{\text{NLO}}\right)_{i,j;k,l}^{\mu \nu \rho \sigma} + \mathcal{O}(k,l) \right] T_n\;. \label{a2.fin2}
\end{align}
We note that as in the single soft graviton result, there are terms that would vanish on contracting with the gravitons either due to the de Donder gauge conditions, or from contracting with terms antisymmetric in the graviton indices. We have however chosen to write the expressions in their entirety for completeness. \ref{a2.fin2}, together with the detailed definitions of 
\ref{mlo.1}--\ref{mnlo.3} constitute the main new results of this paper, namely a generalisation of the previously known double-soft graviton theorem to first subleading order in the momentum expansion. 
We conclude this section by commenting on the individual contributions and our choice of notation. The terms in the top line of \ref{a2.fin2} result from products of single soft factors, and they involve a double sum over the external particle lines. The first two terms in the second line are the double soft graviton terms involving a single sum over the external particle lines, to leading and subleading order. The primed term $\left(\mathcal{M}'_{\text{NLO}}\right)_{i,j;k,l}^{\mu \nu \rho \sigma}$ is a double soft graviton contribution, but involves a double sum over the external particles. This is a piece that is required to satisfy gauge invariance, and does not manifestly vanish by manipulating the derivatives involved in the term, or by momentum and angular momentum conservation of the hard amplitude.

\section{Effective double soft graviton dressing}
\label{sec4}

In the previous section, we derived a universal, gauge-invariant contribution of the double soft graviton factor up to subleading order. 
In this section, we highlight iterative properties of this result and show that it can be reproduced by an exponential (next-to) soft dressing factor acting on the non-radiative amplitude $T_n$, where the exponent has a simpler form than the amplitude itself. That is, we consider writing \ref{amp.se} in the form
\begin{align}
\mathcal{A}_0(\{p_i\}) + \mathcal{A}_1(\{p_i\}\,,h) + \mathcal{A}_2(\{p_i\}\,,h) \equiv \tilde{S}_n(\{p_i\}\,k\,,l\,,h) \; T_n \;  (1 + R)\,,
\label{amp2.dress}
\end{align}
with $R$ a possible remainder, that is required for subleading terms that do not exponentiate in the soft expansion of the left-hand side. Using \ref{a2.fin2} to match the two expressions on both sides of \ref{amp2.dress}, we find that the soft graviton dressing up to NNLO in the momentum expansion is given by
\begin{align}
&\tilde{S}_n(\{p_i\}\,k\,,l\,,h) = \exp \left[  \int_k\, h_{\mu \nu}(-k) \sum_{i= 1}^n \left( \left(\mathcal{M}_{\text{LO}}\right)_{i;k}^{\mu \nu} + \left(\mathcal{M}_{\text{NLO}}\right)_{i;k}^{\mu \nu} + \left(\mathcal{M}_{\text{NNLO}}\right)_{i;k}^{\mu \nu} \right) \right. \notag\\
&\left. + \int_{k l} h_{\mu \nu}(-k) h_{\rho \sigma}(-l) \left(\sum_{i= 1}^n \left(\mathcal{M}_{\text{LO}}\right)_{i;k,l}^{\mu \nu \rho \sigma} + \sum_{i= 1}^n \left(\mathcal{M}_{\text{NLO}}\right)_{i;k,l}^{\mu \nu \rho \sigma}  + \sum_{i,j= 1}^n \left(\mathcal{M}'_{\text{NLO}}\right)_{i,j;k,l}^{\mu \nu \rho \sigma} \right)  \right] \,. 
\label{efd.def}
\end{align}

The graviton dressing from the GWL formalism up to double soft gravitons was noted in \ref{sf.1}. Its single soft graviton contribution is the Weinberg soft graviton factor, contained in $\left(\mathcal{M}_{\text{LO}}\right)_{i;k}^{\mu \nu}$ in \ref{efd.def}. The double soft contributions were a combination of terms due to gravitons emitted from external legs -- the Born and seagull contributions, as well as those from the three-graviton vertex graphs. These terms are contained in the single sum piece 
$\left(\mathcal{M}_{\text{LO}}\right)_{i;k,l}^{\mu \nu \rho \sigma}$ in \ref{efd.def}. The ansatz in \ref{efd.def} now further includes the subleading contributions at single soft and double soft orders that contain angular momentum terms. Notably, on comparing with \ref{a2.fin2}, the soft graviton dressing in \ref{efd.def} does not include the following double sum terms
\be
\int_{k l} h_{\mu \nu}(-k) h_{\rho \sigma}(-l) \sum_{i,j= 1}^n \left[ \left(\mathcal{M}_{\text{LO}}\right)_{i,j;k,l}^{\mu \nu \rho \sigma} + \left(\mathcal{M}_{\text{NLO}}\right)_{i,j;k,l}^{\mu \nu \rho \sigma} \right]\;, \notag
\ee
that are present in the first line of \ref{a2.fin2}. We will now show that these terms result from products of single soft factors exactly following the expansion of the exponential dressing. Ignoring the remainder term in \ref{amp2.dress} and on expanding the soft graviton dressing, we have
\begin{align}
&\tilde{S}_n(\{p_i\}\,k\,,l\,,h) T_n = \left[ 1 + \int_k\, h_{\mu \nu}(-k) \sum_{i= 1}^n \left( \left(\mathcal{M}_{\text{LO}}\right)_{i;k}^{\mu \nu} + \left(\mathcal{M}_{\text{NLO}}\right)_{i;k}^{\mu \nu} + \left(\mathcal{M}_{\text{NNLO}}\right)_{i;k}^{\mu \nu} \right) \right. \notag\\ 
& \left. + \int_{kl} \frac{1}{2}  \, h_{\mu \nu}(-k)  h_{\rho \sigma}(-l)  \sum_{i,j= 1}^n \left( \left(\mathcal{M}_{\text{LO}}\right)_{i;k}^{\mu \nu} + \left(\mathcal{M}_{\text{NLO}}\right)_{i;k}^{\mu \nu} + \left(\mathcal{M}_{\text{NNLO}}\right)_{i;k}^{\mu \nu} \right) \right.\notag\\
&\left.\quad\times
\left( \left(\mathcal{M}_{\text{LO}}\right)_{j;l}^{\rho \sigma} + \left(\mathcal{M}_{\text{NLO}}\right)_{j;l}^{\rho \sigma} + \left(\mathcal{M}_{\text{NNLO}}\right)_{j;l}^{\rho \sigma} \right) \right. \notag\\
& \left. + \int_{k l} h_{\mu \nu}(-k) h_{\rho \sigma}(-l) \left(\sum_{i= 1}^n \left(\mathcal{M}_{\text{LO}}\right)_{i;k,l}^{\mu \nu \rho \sigma} + \sum_{i= 1}^n \left(\mathcal{M}_{\text{NLO}}\right)_{i;k,l}^{\mu \nu \rho \sigma} + \sum_{i,j= 1}^n \left(\mathcal{M}'_{\text{NLO}}\right)_{i,j;k,l}^{\mu \nu \rho \sigma} \right) \right] T_n + \mathcal{O}(h^3) \,.
\label{amp2.exp}
\end{align}
On the right-hand side of \ref{amp2.exp}, the first line recovers $\mathcal{A}_0(\{p_i\})$ and $\mathcal{A}_1(\{p_i\}\,,h)$ (up to sub-subleading order), while the last line recovers $\mathcal{A}_2(\{p_i\}\,,h)$ to subleading order apart from two contributions. These can be derived from the second line of \ref{amp2.exp}, which we expand and collect as the following three terms,
\begin{align}
&\int_{kl} \frac{1}{2}  \, h_{\mu \nu}(-k)  h_{\rho \sigma}(-l)  \sum_{i,j= 1}^n \left[\left(\mathcal{M}_{\text{LO}}\right)_{i;k}^{\mu \nu} \left(\mathcal{M}_{\text{LO}}\right)_{j;l}^{\rho \sigma} \right. \notag\\
&\left. \quad \qquad \qquad\qquad \qquad\qquad \qquad + \left( \left(\mathcal{M}_{\text{LO}}\right)_{i;k}^{\mu \nu} \left(\mathcal{M}_{\text{NLO}}\right)_{j;k}^{\rho \sigma} +  \left(\mathcal{M}_{\text{LO}}\right)_{j;l}^{\rho \sigma} \left(\mathcal{M}_{\text{NLO}}\right)_{i;k}^{\mu \nu} \right) \right] T_n\;, \label{mlo.exp}\\
&\int_{kl} \frac{1}{2}  \, h_{\mu \nu}(-k)  h_{\rho \sigma}(-l)  \sum_{i,j= 1}^n \left[ \left(\mathcal{M}_{\text{NLO}}\right)_{i;k}^{\mu \nu} \left(\mathcal{M}_{\text{NLO}}\right)_{j;l}^{\rho \sigma}  \right. \notag\\
& \left. \quad \qquad \qquad\qquad \qquad\qquad \qquad + \left( \left(\mathcal{M}_{\text{LO}}\right)_{i;k}^{\mu \nu} \left(\mathcal{M}_{\text{NNLO}}\right)_{j;k}^{\rho \sigma} +  \left(\mathcal{M}_{\text{LO}}\right)_{j;l}^{\rho \sigma} \left(\mathcal{M}_{\text{NNLO}}\right)_{i;k}^{\mu \nu} \right) \right] T_n \;, \label{mnlo.exp}\\
&\int_{kl} \frac{1}{2}  \, h_{\mu \nu}(-k)  h_{\rho \sigma}(-l)  \sum_{i,j= 1}^n \left[ \left(\mathcal{M}_{\text{NNLO}}\right)_{i;k}^{\mu \nu} \left(\mathcal{M}_{\text{NNLO}}\right)_{j;l}^{\rho \sigma} \right. \notag\\
& \left. \qquad \qquad \qquad\qquad \qquad \qquad + \left( \left(\mathcal{M}_{\text{NLO}}\right)_{i;k}^{\mu \nu} \left(\mathcal{M}_{\text{NNLO}}\right)_{j;k}^{\rho \sigma} +  \left(\mathcal{M}_{\text{NLO}}\right)_{j;l}^{\rho \sigma} \left(\mathcal{M}_{\text{NNLO}}\right)_{i;k}^{\mu \nu} \right) \right] T_n\;. \label{mnnlo.exp}
\end{align}
The first two of these occur at NNLO order in the soft expansion, and can be simplified as follows. For \ref{mlo.exp}, we find
\begin{align}
&\int_{kl} \frac{1}{2}  \, h_{\mu \nu}(-k)  h_{\rho \sigma}(-l)  \sum_{i,j= 1}^n \left[ \left(\mathcal{M}_{\text{LO}}\right)_{i;k}^{\mu \nu} \left(\mathcal{M}_{\text{LO}}\right)_{j;l}^{\rho \sigma} + \left( \left(\mathcal{M}_{\text{LO}}\right)_{i;k}^{\mu \nu} \left(\mathcal{M}_{\text{NLO}}\right)_{j;k}^{\rho \sigma} +  \left(\mathcal{M}_{\text{LO}}\right)_{j;l}^{\rho \sigma} \left(\mathcal{M}_{\text{NLO}}\right)_{i;k}^{\mu \nu} \right) \right] T_n\;,  \notag\\
&= \int_{kl} \, h_{\mu \nu}(-k)  h_{\rho \sigma}(-l)  \left[\frac{\kappa^2}{2} \sum_{i,j= 1}^n \left(\frac{p_i^{\mu}  p_i^{\nu}}{p_i \cdot k}\frac{p_j^{\rho}  p_j^{\sigma}}{p_j \cdot l}  - i\left(\frac{p_i^{\mu}  p_i^{\nu}}{p_i \cdot k} \frac{p_j^{(\sigma}J_j^{\rho) \alpha}l_{\alpha}}{p_j \cdot l} + \frac{p_j^{\rho}  p_j^{\sigma}}{p_j \cdot l}  \frac{p_i^{(\nu}J_i^{\mu) \alpha}k_{\alpha}}{p_i \cdot k}\right) \right) \right]T_n \notag\\
&\qquad \quad = \int_{kl}  \, h_{\mu \nu}(-k)  h_{\rho \sigma}(-l)  \sum_{i,j= 1}^n \left(\mathcal{M}_{\text{LO}}\right)_{i,j;k,l}^{\mu \nu \rho \sigma} T_n\;.
\label{mid.exp1}
\end{align}
where we made use of \ref{m1.lo}, \ref{m1.nlo} and \ref{mlo.1}. This term thus recovers the leading double sum contribution in the double soft graviton factor, required for the gauge invariance of the result. In the case of \ref{mnlo.exp}, we can use \ref{m1.lo},\ref{m1.nlo} and \ref{m1.nnlo} to find the contribution in \ref{mnlo.1}:
\begin{align}
&\int_{kl} \frac{1}{2}  \, h_{\mu \nu}(-k)  h_{\rho \sigma}(-l)  \sum_{i,j= 1}^n \left[\frac{1}{2} \left( \left(\mathcal{M}_{\text{NLO}}\right)_{i;k}^{\mu \nu} \left(\mathcal{M}_{\text{NLO}}\right)_{j;l}^{\rho \sigma} +  \left(\mathcal{M}_{\text{NLO}}\right)_{j;l}^{\rho \sigma} \left(\mathcal{M}_{\text{NLO}}\right)_{i;k}^{\mu \nu} \right) \right. \notag\\
& \left. \quad \qquad \qquad\qquad \qquad\qquad \qquad + \left( \left(\mathcal{M}_{\text{LO}}\right)_{i;k}^{\mu \nu} \left(\mathcal{M}_{\text{NNLO}}\right)_{j;k}^{\rho \sigma} +  \left(\mathcal{M}_{\text{LO}}\right)_{j;l}^{\rho \sigma} \left(\mathcal{M}_{\text{NNLO}}\right)_{i;k}^{\mu \nu} \right) \right] T_n  \notag\\
&= \int_{kl} \, h_{\mu \nu}(-k)  h_{\rho \sigma}(-l)  \left[ \frac{\kappa^2}{4} \sum_{i,j= 1}^n \left(  - \frac{k_{\alpha}}{p_i \cdot k} \frac{l_{\beta}}{p_j \cdot l} \left( p_i^{(\nu}J_i^{\mu) \alpha}  p_j^{(\sigma}J_j^{\rho) \beta} + p_j^{(\sigma}J_j^{\rho) \beta} p_i^{(\nu}J_i^{\mu) \alpha} \right) \right. \right.\notag\\
& \left. \left. \qquad \quad \qquad \qquad \qquad \qquad \qquad \qquad + \frac{p_i^{\mu}  p_i^{\nu}}{p_i \cdot k} \frac{l_{\alpha}l_{\beta}}{p_j \cdot l} J_j^{\alpha \rho}J_j^{\sigma \beta} + \frac{p_j^{\rho}  p_j^{\sigma}}{p_j \cdot l} \frac{k_{\alpha}k_{\beta}}{p_i \cdot k} J_i^{\alpha \mu}J_i^{\nu \beta} \right) \right] T_n \notag\\
&\qquad \qquad = \int_{kl}  \, h_{\mu \nu}(-k)  h_{\rho \sigma}(-l)  \sum_{i,j= 1}^n \left(\mathcal{M}_{\text{NLO}}\right)_{i,j;k,l}^{\mu \nu \rho \sigma} T_n.
\label{mid.exp2}
\end{align} 
The term \ref{mnnlo.exp} does not appear in the final result at the subleading double soft order, due to being of higher order in the momentum expansion. However, given the relevance of  \ref{mlo.exp} to the leading double soft factor and  \ref{mnlo.exp} at subleading double soft order, we expect \ref{mnnlo.exp} to provide a universal contribution of the sub-subleading double soft graviton factor, and it would be interesting to see if further work confirms this higher-order iterative structure. 

The above results establish that exponentiating the subleading and sub-subleading single soft factors yields relevant terms of the double soft factor, thereby motivating the exponentiation of subleading soft factors derived using the LBK theorem. A more complete test of \ref{efd.def} would require a next-to-soft order graviton calculation, to more specifically verify if products of single soft and double soft factors appear as universal terms in the triple soft graviton factor. 

\subsection{A double soft graviton dressing operator}

We conclude this section with a dressing operator version of \ref{efd.def}, which we will show has the form of a squeezed coherent state. Dressing operators can be derived using the GWL formalism by introducing radiation modes in terms of their creation and annihilation operators, which would replace the Fourier transform \ref{h.mom} throughout the analysis~\cite{Fernandes:2024xqr,Cristofoli:2021jas}. The soft graviton modes \footnote{As noted earlier, these modes are not assumed to be {\emph {strictly soft}}, but rather, there is a cutoff on the energy. With multiple gravitons, the cutoff is placed on the sum over the energies of the soft gravitons, which can be implemented through appropriate Heaviside step functions in the momentum space integrands \cite{Fernandes:2024xqr}.  We do not explicitly denote them in the following discussion.
} are given by
\begin{align}
h_{\mu \nu}(y) &= \sum_{i' = 1}^2\int\limits_{\vec{k}}\left( \epsilon^*_{i'; \mu \nu}(k) a_{i'}^{\dagger}(k) e^{-i k \cdot y} + \epsilon_{i'; \mu \nu}(k) a_{i'}(k) e^{i k \cdot y}\right)
\label{h.mode}
\end{align}
with the graviton creation and annihilation operators satisfying 
\begin{align}
\left[a_{i'}(k) \,, a_{j'}^{\dagger}(k')\right] = \delta_{i'j'} 2 \omega_{k} (2 \pi)^3 \delta^{(3)}(\vec{k} - \vec{k}') \,;,
\label{ca}
\end{align}
where $i'\,,j'\,,\cdots$ denote graviton polarization indices, and we now consider on-shell modes $k^2 = 0$, with $k^{\mu} = (k^0\,, \vec{k}) = \omega_k (1\,, \hat{k})$. The integral measure in \ref{h.mode} is that for on-shell gravitons, which we can consider in our results by the following replacement
\begin{equation}
\int\limits_k \to \int\limits_k 2\pi \delta(k^2) \Theta(k^0) = \int \frac{d^3 k}{(2 \pi)^3 2\omega_k} \equiv \int\limits_{\vec{k}}.
\label{int.meas}
\end{equation}

In \cite{Fernandes:2024xqr}, a generalization of the gravitational dressing operator for single soft emission in \cite{Bonocore:2021qxh} to the double soft emissions was derived, and was shown to be a squeezed coherent state. In the following discussion, we will determine the dressing operator by a simpler argument and leave the detailed derivation to future work. The expression in \ref{efd.def} follows from our choice of \ref{h.mom}, which is as if we had only considered graviton annihilation modes in \ref{h.mode} (after an appropriate change of the integrand and measure). On determining \ref{efd.def} only in terms of graviton annihilation modes, we can subsequently infer the dressing exponent by requiring it to be a unitary operator. Apart from a minor subtlety concerning products of graviton creation and annihilation modes, which we will comment on, this result should agree with the dressing operator had we used \ref{h.mode} throughout.

On using on-shell soft graviton modes with momenta $k^{\mu} = \omega_k (1\,, \hat{k})$ and $l^{\mu} = \omega_l (1\,, \hat{l})$, we consider the following replacements in \ref{efd.def}
\begin{align}
\int\limits_k \to \int\limits_{\vec{k}}\,, \quad&\quad  \int\limits_l \to \int\limits_{\vec{l}}\,,\notag\\
h_{\mu \nu}(k) \to \sum_{i' = 1}^2 \epsilon_{i', \mu \nu}(k) a_{i'}(k)\,, \quad&\quad  h_{\rho \sigma}(l) \to \sum_{j' = 1}^2 \epsilon_{j', \rho \sigma}(k) a_{j'}(l) \,,
\end{align}
and define
\begin{align}
\kappa \, \alpha^*_{i'}(k) &= -\epsilon_{i',\mu \nu}(k) \sum_{i= 1}^n \left( \left(\mathcal{M}_{\text{LO}}\right)_{i;-k}^{\mu \nu} + \left(\mathcal{M}_{\text{NLO}}\right)_{i;-k}^{\mu \nu} + \left(\mathcal{M}_{\text{NNLO}}\right)_{i;-k}^{\mu \nu} \right) \label{alpha.def}\\
\frac{\kappa^2}{2} \,\beta^*_{i'j'}(k\,,l) &= \epsilon_{i',\mu \nu}(k) \epsilon_{j',\rho \sigma}(l)\left(\sum_{i= 1}^n \left(\mathcal{M}_{\text{LO}}\right)_{i;-k,-l}^{\mu \nu \rho \sigma} + \sum_{i= 1}^n \left(\mathcal{M}_{\text{NLO}}\right)_{i;-k,-l}^{\mu \nu \rho \sigma}  + \sum_{i,j= 1}^n \left(\mathcal{M}'_{\text{NLO}}\right)_{i,j;-k,-l}^{\mu \nu \rho \sigma} \right) \label{beta.def}
\end{align}
to find
\begin{align}
&\tilde{S}_n(\{p_i\}\,k\,,l\,,h) = \exp \left[ -\kappa \sum_{i'=1}^2 \int\limits_{\vec{k}} \alpha^*_{i'}(k) a_{i'}(k) + \frac{\kappa^2}{2} \sum_{i',j'=1}^2 \int\limits_{\vec{k}} \int\limits_{\vec{l}} \beta^*_{i'j'}(k\,,l) a_{i'}(k)a_{j'}(l)  \right] \,. 
\label{efd.op1}
\end{align}

By requiring the graviton dressing to be a unitary displacement operator, the integrands in \ref{efd.op1} are expected to involve
\begin{align}
\mathcal{D}^{\alpha}_{i'}(k) &= \alpha_{i'}(k) a_{i'}^{\dagger}(k) - \alpha^*_{i'}(k) a_{i'}(k)\label{d.coh}\\
\mathcal{S}^{\beta}_{i'j'}(k\,,l) &= \beta^*_{i'j'}(k\,,l) a_{i'}(k)a_{j'}(l) - \beta_{i'j'}(k\,,l) a_{i'}^{\dagger}(k)a_{j'}^{\dagger}(l)  \label{d.sque} \,,
\end{align}
where $\alpha_{i'}(k)$ and $\beta_{i'j'}(k\,,l)$ are respectively the complex conjugates of \ref{alpha.def} and \ref{beta.def}.
We can appropriately replace the single and double graviton annihilation operator terms in \ref{efd.op1} with the expressions in \ref{d.coh} and \ref{d.sque} to find
\begin{align}
&\tilde{S}_n(\{p_i\}\,k\,,l\,,h) = \exp \left[ \kappa \sum_{i'=1}^2 \int\limits_{\vec{k}} \mathcal{D}^{\alpha}_{i'}(k) + \frac{\kappa^2}{2} \sum_{i',j'=1}^2 \int\limits_{\vec{k}} \int\limits_{\vec{l}} \mathcal{S}^{\beta}_{i'j'}(k\,,l)\right] \,. 
\label{efd.op2}
\end{align}

Following ref.~\cite{Fernandes:2024xqr}, we would also expect a general two-graviton term in the dressing exponent to contain the following combination
\begin{align}
\mathcal{S}^{\gamma}_{i'j'}(k\,,l) &= \gamma^*_{i'j'}(k\,,l) a_{j'}^{\dagger}(l)a_{i'}(k) - \gamma_{i'j'}(k\,,l) a_{i'}^{\dagger}(k)a_{j'}(l) 
\end{align}
with
\begin{align}
\frac{\kappa^2}{2} \,\gamma^*_{i'j'}(k\,,l) &= \epsilon_{i',\mu \nu}(k) \epsilon^*_{j',\rho \sigma}(l)\left(\sum_{i= 1}^n \left(\mathcal{M}_{\text{LO}}\right)_{i;-k,l}^{\mu \nu \rho \sigma} + \sum_{i= 1}^n \left(\mathcal{M}_{\text{NLO}}\right)_{i;-k,l}^{\mu \nu \rho \sigma}  + \sum_{i,j= 1}^n \left(\mathcal{M}'_{\text{NLO}}\right)_{i,j;-k,l}^{\mu \nu \rho \sigma} \right).
\end{align}
However, since we cannot determine this term from \ref{efd.def} based on our approach (of first expressing in terms of annihilation operators and then requiring the operator be unitary), we leave the derivation of this term to future work.

Proceeding with the expression in \ref{efd.op2}, we can factorize it as a product of exponential operators by using the Baker-Campbell-Hausdorff (BCH) formula to find
\begin{align}
\tilde{S}_n(\{p_i\}\,k\,,l\,,h) = \exp &\left[ \kappa \sum_{i'=1}^2 \int\limits_{\vec{k}} \mathcal{D}^{\alpha}_{i'}(k) + \kappa^3 \sum_{i',j'=1}^2 \int\limits_{\vec{k}} \int\limits_{\vec{l}}  \mathcal{D}^{\alpha\,, \beta}_{i'j'}(k,l)  + \mathcal{O}(\kappa^4)\right] \notag\\
& \qquad \times\exp\left[ \frac{\kappa^2}{2} \sum_{i',j'=1}^2 \int\limits_{\vec{k}} \int\limits_{\vec{l}} \mathcal{S}^{\beta}_{i'j'}(k\,,l)\right] \,. 
\label{efd.fin}
\end{align}
where we defined 
\begin{align}
\mathcal{D}^{\alpha\,, \beta}_{i'j'}(k,l) = \frac{1}{2} \left( - \beta_{i'j'}(k\,,l)\alpha^*_{j'}(l) a_{i'}^{\dagger}(k) + \beta^*_{i'j'}(k\,,l)\alpha_{j'}(l) a_{i'}(k)\right).
\label{coh.corr}
\end{align}

Our result in \ref{efd.fin} is that for a squeezed coherent state. One of the exponentials is simply a squeezing operator and can be determined entirely from the double soft graviton terms. This operator is expected to provide quantum corrections and noise to classical radiative observables. However, the coherent dressing operator now has an exponent with two distinct pieces. The first of these simply arises from the expansion of the single soft graviton factor, to leading order in the gravitational coupling constant, and involves a sum over all the external particles in the scattering process. On account of the BCH formula, there now exist corrections in the coherent dressing exponent to higher orders in the gravitational coupling constant. These arise from contracting the single soft and double soft graviton terms as they appear in the unfactorized Wilson line from the GWL approach. We have noted the leading correction term $\mathcal{D}^{\alpha\,, \beta}_{i'j'}(k,l)$ in \ref{efd.fin}, which follows from contracting single and double graviton vertices, and thus features a double sum over the external lines of the scattering amplitude. 

Squeezed coherent states derived previously from the GWL approach~\cite{Fernandes:2024xqr} were used to investigate low-frequency radiative observables for the waveform, radiated momentum, and angular momentum \footnote{In \cite{Fernandes:2024xqr}, the gravitons in the dressing exponent in the `nearly soft' limit had also been considered, by placing a finite cut-off on the total energies of the two gravitons. Certain radiative observables can vanish in the strict soft limit, and the nearly soft condition provides a leading IR behaviour.}. 
A part of the waveform result was shown to recover the Christodolou non-linear memory effect, when it is sourced by nearly soft gravitons and for collinear emissions \footnote{The non-linear memory results from those terms with all graviton vertices on the same external particles. The correction to the coherent state also involves terms where the contracted single and double graviton vertices are on different external particles, and these could produce additional corrections other than the non-linear memory effect.}.  
We leave the derivation of the radiative observables from the more general double soft graviton dressing operator derived in this subsection, which will address the case beyond the collinear limit, to future work.

\section{Discussion}
\label{sec5}

In this paper, we have utilised the path integral resummation approach of \cite{Laenen:2008gt, White:2011yy, Bonocore:2021qxh} to derive double soft contributions in the GWL of gravitationally interacting theories, which describe the soft function of factorized scattering amplitudes up to next-to-eikonal corrections. We demonstrated in Section 2 that, in addition to the known double soft emissions from external hard particles as derived previously in \cite{Chakrabarti:2017ltl, Distler:2018rwu}, we can also correctly generate the three-graviton vertex contribution that exponentiates and results from considering off-shell contributions in the path integral over the gravitons. Additional double soft contributions were also derived in Section 3 by considering the LBK theorem. These terms restore gauge invariance of the amplitude following an initial displacement of the external hard particles, and after including soft graviton emissions from inside the hard interaction. As a consequence, we demonstrated that the soft function, which now includes three-graviton vertex contributions derived in Section 2, and products of single soft factors up to subleading order, recovers the known gauge invariant double soft graviton factor for scattering amplitudes. We further derived a universal subleading contribution of the double soft graviton factor using the LBK theorem in Section 3. Lastly, we used the result of the double soft graviton factor, up to its subleading order, to motivate the definition of a new soft dressing factor wherein subleading soft factor terms derived by using the LBK theorem are now assumed to be a part of the dressing exponent and not the remainder function. 

Our work is also motivated by effective field theories for binary black hole scattering, which have been particularly relevant over recent years in the derivation of post-Minkowski gravitational waveobservables~\cite{Goldberger:2004jt,Porto:2005ac, Goldberger:2007hy, Porto:2016pyg, Goldberger:2022rqf, Bern:2019crd, Bern:2019nnu, Kosower:2018adc, Maybee:2019jus, AccettulliHuber:2020dal, Aoude:2020ygw, Mogull:2020sak, Kalin:2020mvi, Kalin:2020fhe, Kalin:2020lmz, Kalin:2019rwq,  Bern:2021dqo, Bern:2021yeh, Bini:2021gat, Aoude:2022trd, Aoude:2022thd, Aoude:2021oqj, Bini:2022wrq, Bjerrum-Bohr:2021din, Bjerrum-Bohr:2021wwt, Britto:2021pud, Cristofoli:2021vyo, Cristofoli:2021jas, Chen:2021szm, Damgaard:2021ipf,Dlapa:2021vgp,Dlapa:2021npj, DiVecchia:2021bdo, DiVecchia:2021ndb, Haddad:2021znf, Heissenberg:2021tzo, Herrmann:2021lqe,Herrmann:2021tct, Jakobsen:2021smu,Jakobsen:2021lvp,Jakobsen:2021zvh, Kol:2021jjc, Kosmopoulos:2021zoq, Liu:2021zxr, Mougiakakos:2021ckm, Adamo:2022qci,Barack:2022pde, Bjerrum-Bohr:2022blt,Bjerrum-Bohr:2022ows,Cristofoli:2022phh,Dlapa:2022lmu,DiVecchia:2022owy,DiVecchia:2022nna,DiVecchia:2022piu,FebresCordero:2022jts,Heissenberg:2022tsn,Jakobsen:2022psy,Jakobsen:2022fcj,Jakobsen:2022zsx,Kalin:2022hph,Manohar:2022dea,Mougiakakos:2022sic,Riva:2022fru,Travaglini:2022uwo,Veneziano:2022zwh,Adamo:2023cfp,Aoude:2023dui,Aoude:2023vdk,Barack:2023oqp,Bern:2023ity,Bini:2023fiz,Bohnenblust:2023qmy,Brandhuber:2023hhy,Brandhuber:2023hhl,Caron-Huot:2023vxl,Cheung:2023lnj,Damgaard:2023ttc,Damgaard:2023vnx,DeAngelis:2023lvf,DiVecchia:2023frv,Dlapa:2023hsl,Elkhidir:2023dco,Georgoudis:2023eke,Georgoudis:2023lgf,Georgoudis:2023ozp,Heissenberg:2023uvo,Herderschee:2023fxh,Jakobsen:2023ndj,Jakobsen:2023hig,Jakobsen:2023pvx,Jones:2023ugm,Kosmopoulos:2023bwc,Luna:2023uwd,Riva:2023rcm,Riva:2023xxm,Aoki:2024boe,Aoude:2024jxd,Bini:2024ijq,Bini:2024rsy,Bini:2024tft,Bjerrum-Bohr:2024hul,Buonanno:2024vkx,Buonanno:2024byg,Correia:2024jgr,Correia:2024yfx, Dlapa:2024cje,Driesse:2024feo,Driesse:2024xad,Porto:2024cwd,Haddad:2024ebn,Akpinar:2024meg,Akpinar:2025bkt,Bohnenblust:2025gir,Bini:2025vuk,Caron-Huot:2025tlq,Correia:2025dyy,Dlapa:2025biy,Georgoudis:2025vkk,Haddad:2025cmw,Mogull:2025cfn} and more recently for describing Hawking radiation~\cite{Aoude:2024sve,Aoki:2025ihc,Aoude:2025jvt,Ilderton:2025aql}. Within these approaches, classical radiative observables may either be derived from cuts of scattering amplitudes, or from expectation values of radiative operators with respect to coherent states~\cite{Cristofoli:2021jas,Britto:2021pud,Luna:2023uwd,DiVecchia:2022nna,DiVecchia:2022piu}. 
The gravitational dressing operator from the GWL formalism, due to its multiple soft graviton expansion, provides a generalization of coherent states. This is made clear by expressing the soft gravitons that appear in the Wilson line in terms of their creation and annihilation operators. In Section 4, we inferred that a squeezed coherent state would result as the dressing operator for gravitationally mediated scattering amplitudes. It will be interesting to further explore the dressing operator in further detail, particularly in relation to nonlinear radiative observables.

More recently, there have been efforts to relate worldline QFT approaches with the worldline formalism and eikonal approximations\cite{Du:2024rkf,Ajith:2024fna}, as well as with the KMOC formalism~\cite{Damgaard:2023vnx,Capatti:2024bid}. Gravitational dressings and radiative observables remain open problems to be explored in these approaches, and our results using the GWL formalism might help expand on these results. Another more recent application of the GWL formalism has been in the derivation of spin corrections in PM radiative observables~\cite{Bonocore:2024uxk,Bonocore:2025stf}. In this regard, we note that exponentiated single soft factors to subleading orders have been used in effective field theories of scattered Kerr black holes~\cite{Guevara:2018wpp}. Here, we expect that our results for the gravitational dressing operator could be used to investigate subleading PM corrections and nonlinear radiative effects, as considered previously in the case without angular momentum terms in the soft dressing operator. 

As is clear from the original generalised Wilson line papers in \cite{Laenen:2008gt,White:2011yy}, our analysis can be equally applied to non-abelian gauge theories. This itself suggests an interesting programme of further work. In the case of leading soft behaviour~\cite{Oxburgh:2012zr}, it is known that gauge and gravity results are related by the so-called {\it double copy} relating scattering amplitude in gauge and gravity theories~\cite{Bern:2008qj,Bern:2010ue}, itself inspired by previous work in string theory~\cite{Kawai:1985xq} (see e.g. \cite{Bern:2022wqg,Adamo:2022dcm,White:2024pve} for recent reviews). The approach of this paper could be used to extend these results to subleading orders in the momentum expansion, and indeed, this would provide highly non-trivial evidence for the validity of the double copy at all orders in perturbation theory. 

In summary, the study of next-to-soft radiation in both gauge and gravity theories continues to provide useful insights into novel aspects of field theory, as well as having potential practical applications. We hope that our results make a valuable contribution to the numerous ongoing discussions in this area.

\bigskip
\bigskip

\noindent {\it Acknowledgements.}
The work of KF is supported by Taiwan's National Science and Technology Council (NSTC) with grant number 112-2811-M-003-003-MY3. The work of FLL is supported by
Taiwan's NSTC with grant number 112-2112-M-003-006-MY3. CDW is supported by the U.K. Science and Technology Facilities Council
(STFC) Consolidated Grant ST/P000754/1 ``String theory, gauge theory
and duality''.

\bigskip
\bigskip

 \appendix 

\section{Derivation of the GWL} \label{app1}

In this appendix, we will review the derivation of the generalized Wilson line (GWL) \ref{esf} for scalar field theories minimally coupled to gravity. While our treatment follows~\cite{White:2011yy, Bonocore:2021qxh}, our conventions differ from these references in certain places. We work in the mostly plus convention for the metric, with the weak field expansion
\begin{equation}
g_{\mu \nu} = \eta_{\mu \nu} + 2 \kappa h_{\mu \nu}\,,
\label{met}
\end{equation}
and $\kappa^2 = 8 \pi G$. We will be interested in expressions up to quadratic order in the gravitational fluctuations, for which we have
\begin{align}
g^{\mu \nu} & = \eta^{\mu \nu} - 2 \kappa h^{\mu \nu} + 4 \kappa^2 h^{\mu \alpha} h^{\nu}_{\alpha} + \mathcal{O}(h^3), \label{met.inv} \\
\sqrt{-g} &= 1 + \kappa h + \kappa^2 \left( \frac{h^2}{2} - h_{\alpha \beta}^2\right) + \mathcal{O}(h^3). \label{met.det} 
\end{align}
As in~\cite{White:2011yy, Bonocore:2021qxh} we consider a massive scalar field minimally coupled to gravity
\begin{align}
S &= -\int d^4x \sqrt{-g} \left(g^{\mu \nu} \partial_{\mu}\phi^* \partial_{\nu}\phi + m^2 \phi^* \phi \right) \notag\\
 & = \int d^4x \left( \partial_{\mu}\left( \sqrt{-g} g^{\mu \nu}\right) \partial_{\nu} + \sqrt{-g} g^{\mu \nu} \partial_{\mu} \partial_{\nu} - m^2 \sqrt{-g} \right).
\label{act.sf}
\end{align}
On considering $\partial_{\mu} = - i p_{\mu}$, we then arrive at
\begin{align}
S = - \int d^4x\, \phi^* \left(2 H (x\,,p) \right) \phi\,  ,
\label{SH}
\end{align}
with 
\begin{align}
H (x\,,p) = \frac{1}{2} \left[ i \partial_{\mu}\left( \sqrt{-g} g^{\mu \nu}\right) p_{\nu} + \sqrt{-g} g^{\mu \nu} p_{\mu}p_{\nu} +  m^2 \sqrt{-g}\right].
\label{H.def}
\end{align}

Using \ref{SH}, we can derive the dressed scalar propagator for a single external particle in the amplitude. This follows from promoting $H (x\,,p)$ to an operator and evaluating its inverse using the Schwinger proper time formalism. We will more specifically compute it as worldline path integral from an initial position $x_i (\,\text{with} \; x(0) = x_i)$ to a final momentum $p_i (\, \text{with} \; p(T) = p_i)$ for an external particle of the scattering amplitude~\cite{White:2011yy, Bonocore:2021qxh}
 \begin{align}
\Big \langle p_i \Big \vert &\left(2i \left(\hat{H} - i \varepsilon\right)\right)^{-1} \Big\vert x_i \Big \rangle \notag\\
& = \frac{1}{2} \int\limits_{0}^{\infty} dT \int\limits_{x(0) = x_i}^{p(T) = p_i} \mathcal{D}p \mathcal{D}x \;\exp\left[ - i p(T)\cdot x(T) + i \int\limits_{0}^T dt \left( p\cdot \dot{x} - \hat{H}(x\,, p) + i \varepsilon \right) \right]\,,
\label{prop.sch}
\end{align}
where $T$ is the Schwinger parameter. 

As \ref{prop.sch} cannot be evaluated exactly, the solution is derived perturbatively about a known solution. Consistent with the weak field expansion of the background metric over flat spacetime, we can derive solutions from corrections about asymptotic eikonal trajectories and accordingly assume
\begin{equation}
x(t) =  x_i + p_i t + \tilde{x}(t) \;; \quad p(t) =  p_i + \tilde{p}(t) \,.
\label{xp.sol}
\end{equation}

The path integral in \ref{prop.sch} is now evaluated for fluctuations about the eikonal trajectory solution with limits $\tilde{x}(0) = 0$ and $\tilde{p}(T) = 0$. Additionally, we have
\begin{align}
- i p(T)\cdot x(T) &+i \int\limits_{0}^T dt \left( p\cdot \dot{x} - \hat{H}(x\,, p) \right) \notag\\
& = -i p_i \cdot x_i -i\frac{p_i^2 + m^2}{2} T + i \int\limits_0^T dt \left( - \frac{1}{2}\tilde{p}_{\mu} A^{\mu \nu}\tilde{p}_{\nu} + B^{\mu}\tilde{p}_{\mu} + C \right),
\label{px.int}
\end{align}
with
\begin{align}
A^{\mu \nu} &= \sqrt{-g} g^{\mu \nu}; \label{A.def}\\
B^{\mu} & = \dot{\tilde{x}}^{\mu} - i \frac{\partial_{\nu}\left(\sqrt{-g} g^{\mu \nu}\right)}{2} - \left(\sqrt{-g} g^{\mu \nu} - \eta^{\mu \nu}\right) p_{i \nu}; \label{B.def}\\
C& =  - \frac{\left(\sqrt{-g} g^{\mu \nu} - \eta^{\mu \nu}\right)}{2} p_{i \mu} p_{i \nu} - \frac{\left(\sqrt{-g} - 1\right)}{2}m^2  - i \frac{\partial_{\mu}\left(\sqrt{-g} g^{\mu \nu}\right)}{2}p_{i \nu}.\label{C.def}
\end{align}
Substituting \ref{xp.sol} and \ref{px.int} in \ref{prop.sch} we then find
\begin{align}
\Big \langle p_i \Big \vert \left(2i \left(\hat{H} - i \varepsilon\right)\right)^{-1} \Big\vert x_i \Big \rangle & = \frac{1}{2} \int\limits_{0}^{\infty} dT e^{ -i p_i \cdot x_i -i\frac{p_i^2 + m^2}{2} T - \varepsilon T} f_i(T), \label{prop.eik}
\end{align}
where
\begin{align}
f_i(T) &= \int\limits_{\tilde{x}(0) = 0}^{\tilde{p}(T) = 0} \mathcal{D}\tilde{p} \mathcal{D}\tilde{x} \;\exp\left[i \int\limits_0^T dt \left( - \frac{1}{2}\tilde{p}_{\mu} A^{\mu \nu}\tilde{p}_{\nu} + B^{\mu}\tilde{p}_{\mu} + C \right) \right]\, \notag\\
& = \int\limits_{\tilde{x}(0) = 0} \mathcal{D}\tilde{x} \;\exp\left[i \int\limits_0^T dt \left(\frac{1}{2} B^{\mu} \left(A^{-1}\right)_{\mu \nu}B^{\nu} + C \right)\right],
\label{ft.def}
\end{align}
with $ \left(A^{-1}\right)_{\mu \nu} = (\sqrt{-g})^{-1}g_{\mu \nu}$ the inverse of $A^{\mu \nu}$ in \ref{A.def}. Expanding the result to $\kappa^2$ order, we find
\begin{align}
f_i(T) &= \int\limits_{\tilde{x}(0) = 0}\mathcal{D}\tilde{x} \;\exp\left[i \int\limits_0^T dt \left[ \frac{\dot{\tilde{x}}^2}{2} + \kappa \left(h_{\mu \nu}p_i^{\mu} p_i^{\nu} + i \partial^{\mu}h_{\mu \nu} p_i^{\nu} - \frac{i}{2} \partial_{\mu}h p_i^{\mu} - \frac{h}{2} (p^2 + m^2)\right. \right. \right. \notag\\
& \left. \left. \left. \qquad \qquad  + i \dot{\tilde{x}}^{\mu} \left(\partial^{\nu}h_{\mu \nu}  - \frac{1}{2} \partial_{\mu}h\right) + \left(2 \dot{\tilde{x}}^{\mu} p_i^{\nu} + \dot{\tilde{x}}^{\mu}\dot{\tilde{x}}^{\nu} \right)\left(h_{\mu \nu}  - \frac{h}{2}\eta_{\mu \nu}\right) \right) \right. \right. \notag\\
&\left. \left. \quad + \kappa^2 \left(- h h_{\mu \nu} p_i^{\mu} p_i^{\nu} + \frac{h^2}{2}p^2 + i h^{\alpha \beta} \partial_{\mu} h_{\alpha \beta} p_i^{\mu} - 2i h^{\mu \rho}\left(\partial_{\mu} h_{\rho \nu}\right) p_i^{\nu} - \frac{p_i^2 + m^2}{2} \left(\frac{h^2}{2} - h_{\alpha \beta}^2 \right)  \right. \right. \right. \notag\\
& \left. \left. \left. \qquad \qquad - \left( \frac{1}{2} \left(\partial^{\mu} h_{\mu \alpha}\right) \left(\partial^{\nu} h_{\nu}^{\alpha}\right) + \frac{1}{8} \left(\partial^{\mu} h\right) \left(\partial_{\mu} h\right) - \frac{1}{2} \left(\partial_{\mu} h^{\mu \nu}\right) \left(\partial_{\nu} h\right) \right)  \right. \right. \right. \notag\\ 
& \left. \left. \left. \qquad \qquad \qquad  + i \dot{\tilde{x}}^{\mu} \left(h_{\alpha \beta}\partial_{\mu}h^{\alpha \beta} - 2 h_{\mu \rho}\partial_{\nu}h^{\nu \rho} \right) + \dot{\tilde{x}}^{\mu} \left( \left(h_{\alpha \beta}^2 + \frac{h^2}{2} \right) \eta_{\mu \nu} - 2 h h_{\mu \nu}\right) p_i^{\nu}  \right. \right. \right. \notag\\
& \left. \left. \left. \qquad \qquad \qquad \qquad +  \dot{\tilde{x}}^{\mu}\dot{\tilde{x}}^{\nu} \left( \left(h_{\alpha \beta}^2 - \frac{h^2}{2} \right) \eta_{\mu \nu} - h h_{\mu \nu} \right) \right) + \mathcal{O}(\kappa^3) \right] \right]\;.
\label{ft.res}
\end{align}

The contribution from \ref{prop.eik} to the scattering amplitude in the eikonal approximation follows from truncating the propagator for the external lines, in accordance with the LSZ prescription
\begin{align}
i (p_i^2 + m_i^2) \Big \langle p_i \Big \vert \left(2i \left(\hat{H} - i \varepsilon\right)\right)^{-1} \Big\vert x_i \Big \rangle
& =  \frac{i}{2} (p_i^2 + m_i^2) \left[\int\limits_{0}^{\infty} dT e^{ -i p_i \cdot x_i -i\frac{p_i^2 + m^2}{2} T - \varepsilon T} f_i(T) \right] \notag\\
&= -\left[e^{ -i p_i \cdot x_i}  \int\limits_{0}^{\infty} dT e^{ -i p_i \cdot x_i} \frac{d}{dT}\left( e^{-i\frac{p_i^2 + m^2}{2} T}\right) e^{- \varepsilon T} f_i(T) \right] \notag\\
& = - e^{ -i p_i \cdot x_i}  \left[ -f_i(0) - \int\limits_{0}^{\infty} dT e^{-i\frac{p_i^2 + m^2}{2} T}  \frac{d}{dT}\left( f_i(T)\right) \right].\label{dress.exp0}
\end{align}
Taking the on-shell limit of external particles, we then have the dressing contribution for each external line in the scattering amplitude
\be
\lim_{p_i^2 \to m_i^2} i (p_i^2 + m_i^2) \Big \langle p_i \Big \vert \left(2i \left(\hat{H} - i \varepsilon\right)\right)^{-1} \Big\vert x_i \Big \rangle = e^{ -i p_i \cdot x_i} f(x_i,p_i; h),
\label{dress.exp}
\ee
where we denoted $f(x_i,p_i; h) = f_i(\infty)$ in \ref{dress.exp}, and it refers to the $T \to \infty$ limit of \ref{ft.res}. The terms appearing in \ref{esf} require evaluating the path integral over $\tilde{x}$ in \ref{ft.res}. 

Following \ref{xp.sol}, we now write $x(t) = y_i(t) + \tilde{x}(t)$ and expand all the graviton fields about $y_i(t) = x_i + p_i t$. Thus, for instance
$$h_{\mu \nu}(x(t)) = h_{\mu \nu}(y_i(t)) + \tilde{x}_{\alpha} \partial_i^{\alpha}h_{\mu \nu}(y_i(t)) + \frac{1}{2} \tilde{x}_{\alpha} \tilde{x}_{\beta} \partial_i^{\alpha}\partial_i^{\beta}h_{\mu \nu}(y_i(t)) + \mathcal{O}(\tilde{x}^3), $$
with $\partial_i^{\mu} = \frac{\partial}{\partial y_i^{\mu}}$. From the analysis for the next-to-eikonal Feynman rules in~\cite{White:2011yy,Laenen:2008gt,Bonocore:2021qxh}, we note that \ref{ft.res} will provide the following relevant contributions: 
\begin{align}
&f(x_i,p_i; h)\notag\\
 &= \int\limits_{\tilde{x}(0) = 0}\mathcal{D}\tilde{x} \;\exp\left[i \int\limits_0^T dt  \left[ \frac{\lambda \dot{\tilde{x}}^2}{2} + \kappa \left(\lambda \left(h_{\mu \nu}(y_i(t)) + \tilde{x}_{\alpha}(t) \partial_i^{\alpha}h_{\mu \nu}(y_i(t)) + \frac{1}{2} \tilde{x}_{\alpha}(t) \tilde{x}_{\beta}(t) \partial_i^{\alpha}\partial_i^{\beta}h_{\mu \nu}(y_i(t)) \right) p_i^{\mu} p_i^{\nu} \right. \right. \right.  \notag\\ 
& \left. \left.\left. \qquad \qquad \qquad  \qquad +  2 \lambda \dot{\tilde{x}}^{\mu} p_i^{\nu} h_{\mu \nu}(y_i(t))  +  i \partial^{\mu}h_{\mu \nu}(y_i(t)) p_i^{\nu} - \frac{i}{2} \partial_{\mu}h(y_i(t)) p_i^{\mu} \right)  +  \cdots  \right] \right]\,,
\label{ft.res2}
\end{align}
where in \ref{ft.res2}, we have introduced the book-keeping parameter $\lambda$ as in~\cite{Laenen:2008gt,Bonocore:2021qxh} from rescaling $t \to \lambda^{-1} t$ and $p \to \lambda p$. The terms indicated in \ref{ft.res2} are the leading contributions in the $\lambda$ expansion that will yield \ref{esf}, while the $\cdots$ are terms that will be off-shell at order $\kappa^2$, subleading terms of $\mathcal{O}(\kappa^3)$, and terms which are beyond subleading order in the soft expansion. 

We also need to utilize the following correlators for $\tilde{x}_{\mu}$ and its derivatives
\begin{align}
\langle \tilde{x}_{\mu} (t) \tilde{x}_{\nu}(t') \rangle = \frac{i}{\lambda} \text{min} (t,t')\eta_{\mu \nu} \;, \quad \langle \dot{\tilde{x}}_{\mu} (t) \tilde{x}_{\nu}(t') \rangle = \frac{i}{\lambda} \Theta(t'-t)\eta_{\mu \nu} \;, \quad  \langle \dot{\tilde{x}}_{\mu} (t) \dot{\tilde{x}}_{\nu}(t') \rangle =& \frac{i}{\lambda} \delta(t -t')\eta_{\mu \nu} \,,
\label{x.con}
\end{align}
which are a consequence of the Green's function for $\tilde{x}_{\mu}$ determined from its kinetic term in \ref{ft.res2}. We will seek the solution of the form in \ref{esf}:
\begin{align}
 f(x_i,p_i; h) 
 = \exp\big[\delta_1(x_i,p_i; h) + \delta_2^{(1)}(x_i,p_i; h)+ \delta_2^{(2)}(x_i,p_i; h)+ \delta_2^{(3)}(x_i,p_i; h)\big] \;.
\label{esf1}
\end{align}

On using \ref{x.con}, the term quadratic in $\tilde{x}$ can be readily evaluated to
\be
i \lambda \frac{\kappa}{2}  \int\limits_0^{\infty} dt \langle \tilde{x}_{\alpha}(t) \tilde{x}_{\beta}(t)\rangle \partial_i^{\alpha}\partial_i^{\beta}h_{\mu \nu}(y_i(t))  p_i^{\mu} p_i^{\nu} = \frac{\kappa}{2}  \int\limits_0^{\infty} dt \, t \eta_{\alpha \beta} \partial_{i}^{\alpha} \partial_{i}^{\beta}h_{\mu \nu}(y_i(t))p_i^{\mu}  p_i^{\nu}\,.
\ee
Combining this result with those terms in \ref{ft.res2} involving no $\tilde{x}$, and dropping the parameter $\lambda$, we get the single graviton contribution in the dressing exponent
\begin{align}
&\delta_1(x_i,p_i; h) =   i \kappa  \int\limits_0^{\infty} dt\, \left( h_{\mu \nu}(y_i(t)) p_i^{\mu}  p_i^{\nu}  + i \frac{t}{2} \eta_{\alpha \beta}\partial_{i}^{\alpha} \partial_{i}^{\beta}h_{\mu \nu}(y_i(t))p_i^{\mu}  p_i^{\nu} \right. \notag\\
&\left. \qquad \qquad \qquad \qquad \qquad \qquad +i  \left(\partial_i^{\mu}h_{\mu \nu}(y_i(t)) p_i^{\nu} - \frac{1}{2}  \eta_{\mu \nu}\partial_i^{\mu}h(y_i(t)) p_i^{\nu} \right) \right)  \;, \label{1sf.der}
\end{align}
as in \ref{1sf.int}. 

The double graviton terms arise from contracting the terms that are linear in $\tilde{x}$ or $\dot{\tilde{x}}$ in \ref{ft.res2}. There are three such terms, which on using \ref{x.con}, evaluate to the expressions in \ref{2sf1.int} - \ref{2sf3.int1}:
\begin{align}
\delta_2^{(1)}(x_i,p_i; h) & = -\frac{\kappa^2}{2} \int\limits_0^{\infty} dt\int\limits_0^{\infty} dt' \partial_i^{\alpha}h_{\mu \nu}(y_i(t)) \partial_i^{\beta}h_{\mu \nu}(y_i(t')) p_i^{\mu} p_i^{\nu} p_i^{\rho} p_i^{\sigma} \langle \tilde{x}_{\alpha}(t) \tilde{x}_{\beta}(t')\rangle \notag\\
&= -i \frac{\kappa^2}{2} \int\limits_0^{\infty} dt\int\limits_0^{\infty} dt'  \eta_{\alpha \beta}\partial_i^{\alpha}h_{\mu \nu}(y_i(t)) \partial_i^{\beta}h_{\mu \nu}(y_i(t')) p_i^{\mu} p_i^{\nu} p_i^{\rho} p_i^{\sigma}\text{min} (t,t')\;, \label{ds2.1}\\
\delta_2^{(2)}(x_i,p_i; h)&= -2 \kappa^2  \int\limits_0^{\infty} dt \int\limits_0^{\infty} dt' h_{\mu \nu}(y_i(t)) h_{\rho \sigma}(y_i(t')) p_i^{\mu} p_i^{\rho} \langle \dot{\tilde{x}}^{\nu} (t) \tilde{x}^{\sigma}(t') \rangle \notag\\
& =   - 2 i \kappa^2 \int\limits_0^{\infty} dt \int\limits_0^{\infty} dt'\, h_{\mu \nu}(y_i(t)) h_{\rho \sigma}(y_i(t')) p_i^{\mu} p_i^{\rho} \eta^{\nu \sigma} \delta(t - t')\;,
\label{ds2.2}\\
\delta_2^{(3)}(x_i,p_i; h)&=  -\kappa^2 \int\limits_0^{\infty} dt \int\limits_0^{\infty} dt' \left(\partial_i^{\alpha} h_{\mu \nu}(y_i(t)) h_{\rho \sigma}(y_i(t')) p_i^{\mu} p_i^{\nu} p_i^{\rho} \langle \tilde{x}_{\alpha}(t) \dot{\tilde{x}}^{\sigma}(t') \rangle \right. \notag\\
& \left.  \qquad\qquad\qquad \qquad\qquad \qquad\qquad + h_{\mu \nu}(y_i(t)) \partial_i^{\alpha} h_{\rho \sigma}(y_i(t')) p_i^{\mu} p_i^{\rho} p_i^{\sigma} \langle \tilde{x}_{\alpha}(t') \dot{\tilde{x}}^{\nu}(t) \rangle \right) \notag\\
 &=  -i \kappa^2   \int\limits_0^{\infty} dt \int\limits_0^{\infty} dt'  \, \left(\partial_i^{\sigma} h_{\mu \nu}(y_i(t)) h_{\rho \sigma}(y_i(t')) p_i^{\mu} p_i^{\nu} p_i^{\rho} \Theta(t - t') \right. \notag\\
&\left. \qquad\qquad\qquad \qquad\qquad \qquad\qquad  +  h_{\mu \nu}(y_i(t)) \partial_i^{\nu} h_{\rho \sigma}(y_i(t')) p_i^{\mu} p_i^{\rho} p_i^{\sigma} \Theta(t' - t) \right)\;. \label{ds2.3}
\end{align} 
\ref{ds2.1} and \ref{ds2.2} involve a symmetry factor of $\frac{1}{2}$, while \ref{ds2.3} was evaluated symmetrically with respect to the two gravitons.
 
\section{Derivation of ${\mathcal M}_{\rm NLO}^{\mu \nu \rho \sigma} (\{p_i\} \,, k \,,l)$} \label{app2}
 
This appendix will make use of the definition of angular momentum in \ref{am.def} throughout, as well as the conservation of angular momentum of the hard amplitude:
 \begin{equation}
\sum_{i=1}^n p_i^{\nu} \partial_i^{\alpha} T_n =  \sum_{i=1}^n p_i^{\alpha} \partial_i^{\nu} T_n \,.\label{am.id}
\end{equation}
We begin by considering \ref{m2.nlox} in the following way
\begin{align}
\mathcal{M}_{\text{NLO}}^{\mu \nu \rho \sigma} (\{p_i\}\,, k\,,l) = \mathcal{M}_{\text{NLO1};k,l}^{\mu \nu \rho \sigma} + \mathcal{M}_{\text{NLO2};k,l}^{\mu \nu \rho \sigma} + \bar{L}^{\mu \nu \rho \sigma}_n + \frac{\kappa^2}{2} \sum_{i= 1}^n \frac{\alpha_i^{\mu \nu \rho \sigma}(k\,,l)}{p_i(k+l)}(k+l)_{\alpha} \partial_i^{\alpha}T_n\;, 
\label{m2.dec}
\end{align}
where we have defined
\begin{align}
\mathcal{M}_{\text{NLO1}; k, l}^{\mu \nu \rho \sigma} &= \frac{\kappa}{2} \sum_{j= 1}^n \frac{p_j^{\rho}  p_j^{\sigma}}{p_j \cdot l} \left(\kappa \sum_{i= 1}^n  \frac{p_i^{\mu}  p_i^{\nu}}{p_i \cdot k} \frac{k_{\alpha} k_{\beta}}{2} \partial_i^{\alpha} \partial_i^{\beta} T_n +   k_{\alpha} \partial_k^{\alpha} N_n^{\mu \nu}\right)\notag\\
&\qquad  \qquad + \frac{\kappa}{2} \sum_{i= 1}^n \frac{p_i^{\mu}  p_i^{\nu}}{p_i \cdot k}\left(\kappa \sum_{j= 1}^n \frac{p_j^{\rho}  p_j^{\sigma}}{p_j \cdot l} \frac{l_{\alpha} l_{\beta}}{2} \partial_j^{\alpha} \partial_j^{\beta} T_n  + l_{\beta} \partial_l^{\beta} N_n^{\rho \sigma} \right)\;, \label{m2.nlo1}
 \end{align}
and 
 \begin{align}
\mathcal{M}_{\text{NLO2}; k, l}^{\mu \nu \rho \sigma} &= \frac{\kappa^2}{2}  \sum_{i,j= 1}^n \frac{p_i^{\mu}  p_i^{\nu}}{p_i \cdot k} \frac{p_j^{\rho}  p_j^{\sigma}}{p_j \cdot l} k_{\alpha} l_{\beta} \partial_i^{\alpha} \partial_j^{\beta}T_n +  \frac{\kappa}{2} \sum_{j= 1}^n \frac{p_j^{\rho}  p_j^{\sigma}}{p_j \cdot l} l_{\beta}\partial_j^{\beta}N_{n}^{\mu \nu} + \frac{\kappa}{2} \sum_{i= 1}^n \frac{p_i^{\mu}  p_i^{\nu}}{p_i \cdot k} k_{\alpha}\partial_i^{\alpha}N_{n}^{\rho \sigma} +  N^{\mu\nu\rho\sigma}_n\;.
\label{m2.nlo2}
\end{align}
 
On using \ref{sn.kl} and \ref{an.kl} in \ref{m2.nlo1}, we can readily express it in terms of angular momentum operators 
\begin{align}
&\mathcal{M}_{\text{NLO1}; k, l}^{\mu \nu \rho \sigma} =  \frac{\kappa^2}{4}\sum_{i,j= 1}^n \left(\frac{p_i^{\mu}  p_i^{\nu}}{p_i \cdot k} \frac{l_{\alpha}l_{\beta}}{p_j \cdot l} J_j^{\alpha \rho}J_j^{\sigma \beta} + \frac{p_j^{\rho}  p_j^{\sigma}}{p_j \cdot l} \frac{k_{\alpha}k_{\beta}}{p_i \cdot k} J_i^{\alpha \mu }J_i^{\nu \beta}\right)  T_n \notag\\
& +\frac{\kappa^2}{4} \sum_{i,j= 1}^n \frac{p_j^{\rho}  p_j^{\sigma}}{p_j \cdot l} \left[\eta^{\mu \nu} k_{\alpha} - k^{\mu} \delta^{\nu}_{\alpha} - \frac{p_i^{\mu} k^{\nu}}{p_i \cdot k} k_{\alpha}\right] \partial_i^{\alpha} T_n + \frac{\kappa^2}{4} \sum_{i,j= 1}^n \frac{p_i^{\mu}  p_i^{\nu}}{p_i \cdot k} \left[\eta^{\rho \sigma} l_{\alpha} - l^{\rho} \delta^{\sigma}_{\alpha} - \frac{p_j^{\rho} l^{\sigma}}{p_j \cdot l} l_{\alpha}\right] \partial_j^{\alpha} T_n.
\label{m2.nlo1f}
\end{align}

We next consider \ref{m2.nlo2}, which has the following result on using \ref{n.kl} and \ref{n.sol}:
\begin{align}
&\mathcal{M}_{\text{NLO2};k, l}^{\mu \nu \rho \sigma}  = -\frac{\kappa^2}{4}\sum_{i,j= 1}^n \frac{k_{\alpha}}{p_i \cdot k} \frac{l_{\beta}}{p_j \cdot l} \left[ p_i^{\nu}J_i^{\mu \alpha}  p_j^{\sigma}J_j^{\rho \beta} + p_j^{\sigma}J_j^{\rho \beta} p_i^{\nu}J_i^{\mu \alpha} \right] T_n + \frac{\kappa^2}{4}\sum_{i= 1}^n \left[  \frac{p_i^{\mu}  p_i^{\nu}}{p_i \cdot k}  k^{\sigma} \partial_i^{\rho}   + \frac{p_i^{\rho}  p_i^{\sigma}}{p_i \cdot l} l^{\nu} \partial_i^{\mu}  \right] T_n  \notag\\
& \quad - \frac{\kappa^2}{4} \sum_{i= 1}^n \left[ 2 \frac{p_i^{\mu} p_i^{\nu}}{p_i \cdot k} \frac{p_i^{(\sigma} k^{\rho)}}{p_i \cdot l}  l_{\beta} \partial_i^{\beta}   +   2 \frac{p_i^{\rho} p_i^{\sigma}}{p_i \cdot l} \frac{p_i^{(\mu} l^{\nu)}}{p_i \cdot k}  k_{\beta} \partial_i^{\beta}  - \frac{p_i^{\mu} p_i^{\nu}}{p_i \cdot k} \frac{k \cdot l}{p_i \cdot l} p_i^{\sigma} \partial_i^{\rho}  -  \frac{p_i^{\rho} p_i^{\sigma}}{p_i \cdot l}\frac{k \cdot l}{p_i \cdot k} p_i^{\nu} \partial_i^{\mu} \right]T_n\notag\\
&\qquad \quad - \frac{\kappa^2}{4} \sum_{i= 1}^n \left[ p_i^{\nu} \eta^{\mu \sigma} \partial_i^{\rho} - 2 \frac{p_i^{\nu} \eta^{\mu (\rho} p_i^{\sigma)} }{p_i \cdot l} l_{\beta} \partial_i^{\beta} + p_i^{\sigma} \eta^{\rho \nu} \partial_i^{\mu} - 2 \frac{p_i^{\sigma} \eta^{\rho (\mu} p_i^{\nu)} }{p_i \cdot k} k_{\beta} \partial_i^{\beta} \right]T_n \notag\\
&\qquad \qquad \qquad - \frac{\kappa^2}{4} \sum_{i= 1}^n p_i^{\sigma} p_i^{\nu} \left[\frac{l^{\mu} \partial_i^{\rho}}{p_i \cdot l} + \frac{k^{\rho} \partial_i^{\mu}}{p_i \cdot k} \right] T_n \;.
\label{n.simp}
\end{align}

We can first simplify the following three terms that appear in \ref{n.simp}:
\begin{align}
&- \frac{\kappa^2}{4} \sum_{i= 1}^n p_i^{\sigma} p_i^{\nu} \left[\frac{l^{\mu} \partial_i^{\rho}}{p_i \cdot l} + \frac{k^{\rho} \partial_i^{\mu}}{p_i \cdot k} \right] T_n \notag\\
&\quad  = - i \frac{\kappa^2}{4} \sum_{i= 1}^n  \left[  \frac{1}{p_i \cdot l} p_i^{\sigma}J_i^{\rho \nu}  l^{\mu} + \frac{1}{p_i \cdot k} p_i^{\nu} J_i^{\mu \sigma} k^{\rho} \right] T_n  - \frac{\kappa^2}{4} \sum_{i= 1}^n \left[ \frac{p_i^{\mu} p_i^{\nu}}{p_i \cdot k} k^{\rho} \partial_i^{\sigma} +   \frac{p_i^{\rho} p_i^{\sigma}}{p_i \cdot l} l^{\mu} \partial_i^{\nu} \right] T_n  \,,\label{t.1} 
\end{align}
\begin{align}
&- \frac{\kappa^2}{2} \sum_{i= 1}^n \left[ \frac{p_i^{\mu} p_i^{\nu}}{p_i \cdot k} \frac{p_i^{(\sigma} k^{\rho)}}{p_i \cdot l}  l_{\beta} \partial_i^{\beta}   +  \frac{p_i^{\rho} p_i^{\sigma}}{p_i \cdot l} \frac{p_i^{(\mu} l^{\nu)}}{p_i \cdot k}  k_{\beta} \partial_i^{\beta} \right] T_n \notag\\
&\quad  = i \frac{\kappa^2}{2} \sum_{i= 1}^n  \left[  \frac{p_i^{\mu} p_i^{\nu}}{p_i \cdot k} \frac{l_{\beta}}{p_i \cdot l}k^{(\sigma} J_i^{\rho) \beta}  + \frac{p_i^{\rho} p_i^{\sigma}}{p_i \cdot l} \frac{k_{\alpha}}{p_i \cdot k} l^{(\mu} J_i^{\nu) \alpha} \right] T_n  - \frac{\kappa^2}{2} \sum_{i= 1}^n \left[ \frac{p_i^{\mu} p_i^{\nu}}{p_i \cdot k} k^{(\sigma} \partial_i^{\rho)}   +   \frac{p_i^{\rho} p_i^{\sigma}}{p_i \cdot l} l^{(\mu} \partial_i^{\nu)} \right] T_n  \label{t.2}\,,
\end{align}
and 
\begin{align}
& \frac{\kappa^2}{2} \sum_{i= 1}^n \left[ \frac{p_i^{\nu} \eta^{\mu (\rho} p_i^{\sigma)} }{p_i \cdot l} l_{\beta} \partial_i^{\beta} + \frac{p_i^{\sigma} \eta^{\rho (\mu} p_i^{\nu)} }{p_i \cdot k} k_{\alpha} \partial_i^{\alpha} \right]T_n \notag\\
& = -i  \frac{\kappa^2}{2} \sum_{i= 1}^n \left[ \frac{l_{\beta}}{p_i \cdot l}  p_i^{\nu} \eta^{\mu (\rho} J_i^{\sigma) \beta} + \frac{k_{\alpha}}{p_i \cdot k} p_i^{\sigma} \eta^{\rho (\mu} J_i^{\nu) \alpha} \right]T_n + \frac{\kappa^2}{2} \sum_{i= 1}^n \left[ p_i^{\nu} \eta^{\mu (\rho} \partial_i^{\sigma)} + p_i^{\sigma} \eta^{\rho (\mu} \partial_i^{\nu)} \right]T_n \;. \label{t.3}
\end{align}

On substituting \ref{t.1}, \ref{t.2} and \ref{t.3} in \ref{n.simp}, and separating contributions into those symmetric in the pair of graviton indices, and those that are not, we arrive at the expression
\begin{align}
&\mathcal{M}_{\text{NLO2};k,l}^{\mu \nu \rho \sigma}  = \mathcal{E}^{\mu \nu \rho \sigma}_{\text{NLO2};k,l} -\frac{\kappa^2}{4}\sum_{i,j= 1}^n \frac{k_{\alpha}}{p_i \cdot k} \frac{l_{\beta}}{p_j \cdot l} \left[ p_i^{\nu}J_i^{\mu \alpha}  p_j^{\sigma}J_j^{\rho \beta} + p_j^{\sigma}J_j^{\rho \beta}  p_i^{\nu}J_i^{\mu \alpha} \right] T_n  \notag\\
& - \frac{i \kappa^2}{2}\sum_{i=1}^n \left[ \frac{1}{p_i \cdot l} \left(\frac{p_i^{\sigma} J_i^{\rho \nu} l^{\mu}}{2} + p_i^{\nu} \eta^{\mu (\rho} J_i^{\sigma)  \beta} l_{\beta} \right)  +  \frac{1}{p_i \cdot k} \left(\frac{ p_i^{\nu} J_i^{\mu \sigma} k^{\rho}}{2} +  p_i^{\sigma} \eta^{\rho (\mu} J_i^{\nu) \alpha} k_{\alpha} \right) \right. \notag\\
& \left.  \qquad  \qquad \qquad \qquad  \qquad \qquad \qquad   - \frac{p_i^{\mu}  p_i^{\nu}}{p_i \cdot k} \frac{1}{p_i \cdot l}k^{(\sigma} J_i^{\rho) \beta} l_{\beta} - \frac{p_i^{\rho}  p_i^{\sigma}}{p_i \cdot l}\frac{1}{p_i \cdot k} l^{(\nu} J_i^{\mu) \alpha} k_{\alpha} \right] T_n \notag\\
%&  - \frac{i \kappa^2}{2}\sum_{i=1}^n \left[ \frac{1}{p_i \cdot l} \frac{p_i^{(\sigma} J_i^{\rho)(\mu} l^{\nu)}}{2}  +  \frac{1}{p_i \cdot k} \frac{ p_i^{(\nu} J_i^{\mu)(\sigma} k^{\rho)}}{2}  \right] T_n \notag\\
&- \frac{\kappa^2}{4} \sum_{i= 1}^n \left[2 \left( \frac{p_i^{\mu} p_i^{\nu}}{p_i \cdot k} k^{(\sigma} \partial_i^{\rho)}   +   \frac{p_i^{\rho} p_i^{\sigma}}{p_i \cdot l} l^{(\mu} \partial_i^{\nu)} \right)  - \frac{p_i^{\mu} p_i^{\nu}}{p_i \cdot k} \frac{k \cdot l}{p_i \cdot l} p_i^{(\sigma} \partial_i^{\rho)}  -  \frac{p_i^{\rho} p_i^{\sigma}}{p_i \cdot l}\frac{k \cdot l}{p_i \cdot k} p_i^{(\nu} \partial_i^{\mu)} \right]T_n\notag\\
& \qquad + \frac{\kappa^2}{4} \sum_{i= 1}^n \left[ p_i^{(\nu} \eta^{\mu) (\sigma} \partial_i^{\rho)} +  p_i^{(\sigma} \eta^{\rho) (\nu} \partial_i^{\mu)} \right]T_n\;,
% \notag\\
%&- \frac{\kappa^2}{4} \sum_{i= 1}^n p_i^{\sigma} p_i^{\nu} \left[\frac{l^{\mu} \partial_i^{\rho}}{p_i \cdot l} + \frac{k^{\rho} \partial_i^{\mu}}{p_i \cdot k} \right] T_n + \frac{\kappa^2}{4}\sum_{i= 1}^n \left[  \frac{p_i^{\mu}  p_i^{\nu}}{p_i \cdot k}  k^{\sigma} \partial_i^{\rho}   + \frac{p_i^{\rho}  p_i^{\sigma}}{p_i \cdot l} l^{\nu} \partial_i^{\mu}  \right] T_n
\label{m2.nlo2f}
\end{align}
where $\mathcal{E}^{\mu \nu \rho \sigma}_{\text{NLO2};k,l}$ is given by
\begin{align}
\mathcal{E}^{\mu \nu \rho \sigma}_{\text{NLO2};k,l} &= - \frac{\kappa^2}{2} \sum_{i= 1}^n \left[ \frac{p_i^{\mu} p_i^{\nu}}{p_i \cdot k} k^{[\rho} \partial_i^{\sigma]} +   \frac{p_i^{\rho} p_i^{\sigma}}{p_i \cdot l} l^{[\mu} \partial_i^{\nu]} \right] T_n   - \frac{\kappa^2}{4} \sum_{i= 1}^n \left[  - p_i^{(\nu} \eta^{\mu) (\sigma} \partial_i^{\rho)}  - p_i^{(\sigma} \eta^{\rho) (\nu} \partial_i^{\mu)} \right. \notag\\
& \left. \qquad + p_i^{\nu} \eta^{\mu \sigma} \partial_i^{\rho} + p_i^{\sigma} \eta^{\rho \nu} \partial_i^{\mu} -2 p_i^{[\nu} \eta^{\mu] (\rho} \partial_i^{\sigma)} - 2 p_i^{[\sigma} \eta^{\rho] (\mu} \partial_i^{\nu)} \right]T_n \;.
\label{enlo2}
\end{align}

We will next consider $L^{\mu \nu \rho \sigma}$ and simplify the following terms in \ref{l.sol}:
\begin{align}
&- \frac{\kappa^2}{4 k \cdot l} \sum_{i=1}^n\left[ \frac{p_i^{\rho}p_i^{\sigma}}{p_i \cdot l} l^{\nu} l^{\mu} k_{\alpha} \partial_i^{\alpha} - 2 p_{i}^{(\rho} \eta^{\sigma) (\nu} l^{\mu)} k_{\alpha} \partial_i^{\alpha} + (p_i\cdot l) \eta^{\nu (\rho} k^{\sigma)}\partial_i^{\mu} \right] T_n \notag\\
&\qquad =  -\frac{\kappa^2}{4 k \cdot l} \sum_{i=1}^n \frac{p_i^{\rho}p_i^{\sigma}}{p_i \cdot l} l^{\nu} l^{\mu} k_{\alpha} \partial_i^{\alpha} T_n + \frac{\kappa^2}{4 k \cdot l} \sum_{i=1}^n  p_{i}^{(\rho} \eta^{\sigma) (\nu} l^{\mu)} k_{\alpha} \partial_i^{\alpha} T_n  \notag\\
& \qquad \qquad + \frac{\kappa^2}{4 k \cdot l} \sum_{i=1}^n \left( p_{i}^{(\rho} \eta^{\sigma) (\nu} l^{\mu)} k_{\alpha} \partial_i^{\alpha}  - (p_i\cdot l) k^{(\rho}\eta^{\sigma) \nu} \partial_i^{\mu} \right) T_n\,,
\label{ql.rel}
\end{align}
and
\begin{align}
&- \frac{\kappa^2}{4 k \cdot l} \sum_{i=1}^n\left[ \frac{p_i^{\mu}p_i^{\nu}}{p_i \cdot k} k^{\rho} k^{\sigma} l_{\beta} \partial_i^{\beta} - 2 p_{i}^{(\mu} \eta^{\nu) (\sigma} k^{\rho)} l_{\beta} \partial_i^{\beta} + (p_i\cdot k) \eta^{\rho (\mu} l^{\nu)}\partial_i^{\sigma} \right] T_n \notag\\
&=  -\frac{\kappa^2}{4 k \cdot l} \sum_{i=1}^n \frac{p_i^{\mu}p_i^{\nu}}{p_i \cdot k} k^{\rho} k^{\sigma} l_{\beta} \partial_i^{\beta} T_n + \frac{\kappa^2}{4 k \cdot l} \sum_{i=1}^n  p_{i}^{(\mu} \eta^{\nu) (\sigma} k^{\rho)} l_{\beta} \partial_i^{\beta} T_n\notag\\
& \qquad + \frac{\kappa^2}{4 k \cdot l} \sum_{i=1}^n \left(p_{i}^{(\mu} \eta^{\nu) (\sigma} k^{\rho)} l_{\beta} \partial_i^{\beta} - (p_i\cdot k)  l^{(\nu}\eta^{\mu) \sigma} \partial_i^{\rho} \right) T_n\,.
\label{qk.rel}
\end{align}

The first  term on the RHS of \ref{ql.rel} is 
\begin{align}
& -\frac{\kappa^2}{4 k \cdot l} \sum_{i=1}^n \frac{p_i^{\rho}p_i^{\sigma}}{p_i \cdot l} l^{\nu} l^{\mu} k_{\alpha} \partial_i^{\alpha} T_n \notag\\
& = -\frac{\kappa^2}{4 k \cdot l} \sum_{j=1}^n \frac{p_j^{\rho}p_j^{\sigma}}{p_j \cdot l}  l_{\beta} l^{(\mu} \partial_j^{|\beta|} \left( \sum_{i=1}^n p_i^{\nu)} k_{\alpha} \partial_i^{\alpha} T_n \right) + \frac{\kappa^2}{4 k \cdot l} \sum_{i,j=1}^n \frac{p_j^{\rho}p_j^{\sigma}}{p_j \cdot l}  l^{(\mu}  p_i^{\nu)} k_{\alpha}  l_{\beta} \partial_j^{\beta} \partial_i^{\alpha} T_n\;, \notag\\
& = -\frac{\kappa^2}{4 k \cdot l} \sum_{j=1}^n \frac{p_j^{\rho}p_j^{\sigma}}{p_j \cdot l}  l_{\beta} l^{(\mu} \partial_j^{|\beta|} \left( \sum_{i=1}^n (p_i \cdot k) \partial_i^{\nu)} T_n \right) + \frac{\kappa^2}{4 k \cdot l} \sum_{i,j=1}^n \frac{p_j^{\rho}p_j^{\sigma}}{p_j \cdot l}  l^{(\mu}  p_i^{\nu)} k_{\alpha}  l_{\beta} \partial_j^{\beta} \partial_i^{\alpha} T_n\;, \notag\\
&= -\frac{\kappa^2}{4} \sum_{i=1}^n \frac{p_i^{\rho}p_i^{\sigma}}{p_i \cdot l} l^{(\mu} \partial_i^{\nu)} T_n - \frac{i \kappa^2}{4 k \cdot l} \sum_{i,j=1}^n \frac{p_j^{\rho}p_j^{\sigma}}{p_j \cdot l}  l^{(\mu}  J_i^{\nu) \alpha} k_{\alpha}  l_{\beta} \partial_j^{\beta} T_n\;,
\end{align} 
where we made use of \ref{am.id}  in the third line, and \ref{am.def} in the last line. Hence the first terms on the RHS of \ref{ql.rel} and \ref{qk.rel} are 
\begin{align}
& -\frac{\kappa^2}{4 k \cdot l} \sum_{i=1}^n \frac{p_i^{\rho}p_i^{\sigma}}{p_i \cdot l} l^{\nu} l^{\mu} k_{\alpha} \partial_i^{\alpha} T_n = -\frac{\kappa^2}{4} \sum_{i=1}^n \frac{p_i^{\rho}p_i^{\sigma}}{p_i \cdot l} l^{(\mu} \partial_i^{\nu)} T_n - \frac{i \kappa^2}{4 k \cdot l} \sum_{i,j=1}^n \frac{p_j^{\rho}p_j^{\sigma}}{p_j \cdot l}  l^{(\mu}  J_i^{\nu) \alpha} k_{\alpha}  l_{\beta} \partial_j^{\beta} T_n \;, \notag\\
& -\frac{\kappa^2}{4 k \cdot l} \sum_{i=1}^n \frac{p_i^{\mu}p_i^{\nu}}{p_i \cdot k} k^{\rho} k^{\sigma} l_{\beta} \partial_i^{\beta} T_n = -\frac{\kappa^2}{4} \sum_{i=1}^n \frac{p_i^{\mu}p_i^{\nu}}{p_i \cdot k} k^{(\rho} \partial_i^{\sigma)}  T_n - \frac{i \kappa^2}{4 k \cdot l} \sum_{i,j=1}^n \frac{p_i^{\mu}p_i^{\nu}}{p_i \cdot k}  k^{(\rho}  J_j^{\sigma) \beta} l_{\beta}  k_{\alpha} \partial_i^{\alpha} T_n\;.
\label{rel.l1}
\end{align} 

Using \ref{am.id}, we can simplify the second term on the RHS of \ref{ql.rel}
\begin{align}
&  \frac{\kappa^2}{4 k \cdot l} \sum_{i=1}^n  p_{i}^{(\rho} \eta^{\sigma) (\nu} l^{\mu)} k_{\alpha} \partial_i^{\alpha} T_n \notag\\
& =  \frac{\kappa^2}{4 k \cdot l} \sum_{j=1}^n  p_{j}^{(\rho} \partial_j^{\sigma)} \left( \sum_{i=1}^n p_i^{(\nu} l^{\mu)} k_{\alpha} \partial_i^{\alpha} T_n \right) - \frac{\kappa^2}{2 k \cdot l} \sum_{i,j=1}^n p_i^{(\nu} l^{\mu)} p_j^{(\rho} \partial_j^{\sigma)} k_{\alpha} \partial_i^{\alpha} T_n\;, \notag\\
& =  \frac{\kappa^2}{4 k \cdot l} \sum_{i=1}^n  p_{i}^{(\rho} k^{\sigma)} l^{(\mu} \partial_i^{\nu)} T_n + \frac{i \kappa^2}{8 k \cdot l} \sum_{i=1}^n  p_{j}^{\rho} \left(l^{(\mu} J_i^{\nu) \alpha} k_{\alpha}\right) \partial_j^{\sigma} T_n  + \frac{i \kappa^2}{8 k \cdot l} \sum_{i=1}^n  p_{j}^{\sigma} \left(l^{(\mu} J_i^{\nu) \alpha} k_{\alpha}\right) \partial_j^{\rho} T_n\;.
\end{align}

Hence the second terms appearing on the RHS of \ref{ql.rel} and \ref{qk.rel} are
\begin{align}
&  \frac{\kappa^2}{4 k \cdot l} \sum_{i=1}^n  p_{i}^{(\rho} \eta^{\sigma) (\nu} l^{\mu)} k_{\alpha} \partial_i^{\alpha} T_n \notag\\
& =  \frac{\kappa^2}{4 k \cdot l} \sum_{i=1}^n  p_{i}^{(\rho} k^{\sigma)} l^{(\mu} \partial_i^{\nu)} T_n + \frac{i \kappa^2}{8 k \cdot l} \sum_{i=1}^n  p_{j}^{\rho} \left(l^{(\mu} J_i^{\nu) \alpha} k_{\alpha}\right) \partial_j^{\sigma} T_n  + \frac{i \kappa^2}{8 k \cdot l} \sum_{i=1}^n  p_{j}^{\sigma} \left(l^{(\mu} J_i^{\nu) \alpha} k_{\alpha}\right) \partial_j^{\rho} T_n\;, \notag\\
&  \frac{\kappa^2}{4 k \cdot l} \sum_{i=1}^n  p_{i}^{(\mu} \eta^{\nu) (\sigma} k^{\rho)} l_{\beta} \partial_i^{\beta} T_n \notag\\
& =  \frac{\kappa^2}{4 k \cdot l} \sum_{i=1}^n  p_{i}^{(\mu} l^{\nu)} k^{(\rho} \partial_i^{\sigma)} T_n + \frac{i \kappa^2}{8 k \cdot l} \sum_{i=1}^n  p_{i}^{\nu} \left(k^{(\rho} J_j^{\sigma) \beta} l_{\beta}\right) \partial_i^{\mu} T_n  + \frac{i \kappa^2}{8 k \cdot l} \sum_{i=1}^n  p_{i}^{\mu} \left(k^{(\rho} J_j^{\sigma) \beta} l_{\beta}\right) \partial_i^{\nu} T_n\;.
\label{rel.l2}
\end{align}

The last lines of \ref{ql.rel} and \ref{qk.rel} can be readily combined and simplified to
\begin{align}
&\frac{\kappa^2}{4 k \cdot l} \sum_{i=1}^n \left( p_{i}^{(\rho} \eta^{\sigma) (\nu} l^{\mu)} k_{\alpha} \partial_i^{\alpha} + p_{i}^{(\mu} \eta^{\nu) (\sigma} k^{\rho)} l_{\beta} \partial_i^{\beta} - (p_i\cdot k)  l^{(\nu}\eta^{\mu) \sigma} \partial_i^{\rho}  - (p_i\cdot l) k^{(\rho}\eta^{\sigma) \nu} \partial_i^{\mu} \right) T_n   \notag\\
&= \frac{\kappa^2}{4 k \cdot l} \sum_{i=1}^n \left( p_{i}^{(\rho} \eta^{\sigma) (\nu} l^{\mu)} k_{\alpha} \partial_i^{\alpha}  - p_{i}^{\rho} \eta^{\sigma (\nu} l^{\mu)} k_{\alpha} \partial_i^{\alpha} + p_{i}^{(\mu} \eta^{\nu) (\sigma} k^{\rho)} l_{\beta} \partial_i^{\beta} - p_{i}^{\mu} \eta^{\nu (\sigma} k^{\rho)} l_{\beta} \partial_i^{\beta} \right) T_n\;, \notag\\
&= -\frac{\kappa^2}{4 k \cdot l} \sum_{i=1}^n \left( p_{i}^{[\rho} \eta^{\sigma] (\nu} l^{\mu)} k_{\alpha} \partial_i^{\alpha} + p_{i}^{[\mu} \eta^{\nu] (\sigma} k^{\rho)} l_{\beta} \partial_i^{\beta}\right) T_n\;.
\label{rel.l3}
\end{align}

On using \ref{rel.l1}, \ref{rel.l2} and \ref{rel.l3} in \ref{l.sol}, we find
\begin{align}
 \bar{L}^{\mu \nu \rho \sigma}_n &=  \mathcal{E}_L^{\mu \nu \rho \sigma}  -\frac{i \kappa^2}{4} \frac{1}{k \cdot l} \sum_{i,j= 1}^n \left[k_{\alpha} l_{\beta} \left(\frac{p_i^{\mu}  p_i^{\nu}}{p_i \cdot k} k^{(\rho}J_j^{\sigma) \beta} \partial_i^{\alpha} +  \frac{p_j^{\rho}  p_j^{\sigma}}{p_j \cdot l} l^{(\mu}J_i^{\nu) \alpha} \partial_j^{\beta}\right)   \right. \notag\\
&\left. \qquad \qquad - \frac{k_{\alpha}}{2} \left(p_j^{\rho} l^{(\mu}J_i^{\nu) \alpha} \partial_j^{\sigma} +  p_j^{\sigma} l^{(\mu}J_i^{\nu) \alpha} \partial_j^{\rho}\right)  - \frac{l_{\beta}}{2} \left(p_i^{\mu} k^{(\rho}J_j^{\sigma) \beta} \partial_i^{\nu} +  p_i^{\nu} k^{(\rho}J_j^{\sigma) \beta} \partial_i^{\mu}\right) \right] T_n
\notag\\
& \quad  -\frac{\kappa^2}{2}\sum_{i= 1}^n \left[  \frac{p_i^{\mu}  p_i^{\nu}}{p_i \cdot k}  k^{(\sigma} \partial_i^{\rho)}   + \frac{p_i^{\rho}  p_i^{\sigma}}{p_i \cdot l} l^{(\nu} \partial_i^{\mu)}   \right] T_n + \frac{\kappa^2}{4}\sum_{i=1}^n \left[ p_{i}^{(\rho} \eta^{\sigma) (\nu} \partial_i^{\mu)} + p_{i}^{(\mu} \eta^{\nu) (\sigma} \partial_i^{\rho)}\right] T_n \notag\\
 & \qquad  + \frac{\kappa^2}{2 k \cdot l} \sum_{i=1}^n  \left[p_{i}^{(\mu} l^{\nu)} k^{(\rho} \partial_i^{\sigma)} + p_{i}^{(\rho} k^{\sigma)} l^{(\mu} \partial_i^{\nu)} - k^{\rho} k^{\sigma} p_i^{(\mu} \partial_i^{\nu)}- l^{\mu} l^{\nu} p_i^{(\rho} \partial_i^{\sigma)} \right] T_n \notag\\
& \qquad \quad - \frac{\kappa^2}{2 k \cdot l} \sum_{i=1}^n \eta^{\nu (\rho} \eta^{\sigma) \mu} \left[\left( (p_i\cdot k) + \frac{1}{2}(p_i\cdot l) \right) k_{\alpha}\partial_i^{\alpha} + \left( (p_i \cdot l) + \frac{1}{2} (p_i\cdot k) \right) l_{\beta} \partial_i^{\beta} \right] T_n \notag\\
 & \qquad \qquad \qquad + \frac{\kappa^2}{k \cdot l} \sum_{i=1}^n\left[ l^{(\mu} \eta^{\nu) (\rho} p_i^{\sigma)} l_{\beta}\partial_i^{\beta}  + k^{(\rho} \eta^{\sigma) (\mu} p_i^{\nu)} k_{\alpha}\partial_i^{\alpha} \right] T_n \;, 
\label{m2.lsol}
\end{align}
where $\mathcal{E}_L^{\mu \nu \rho \sigma}$ has the expression
\begin{align}
\mathcal{E}_L^{\mu \nu \rho \sigma} &=  \frac{\kappa^2}{4}\sum_{i=1}^n \left[ p_{i}^{(\rho} \eta^{\sigma) [\nu} \partial_i^{\mu]} + p_{i}^{(\mu} \eta^{\nu) [\sigma} \partial_i^{\rho]}\right] T_n -\frac{\kappa^2}{4 k \cdot l} \sum_{i=1}^n \left( p_{i}^{[\rho} \eta^{\sigma] (\nu} l^{\mu)} k_{\alpha} \partial_i^{\alpha} + p_{i}^{[\mu} \eta^{\nu] (\sigma} k^{\rho)} l_{\beta} \partial_i^{\beta}\right) T_n \notag\\
&+ \frac{\kappa^2}{4 k \cdot l}\sum_{i=1}^n \left[ p_{i}^{(\rho} k^{\sigma)}l^{[\nu} \partial_i^{\mu]} + p_{i}^{(\mu} l^{\nu)} k^{[\sigma} \partial_i^{\rho]}\right] T_n -\frac{\kappa^2}{4}\sum_{i= 1}^n \left[  \frac{p_i^{\mu}  p_i^{\nu}}{p_i \cdot k}  k^{[\sigma} \partial_i^{\rho]}   + \frac{p_i^{\rho}  p_i^{\sigma}}{p_i \cdot l} l^{[\nu} \partial_i^{\mu]}   \right] T_n\;.
\label{el}
\end{align}

Before proceeding, we combine the terms in \ref{el} and \ref{enlo2} to find
\begin{align}
&\mathcal{E}^{\mu \nu \rho \sigma}_{\text{NLO2};k,l} + \mathcal{E}_L^{\mu \nu \rho \sigma}= \frac{\kappa^2}{4} \sum_{i=1}^n \left( p_i^{\sigma} \eta^{\rho [\mu} \partial_i^{\nu]} + p_i^{\nu} \eta^{\mu [\rho} \partial_i^{\sigma]} + \frac{p_i^{\mu}p_i^{\nu}}{p_i \cdot k} k^{[\sigma} \partial_i^{\rho]} + \frac{p_i^{\rho}p_i^{\sigma}}{p_i \cdot l} l^{[\nu} \partial_i^{\mu]} \right) T_n \notag\\
& +\frac{\kappa^2}{4 k\cdot l}\sum_{i=1}^n \left(p_i^{(\rho}k^{\sigma)} l^{[\nu} \partial_i^{\mu]} + p_i^{(\mu}l^{\nu)} k^{[\sigma} \partial_i^{\rho]} - (p_i\cdot k) l^{(\mu} \eta^{\nu) [\sigma} \partial_i^{\rho]} - (p_i\cdot l) k^{(\rho}\eta^{\sigma ) [\nu} \partial_i^{\mu]}\right) T_n.
\label{enlo.t}
\end{align}

We can now make repeated use of the conservation of angular momentum for the hard amplitude to write the above terms as a double sum over particles with angular momentum operators. For instance, we have
\begin{align}
\sum_{i=1}^n p_i^{\sigma} \eta^{\rho [\mu} \partial_i^{\nu]} T_n &= \sum_{j=1}^n p_j^{\sigma} \partial_j^{\rho} \left(\sum_{i=1}^n p_i^{[\mu} \partial_i^{\nu]} T_n\right) + \frac{i}{2} \sum_{i,j=1}^n p_j^{\sigma} J_i^{\mu \nu} \partial_j^{\rho} T_n\notag\\
& = \frac{i}{2} \sum_{i,j=1}^n p_j^{\sigma} J_i^{\mu \nu} \partial_j^{\rho} T_n, \notag\\
\sum_{i=1}^n p_i^{(\rho}k^{\sigma)} l^{[\nu} \partial_i^{\mu]} T_n &= \sum_{j=1}^n p_j^{(\rho}k^{\sigma)} l_{\beta}\partial_j^{\beta} \left(\sum_{i=1}^n p_i^{[\nu} \partial_i^{\mu]} T_n\right) - \frac{i}{2} \sum_{i,j=1}^n p_j^{(\rho}k^{\sigma)} J_i^{\mu \nu} l_{\beta}\partial_j^{\beta} T_n\notag\\
& = - \frac{i}{2} \sum_{i,j=1}^n p_j^{(\rho}k^{\sigma)} J_i^{\mu \nu} l_{\beta}\partial_j^{\beta} T_n,
\label{ids}
\end{align}
where we used the relation
\be
p_i^{[\mu} \partial_i^{\nu]} = -\frac{i}{2} J_i^{\mu \nu} = i k_{\alpha} \frac{p_i^{[\mu}J_i^{\nu] \alpha}}{p_i \cdot k}.
\ee

Proceeding similarly for all other terms in \ref{enlo.t} as we did in \ref{ids}, we then find
\begin{align}
&\mathcal{E}^{\mu \nu \rho \sigma}_{\text{NLO2};k,l} + \mathcal{E}_L^{\mu \nu \rho \sigma}= \frac{i \kappa^2}{8}  \sum_{i,j=1}^n \left(p_j^{\sigma} J_i^{\mu \nu} \partial_j^{\rho} - \frac{p_j^{\rho} p_j^{\sigma}}{p_j \cdot l} J_i^{\mu \nu} l_{\beta}\partial_j^{\beta} + p_i^{\nu} J_j^{\rho \sigma} \partial_i^{\mu} - \frac{p_i^{\mu} p_i^{\nu}}{p_i \cdot k} J_j^{\rho \sigma} k_{\alpha}\partial_i^{\alpha}\right) T_n \notag\\
&-\frac{i \kappa^2}{8 k \cdot l} \sum_{i,j=1}^n \left( l^{(\mu} p_i^{\nu)} J_i^{\rho \sigma} k_{\alpha}\partial_i^{\alpha} - p_i \cdot k  J_j^{\rho \sigma}  l^{(\mu} \partial_i^{\nu)} + k^{(\rho} p_j^{\sigma)} J_i^{\mu \nu} l_{\beta}\partial_j^{\beta} - p_j \cdot l  J_i^{\mu \nu} k^{(\rho} \partial_j^{\sigma)} \right) T_n.
\label{esol}
\end{align}

Hence on combining \ref{m2.nlo1f}, \ref{m2.nlo2f} and \ref{m2.lsol}, and on using \ref{esol}, we find 
\begin{align}
&\mathcal{M}_{\text{NLO1};k,l}^{\mu \nu \rho \sigma} + \mathcal{M}_{\text{NLO2};k,l}^{\mu \nu \rho \sigma} + \bar{L}^{\mu \nu \rho \sigma}_n =  \left[\sum_{i,j= 1}^n \left(\mathcal{M}_{\text{NLO}}\right)_{i,j;k,l}^{\mu \nu \rho \sigma} + \sum_{i,j= 1}^n \left(\mathcal{M}'_{\text{NLO}}\right)_{i,j;k,l}^{\mu \nu \rho \sigma} \right] T_n \notag\\
& - \frac{i \kappa^2}{2}\sum_{i=1}^n \left[ \frac{1}{p_i \cdot l} \left(\frac{p_i^{\sigma} J_i^{\rho \nu} l^{\mu}}{2} + p_i^{\nu} \eta^{\mu (\rho} J_i^{\sigma)  \beta} l_{\beta} \right)  +  \frac{1}{p_i \cdot k} \left(\frac{ p_i^{\nu} J_i^{\mu \sigma} k^{\rho}}{2} +  p_i^{\sigma} \eta^{\rho (\mu} J_i^{\nu) \alpha} k_{\alpha} \right) \right. \notag\\
& \left.  \qquad  \qquad \qquad \qquad  \qquad \qquad \qquad   - \frac{p_i^{\mu}  p_i^{\nu}}{p_i \cdot k} \frac{1}{p_i \cdot l}k^{(\sigma} J_i^{\rho) \beta} l_{\beta} - \frac{p_i^{\rho}  p_i^{\sigma}}{p_i \cdot l}\frac{1}{p_i \cdot k} l^{(\nu} J_i^{\mu) \alpha} k_{\alpha} \right] T_n \notag\\
&- \kappa^2 \sum_{i= 1}^n \left[ \frac{p_i^{\mu} p_i^{\nu}}{p_i \cdot k} k^{(\sigma} \partial_i^{\rho)}   +   \frac{p_i^{\rho} p_i^{\sigma}}{p_i \cdot l} l^{(\mu} \partial_i^{\nu)} - \frac{1}{4} \left( \frac{p_i^{\mu} p_i^{\nu}}{p_i \cdot k} \frac{k \cdot l}{p_i \cdot l} p_i^{(\sigma} \partial_i^{\rho)}  +  \frac{p_i^{\rho} p_i^{\sigma}}{p_i \cdot l}\frac{k \cdot l}{p_i \cdot k} p_i^{(\nu} \partial_i^{\mu)}\right)  \right]T_n\notag\\
& \quad + \frac{\kappa^2}{2} \sum_{i= 1}^n \left[ p_i^{(\nu} \eta^{\mu) (\sigma} \partial_i^{\rho)} +  p_i^{(\sigma} \eta^{\rho) (\nu} \partial_i^{\mu)} \right]T_n + \frac{\kappa^2}{k \cdot l} \sum_{i=1}^n\left[ l^{(\mu} \eta^{\nu) (\rho} p_i^{\sigma)} l_{\beta}\partial_i^{\beta}  + k^{(\rho} \eta^{\sigma) (\mu} p_i^{\nu)} k_{\alpha}\partial_i^{\alpha} \right] T_n  \notag\\
& \qquad  + \frac{\kappa^2}{2 k \cdot l} \sum_{i=1}^n  \left[p_{i}^{(\mu} l^{\nu)} k^{(\rho} \partial_i^{\sigma)} + p_{i}^{(\rho} k^{\sigma)} l^{(\mu} \partial_i^{\nu)} - k^{\rho} k^{\sigma} p_i^{(\mu} \partial_i^{\nu)}- l^{\mu} l^{\nu} p_i^{(\rho} \partial_i^{\sigma)} \right] T_n \notag\\
& \qquad \quad - \frac{\kappa^2}{2 k \cdot l} \sum_{i=1}^n \eta^{\nu (\rho} \eta^{\sigma) \mu} \left[\left( (p_i\cdot k) + \frac{1}{2}(p_i\cdot l) \right) k_{\alpha}\partial_i^{\alpha} + \left( (p_i \cdot l) + \frac{1}{2} (p_i\cdot k) \right) l_{\beta} \partial_i^{\beta} \right] T_n,
\label{m2.nlol}
\end{align}

where the double sum terms in the first line of \ref{m2.nlol} are as they appear in \ref{mnlo.1} and \ref{mnlo.2}. To recover the final single sum term in \ref{m2.nlo}, we need to include the contribution from 
\begin{align}
&\frac{\kappa^2}{2}\sum_{i=1}^n\frac{\alpha_i^{\mu \nu \rho \sigma}(p_i, k, l)}{p_i \cdot (k+l)} (k+l)_{\alpha}\partial_i^{\alpha} T_n \notag\\
& \quad =  \frac{\kappa^2}{2} \sum_{i=1}^n \frac{(k+l)_{\alpha}}{p_i \cdot (k+l)} \left[- \frac{k\cdot l}{(p_i\cdot k) (p_i\cdot l)} p_i^{\mu} p_i^{\nu} p_i^{\rho} p_i^{\sigma}  - 2 p_i^{(\mu}\eta^{\nu) (\rho} p_i^{\sigma)}  + 2 \frac{p_i^{\mu} p_i^{\nu}}{p_i\cdot k} p_i^{(\rho} k^{\sigma)} + 2 \frac{p_i^{\rho} p_i^{\sigma}}{p_i\cdot l} p_i^{(\mu} l^{\nu)} \right.  \notag\\
& \left. \qquad + \frac{1}{k\cdot l}  \left( \eta^{\rho (\mu} \eta^{\nu) \sigma} \left((p_i\cdot k)^2 + (p_i\cdot l)^2+ (p_i\cdot k)(p_i\cdot l) \right)   +   p_i^{\mu} p_i^{\nu} k^{\rho} k^{\sigma}      + l^{\mu} l^{\nu} p_i^{\rho} p_i^{\sigma}  \phantom{ \frac{p_i^{\mu} }{p_i \cdot k}} \right. \right.\notag\\
& \left. \left. \phantom{ \frac{p_i^{\mu} }{p_i \cdot k}}   - 2 p_i^{(\mu}l^{\nu)} p_i^{(\rho}  k^{\sigma)}  - 2 (p_i \cdot l) l^{(\mu} \eta^{\nu) (\rho}  p_i^{\sigma)} - 2 (p_i \cdot k) p_i^{(\mu} \eta^{\nu) (\rho}  k^{\sigma)} \right) \right] \partial_i^{\alpha} T_n \,.
\label{alph.def1}
\end{align}

Among the unresolved terms in \ref{m2.nlol} and \ref{alph.def1}, we have contributions from two Born terms, one seagull term, and three-graviton pieces. The two Born terms in \ref{m2.nlol} and \ref{alph.def1} combine to give
\begin{align}
&\frac{\kappa^2}{4} \sum_{i=1}^n  \left[\frac{p_i^{\mu} p_i^{\nu}}{p_i \cdot k} \frac{k \cdot l}{p_i \cdot l} p_i^{(\sigma} \partial_i^{\rho)}  +  \frac{p_i^{\rho} p_i^{\sigma}}{p_i \cdot l}\frac{k \cdot l}{p_i \cdot k} p_i^{(\nu} \partial_i^{\mu)}   - 2 \frac{(k\cdot l) (k+l)_{\alpha}}{(p_i\cdot k) (p_i\cdot l) (p_i \cdot (k+l))} p_i^{\mu} p_i^{\nu} p_i^{\rho} p_i^{\sigma} \partial_i^{\alpha} \right] T_n\notag\\
& = i \frac{\kappa^2}{4} \sum_{i=1}^n \frac{k\cdot l}{(p_i\cdot k) (p_i\cdot l) (p_i \cdot (k+l))}(k+l)_{\alpha} \left(p_i^{\mu} p_i^{\nu}  p_i^{(\sigma} J_i^{\rho) \alpha}  +  p_i^{\rho} p_i^{\sigma} p_i^{(\nu} J_i^{\mu) \alpha} \right) T_n\;,  
\label{b.1}
\end{align}
and 
\begin{align}
&- \kappa^2 \sum_{i= 1}^n \left[ \frac{p_i^{\mu} p_i^{\nu}}{p_i \cdot k} k^{(\sigma} \partial_i^{\rho)}   +   \frac{p_i^{\rho} p_i^{\sigma}}{p_i \cdot l} l^{(\mu} \partial_i^{\nu)} - \frac{(k+l)_{\alpha}}{p_i \cdot (k+l)} \left(\frac{p_i^{\mu} p_i^{\nu}}{p_i\cdot k} p_i^{(\rho} k^{\sigma)} + \frac{p_i^{\rho} p_i^{\sigma}}{p_i\cdot l} p_i^{(\mu} l^{\nu)} \right) \partial_i^{\alpha} \right] T_n \notag\\
&\qquad \qquad = - i \kappa^2 \sum_{i= 1}^n  \frac{(k+l)_{\alpha}}{p_i \cdot (k+l)} \left(\frac{p_i^{\mu} p_i^{\nu}}{p_i\cdot k} k^{(\rho} J_i^{\sigma) \alpha} + \frac{p_i^{\rho} p_i^{\sigma}}{p_i\cdot l} l^{(\mu} J_i^{\nu) \alpha} \right) T_n\;.
\label{b.2}
\end{align}

Likewise, from the seagull terms in \ref{m2.nlol} and \ref{alph.def1}, we get
\begin{align}
& \frac{\kappa^2}{2} \sum_{i= 1}^n \left[ p_i^{(\nu} \eta^{\mu) (\sigma} \partial_i^{\rho)} +  p_i^{(\sigma} \eta^{\rho) (\nu} \partial_i^{\mu)} - 2 \frac{(k+l)_{\alpha}}{p_i \cdot (k+l)} p_i^{(\mu}\eta^{\nu) (\rho} p_i^{\sigma)} \partial_i^{\alpha} \right]T_n \notag\\
&\qquad \quad  = i \frac{\kappa^2}{2} \sum_{i= 1}^n \frac{(k+l)_{\alpha}}{p_i \cdot (k+l)}  \left( p_i^{(\nu} \eta^{\mu) (\sigma} J_i^{\rho) \alpha} +  p_i^{(\sigma} \eta^{\rho) (\nu} J_i^{\mu) \alpha}\right]T_n\;.
\label{sgl}
\end{align}

All other terms from combining \ref{m2.nlol} and \ref{alph.def1} correspond to the following three-graviton vertex pieces:
\begin{align}
& \frac{\kappa^2}{k \cdot l} \sum_{i=1}^n\left[ l^{(\mu} \eta^{\nu) (\rho} p_i^{\sigma)} l_{\beta}\partial_i^{\beta}  + k^{(\rho} \eta^{\sigma) (\mu} p_i^{\nu)} k_{\alpha}\partial_i^{\alpha} -  \frac{(k+l)_{\alpha}}{p_i \cdot (k+l)}  \left( (p_i \cdot l) l^{(\mu} \eta^{\nu) (\rho}  p_i^{\sigma)} + (p_i \cdot k) p_i^{(\mu} \eta^{\nu) (\rho}  k^{\sigma)} \right)  \partial_i^{\alpha} \right] T_n \notag\\
& \qquad \quad  = i\frac{\kappa^2}{k \cdot l} \sum_{i=1}^n \frac{(k+l)_{\alpha}}{p_i \cdot (k+l)} \left[ l^{(\mu} \eta^{\nu) (\rho} p_i^{\sigma)} l_{\beta} J_i^{\beta \alpha}  + k^{(\rho} \eta^{\sigma) (\mu} p_i^{\nu)} k_{\beta} J_i^{\beta \alpha} \right]  T_n \;, \notag\\
& \qquad \qquad \qquad   = - i \kappa^2 \sum_{i= 1}^n \frac{k_{\alpha} l_{\beta}}{p_i \cdot (k+l)} \frac{1}{k \cdot l} \left( l^{(\mu} \eta^{\nu)(\rho} p_i^{\sigma)} -  k^{(\rho} \eta^{\sigma) (\mu} p_i^{\nu)} \right) J_i^{\alpha \beta} T_n\;,
\label{gp.1}
\end{align}
\begin{align}
 &\frac{\kappa^2}{2 k \cdot l} \sum_{i=1}^n  \left[p_{i}^{(\mu} l^{\nu)} k^{(\rho} \partial_i^{\sigma)} +  p_{i}^{(\rho} k^{\sigma)} l^{(\mu} \partial_i^{\nu)}  - \frac{(k+l)_{\alpha}}{p_i \cdot (k+l)} 2 p_i^{(\mu}l^{\nu)} p_i^{(\rho}  k^{\sigma)} \partial_i^{\alpha} \right] T_n \notag\\
 & \qquad = i \frac{\kappa^2}{2 k \cdot l} \sum_{i=1}^n \frac{(k+l)_{\alpha}}{p_i \cdot (k+l)} \left[p_{i}^{(\mu} l^{\nu)} k^{(\rho} J_i^{\sigma) \alpha} +  p_{i}^{(\rho} k^{\sigma)} l^{(\mu} J_i^{\nu) \alpha} \right] T_n\;,
 \label{gp.2}
\end{align}
\begin{align}
&- \frac{\kappa^2}{2 k \cdot l} \sum_{i=1}^n  \left[k^{\rho} k^{\sigma} p_i^{(\mu} \partial_i^{\nu)} + l^{\mu} l^{\nu} p_i^{(\rho} \partial_i^{\sigma)} - \frac{(k+l)_{\alpha}}{p_i \cdot (k+l)} \left(p_i^{\mu} p_i^{\nu} k^{\rho} k^{\sigma} + l^{\mu} l^{\nu} p_i^{\rho} p_i^{\sigma}  \right) \partial_i^{\alpha} \right] T_n  \notag\\
& \qquad = - i\frac{\kappa^2}{2 k \cdot l} \sum_{i=1}^n  \frac{(k+l)_{\alpha}}{p_i \cdot (k+l)} \left(k^{\rho} k^{\sigma} p_i^{(\mu} J_i^{\nu) \alpha} + l^{\mu} l^{\nu} p_i^{(\rho} J_i^{\sigma) \alpha} \right) T_n\;, 
\label{gp.3}
\end{align}
and
\begin{align}
&- \frac{\kappa^2}{2 k \cdot l} \sum_{i=1}^n \eta^{\nu (\rho} \eta^{\sigma) \mu} \left[\left(p_i\cdot k + \frac{p_i \cdot l}{2} \right) k_{\beta} \partial_i^{\beta} + \left(p_i\cdot l + \frac{p_i \cdot k}{2} \right) l_{\beta} \partial_i^{\beta} \right. \notag\\
&\left. \qquad \qquad - \frac{ (k+l)_{\alpha}}{p_i \cdot (k+l)}  \left( (p_i\cdot k) \left(p_i\cdot k + \frac{p_i \cdot l}{2} \right)  + (p_i\cdot l) \left(p_i\cdot l + \frac{p_i \cdot k}{2} \right) \right) \partial_i^{\alpha} \right]  T_n  \notag\\
& \quad =  -i \frac{\kappa^2}{2 k\cdot l} \sum_{i=1}^n  \eta^{\nu (\rho} \eta^{\sigma) \mu}  \frac{ (k+l)_{\alpha}}{p_i \cdot (k+l)} \left[ \left(p_i\cdot l + \frac{p_i \cdot k}{2} \right) l_{\beta} J_i^{\beta \alpha} +  \left(p_i\cdot k + \frac{p_i \cdot l}{2} \right) k_{\beta} J_i^{\beta \alpha} \right]  T_n \;, \notag\\
& \qquad \quad =  -i \frac{\kappa^2}{4 k\cdot l} \sum_{i=1}^n  \eta^{\nu (\rho} \eta^{\sigma) \mu}  \frac{ k_{\alpha}l_{\beta}}{p_i \cdot (k+l)} p_i\cdot (k-l) J_i^{\alpha \beta} T_n\;. 
\label{gp.4}
\end{align}
Finally, on combining \ref{m2.nlol} and \ref{alph.def1}, and using \ref{b.1} - \ref{gp.4}, we find \ref{m2.nlo}.

\bibliographystyle{jhep}

\bibliography{refs.bib}	

\end{document}